\documentclass[a4paper,twocolumn,showpacs,notitlepage,floatfix,nofootinbib,superscriptaddress,prc]{revtex4-1}

\usepackage{graphicx}
\usepackage[hypertexnames=false]{hyperref}
\usepackage{amsfonts}
\usepackage{amsmath}
\usepackage{float}
\usepackage{amssymb}
\usepackage{epstopdf}
\usepackage{xspace}
\usepackage{color}
\usepackage[utf8]{inputenc}

\newcommand{\dd}{{\rm d}}

\newcommand{\UNIT}[1]{\ensuremath{\,{\rm #1}}\xspace}

\newcommand{\MeV}{\UNIT{MeV}}
\newcommand{\GeV}{\UNIT{GeV}}

\newcommand{\fm}{\UNIT{fm}}
\newcommand{\mb}{\UNIT{mb}}
\newcommand{\proz}{\UNIT{\%}}

\definecolor{magenta}{cmyk}{0,1,0,0}

\begin{document}

\title{Exploring the applicability of dissipative fluid dynamics to small systems by comparison to the Boltzmann equation}
\author{K.\ Gallmeister}
\affiliation{Institut f\"ur Theoretische Physik, Johann Wolfgang Goethe-Universit\"at,
Max-von-Laue-Str.\ 1, D-60438 Frankfurt am Main, Germany}

\author{H.\ Niemi}
\affiliation{Institut f\"ur Theoretische Physik, Johann Wolfgang Goethe-Universit\"at,
Max-von-Laue-Str.\ 1, D-60438 Frankfurt am Main, Germany}
\affiliation{Department of Physics, University of Jyv\"askyl\"a, P.O.\ Box 35, FI-40014 University of Jyv\"askyl\"a, Finland}
\affiliation{Helsinki Institute of Physics, P.O.\ Box 64, FI-00014 University of Helsinki, Finland}

\author{C.\ Greiner}
\affiliation{Institut f\"ur Theoretische Physik, Johann Wolfgang Goethe-Universit\"at,
Max-von-Laue-Str.\ 1, D-60438 Frankfurt am Main, Germany}

\author{D.H.\ Rischke}
\affiliation{Institut f\"ur Theoretische Physik, Johann Wolfgang Goethe-Universit\"at,
Max-von-Laue-Str.\ 1, D-60438 Frankfurt am Main, Germany}
\affiliation{Department of
Modern Physics, University of Science and Technology of China, Hefei, Anhui 230026, China}


\begin{abstract}
\begin{description}
\item[Background] Experimental data from heavy-ion experiments at RHIC-BNL and LHC-CERN are quantitatively
described using relativistic fluid dynamics. Even p+A and p+p collisions show signs of collective behavior describable
in the same manner. Nevertheless, small system sizes and large gradients strain the limits of applicability of
fluid-dynamical methods.
\item[Purpose] The range of applicability of fluid dynamics for the description of the collective behavior,
and in particular of the elliptic flow, of small systems needs to be explored.
\item[Method] Results of relativistic fluid-dynamical simulations are compared with solutions of the Boltzmann equation
in a longitudinally boost-invariant picture. As initial condition, several different transverse energy-density profiles
for equilibrated matter are investigated.
\item[Results] While there is overall a fair agreement of energy- and particle-density profiles, components of the
shear-stress tensor are more sensitive to details of the implementation. The highest sensitivity
is exhibited by quantities influenced by properties of the medium at freeze-out.
\item[Conclusions] For some quantities, like the shear-stress tensor, agreement between fluid dynamics and transport 
theory extends into regions of
Knudsen numbers and inverse Reynolds numbers where relativistic fluid dynamics is believed to fail.
\end{description}
\end{abstract}

\date{\today }
\maketitle

\section{Introduction}
\label{sec:Introduction}

One of the main goals of heavy-ion experiments at RHIC-BNL and LHC-CERN is to create hot and dense
strong-interaction matter, i.e., a quark-gluon plasma (QGP), and study its properties. The matter properties cannot,
however, be directly measured, but need to be inferred indirectly from the experimental data through dynamical models
of the collision. One of the main tools for understanding the dynamics and extracting the matter properties is
relativistic fluid dynamics. It is a natural framework for such a task, as it is the limit where the dynamics is governed
entirely by a few macroscopic properties, such as the equation of state and the transport coefficients, which are the
very properties one is interested in.

Comparisons of fluid-dynamical model calculations with experimental data, especially for the azimuthal structure of
hadron spectra, have revealed that the ratios of shear and bulk viscosity to entropy density must be small in order 
to be consistent with the data~\cite{Romatschke:2007mq, Huovinen:2013wma, Heinz:2013th, Gale:2012rq, Ryu:2017qzn, Niemi:2015qia, Bernhard:2016tnd}.
The good agreement between fluid-dynamical models and experimental data is one of the
strongest evidences that a small droplet of QGP is indeed formed in these collisions.
However, such a system is extremely small, of the order of the size of an atomic nucleus, and it is
non-trivial that fluid dynamics can be applied for such small systems.

The fluid-dynamical equations of motion can be written as a power series in two types of quantities,
the Knudsen number $\rm Kn$ and the inverse Reynolds number $\mathrm{R}^{-1}$ \cite{Denicol:2012cn}.
The Knudsen number is a ratio of a microscopic and a macroscopic length or time scale.
For example, in the Boltzmann equation the microscopic scale is essentially the mean free path of the
particles $\lambda_{\rm mfp} = 1/ (n \sigma)$, where $n$ is the particle density in the local rest frame and
$\sigma$ is the cross section. The (local) macroscopic scale is given by gradients of fluid-dynamical variables,
such as energy density or flow velocity. On the other hand, the inverse Reynolds number measures the deviation
from local thermal
equilibrium, and can be expressed as a ratio of a dissipative quantity, such as the shear-stress tensor $\pi^{\mu\nu}$,
and an equilibrium quantity, e.g.~the thermodynamic pressure. When both measures are sufficiently small, i.e.,
gradients are sufficiently weak, and the system is sufficiently close to local thermal equilibrium, one expects that fluid
dynamics becomes a good approximation to describe the space-time evolution of the matter.

The Knudsen and inverse Reynolds numbers assumed in heavy-ion and proton-nucleus (p+A) 
collisions were estimated in
Ref.~\cite{Niemi:2014wta}. Even for heavy-ion collisions the corresponding values can be so large that one may
doubt the applicability of fluid dynamics.
On the other hand, even smaller systems like those formed in p+A or event-by-event
proton-proton (p+p) collisions experimentally exhibit signs of collective behavior that can be at least 
qualitatively described by
fluid-dynamical models~\cite{Mantysaari:2017cni, Shen:2016zpp, Bozek:2014era, Bozek:2013ska, Werner:2013ipa, Romatschke:2015gxa, Qin:2013bha, Bzdak:2013zma}.
All these observations raise the question about the conditions when fluid dynamics is a sufficiently good approximation
for describing the space-time evolution of these systems, so that the matter properties can be reliably extracted by
comparing fluid-dynamical models to experimental data.

One way to address this question is to test the validity of fluid dynamics in situations where the underlying microscopic
theory can be explicitly solved. The Boltzmann equation represents such a theory, as it can be numerically solved in
situations that resemble those formed in heavy-ion collisions~\cite{Xu:2004mz}. Furthermore, the fluid-dynamical limit of
the Boltzmann equation is known, i.e., fluid-dynamical equations of motion and the pertinent transport coefficients can be
derived in various approximations by employing the method of moments
\cite{Israel:1979wp, Betz:2010cx, Denicol:2012cn, Denicol:2012vq, Molnar:2013lta}.

Currently, there are several works that compare the solutions of fluid dynamics to the solutions of the Boltzmann equation
in simple systems. In Ref.~\cite{Molnar:2004yh} the second Fourier coefficient of the azimuthal particle spectrum,
the so-called elliptic-flow coefficient, was compared from calculations within ideal fluid dynamics and the Boltzmann
equation starting from initial conditions resembling a heavy-ion collision. It was concluded that the cross section must be
unrealistically large in order to obtain an agreement between the two approaches. However, in that work, viscous effects
were neglected, and within the current understanding of heavy-ion collisions, ideal fluid dynamics is not sufficient to
describe experimental data. The Israel-Stewart formulation of fluid dynamics has been compared to the Boltzmann
equation in the $(0+1)$ dimensional boost-invariant expansion scenario in
Refs.~\cite{Huovinen:2008te, Denicol:2010xn, Kurkela:2015qoa}, and for $(1+1)$ dimensional shock waves in
Refs.~\cite{Bouras:2009nn, Bouras:2010hm, Denicol:2012vq}. Furthermore, in the relaxation-time approximation (RTA),
both the Boltzmann equation becomes much easier to solve and also the derivation of fluid dynamics
from the underlying microscopic theory, in this case the Boltzmann equation, becomes much
simpler. Here, many comparisons between the Boltzmann equation in RTA and Israel-Stewart fluid dynamics, and also
its extension, so-called anisotropic fluid dynamics, exist
\cite{Jaiswal:2013vta, Florkowski:2013lya, Denicol:2014xca, Denicol:2014tha, Heinz:2015gka, Tinti:2015xwa, Molnar:2016gwq, Chattopadhyay:2018apf}.

While in the above mentioned cases the numerical solution of the Boltzmann equation is readily obtained, it is not
clear how the results of these comparisons can be translated to more realistic scenarios and in particular to
experimental observables. The comparisons between viscous fluid dynamics and the Boltzmann equation were made in
azimuthally symmetric cases, where the transverse flow has only a radial component. However, the most direct constraints
for the viscosity coefficients come from the azimuthal structure of the transverse-momentum spectra in heavy-ion collisions,
which in turn result from the azimuthal asymmetries of the transverse-flow field.

The aim of this work is to compare the solutions of the full Boltzmann equation with the
corresponding fluid-dynamical solution in situations that resemble the system formed in heavy-ion or p+A collisions,
allowing to access experimental observables like transverse-momentum spectra and flow coefficients. In particular, in this
work, the Boltzmann equation is solved by the ``Boltzmann Approach to Multiparticle collisions (BAMPS)'' framework
\cite{Xu:2004mz} in several different boost-invariant geometries with different values for a constant isotropic cross section.
These solutions are compared to numerical solutions of the fluid-dynamical equations of motion within a theory derived by
Israel and Stewart using the 14-moment approximation \cite{Israel:1979wp}.
For convenience and simplicity, the initial conditions are taken to be thermal at some initial proper time,
although this is not a realistic assumption for the initial stage of heavy-ion collisions.

The paper is organized as follows: In Sec.~\ref{sec:fluids} a short introduction to the Boltzmann equation and relativistic
fluid dynamics is given, followed by Sec.~\ref{sec:numerical}, where the numerical methods to solve the fluid-dynamical
equations of motion and the Boltzmann equation are described. The initial conditions and the set-up to compare the
space-time evolution are given in Sec.~\ref{sec:setup}, and the comparisons of fluid-dynamical quantities from both
approaches are shown in Sec.~\ref{sec:Comparisons}. In Sec.~\ref{sec:Freezeout}, comparisons of transverse-momentum
spectra and elliptic flow follow. A discussion and conclusions are given in Sec.~\ref{sec:Conclusions}.
Units are $\hbar = c = k_B=1$, the metric tensor is  $g_{\mu \nu} = {\rm diag} (+,-,-,-)$.

\section{Relativistic fluid dynamics from the Boltzmann equation}
\label{sec:fluids}

Fluid dynamics is an effective theory describing the long-wavelength, small-frequency dynamics of a system. The basic
equations of fluid dynamics are the conservation of energy, momentum, and charge number,
\begin{align}
 \partial_\mu T^{\mu\nu} &= 0\; ,
 \label{eq:econs}\\
 \partial_\mu N^{\mu} &= 0\;,
 \label{eq:ncons}
\end{align}
where the energy-momentum tensor $T^{\mu\nu}$ and the charge 4-current $N^\mu$ can be decomposed
with respect to the 4-velocity $u^\mu$ as
\begin{align}
 T^{\mu\nu} &= e u^\mu u^\nu - P\Delta^{\mu\nu} + \pi^{\mu\nu}\;,
 \label{eq:Tmunu_decomposition}\\
 N^{\mu} &= n u^\mu + n^\mu\;,
 \label{eq:Nmu_decomposition}
\end{align}
where $e=u_\mu u_\nu T^{\mu\nu}$ is the energy density in the local rest frame (LRF),
$P = -\frac{1}{3}\Delta_{\mu\nu}T^{\mu\nu}$ is the isotropic pressure, $\pi^{\mu\nu} = T^{\langle\mu\nu\rangle}$ is the
shear-stress tensor, $n = u_\mu N^\mu$ is the charge density in the LRF, and
$n^\mu = \Delta^{\mu}_{\nu}N^\nu$ is the charge-diffusion current. In this work, the 4-velocity is defined
in the Landau frame, i.e., it is the time-like eigenvector of the energy-momentum tensor,
$T^{\mu\nu}u_\nu = e u^\mu$. Furthermore,
$\Delta^{\mu\nu} = g^{\mu\nu} - u^\mu u^\nu$ is the projection operator onto the 3-space orthogonal to $u^\mu$.
Usually, the isotropic pressure is decomposed as $P=p_0 + \Pi$, with $p_0 = p_0(e_0,n_0)$ being the thermodynamic
pressure of a fictitious system in local equilibrium that has the
same energy density, $e_0 =e$, and charge density, $n_0 = n$, as
the system under consideration (the so-called Landau matching conditions),
and with $\Pi$ being the bulk viscous pressure.
Angular brackets denote the traceless and symmetric part of a tensor that is orthogonal to $u^\mu$, i.e.,
\begin{equation}
 T^{\langle\mu\nu\rangle} = \frac{1}{2}\left(\Delta^{\mu}_{\alpha} \Delta^{\nu}_{\beta} +
 \Delta^{\mu}_{\beta} \Delta^{\nu}_{\alpha} - \frac{2}{3}\Delta^{\mu\nu}\Delta_{\alpha\beta}\right)T^{\alpha\beta}\; .
 \label{eq:transverse_traceless_projector}
\end{equation}

The conservation laws do not form a closed system; additional assumptions are needed in order to close the set of
equations. In principle, the simplest dissipative theory is a relativistic generalization of the Navier-Stokes theory, where
the dissipative currents are proportional to the first-order gradients of 4-velocity and the so-called
thermal potential $\alpha_0$,
\begin{align}
 \pi_{NS}^{\mu\nu} &= 2\eta \sigma^{\mu\nu} \equiv 2\eta \nabla^{\left\langle \mu \right.} u^{\left.\nu\right\rangle}\;, \\
 n_{NS}^{\mu} &= \kappa_n \nabla^{\mu} \alpha_0\;.
 \label{eq:NavierStokes}
\end{align}
Here, $\nabla^{\mu} = \Delta^{\mu\nu} \partial_\nu$ and the thermal potential is defined as
$\alpha_0 \equiv \mu_0/T_0$, where $\mu_0$ and $T_0$ are the charge
chemical potential and temperature of the fictitious
local equilibrium system, respectively (i.e., the entropy density of that system is
$s_0 = \partial p_0/\partial T_0$, while the charge density is $n_0 = \partial p_0/\partial \mu_0$).
The proportionality constants $\eta$ and $\kappa_n$ are the shear-viscosity and charge-diffusion
coefficients, respectively. It is well known that this theory suffers from the problem that the signal-propagation speed is not
limited by the speed of light, and that this acausality leads to the instability of an equilibrated system when viewed from
a moving reference frame
\cite{Hiscock:1983zz, Hiscock:1985zz, Hiscock:1987zz, Denicol:2008ha, Pu:2009fj}. This cannot be cured by
accounting for higher-order terms in gradients. Rather, the gradient expansion itself
is not guaranteed to converge~\cite{Heller:2013fn, Denicol:2016bjh}.

The simplest well-behaved relativistic fluid-dynamical
theory is that of transient or second-order fluid dynamics~\cite{Israel:1979wp, Pu:2009fj}. In these type of
theories the basic equations governing the evolution of $n^\mu$ and $\pi^{\mu\nu}$ can, in the absence of
bulk viscosity, be written as
\begin{align}
\label{eq:IS_equations_diffusion}
\tau _{n}\dot{n}^{\left\langle \mu \right\rangle }+n^{\mu }& =\kappa_{n}\nabla ^{\mu }\alpha_{0}
- n_{\nu }\omega ^{\nu \mu } - \delta _{nn}n^{\mu }\theta \notag \\
&+\ell _{n\pi }\Delta ^{\mu \nu }\nabla_{\lambda }\pi _{\nu }^{\lambda }
-\tau _{n\pi }\pi^{\mu \nu }\nabla_{\nu }p_{0} \notag \\
& -\lambda _{nn}n_{\nu }\sigma ^{\mu \nu } -\lambda _{n\pi }\pi ^{\mu \nu }\nabla _{\nu }\alpha_{0}\;, \\
\label{eq:IS_equations_shear}
\tau _{\pi }\dot{\pi}^{\left\langle \mu \nu \right\rangle }+\pi ^{\mu \nu }&
=2\eta \sigma ^{\mu \nu } - 2\pi _{\lambda }^{\left\langle \mu \right. }\omega
^{\left. \nu \right\rangle \lambda }-\delta _{\pi \pi }\pi ^{\mu \nu }\theta \notag \\
&-\tau _{\pi \pi }\pi ^{\lambda \left\langle \mu \right. }\sigma _{\lambda}^{\left. \nu \right\rangle }
-\tau _{\pi n}n^{\left\langle \mu \right. }\nabla ^{\left. \nu \right\rangle}p_{0} \notag \\
&+\ell _{\pi n}\nabla ^{\left\langle \mu \right. }n^{\left. \nu\right\rangle }
+\lambda _{\pi n}n^{\left\langle \mu \right. }\nabla ^{\left. \nu\right\rangle }\alpha_{0}\;,
\end{align}
where the expansion scalar is
$\theta = \nabla_{\mu}u^{\mu}$. The vorticity is defined as $\omega^{\mu\nu}
= (\nabla^{\mu}u^{\nu} - \nabla^{\nu}u^{\mu})/2$. In principle, this theory comprises the contributions
up to second order in gradients to the evolution of
$\pi^{\mu\nu}$ and $n^{\mu}$. The main feature of the theory is that the dissipative quantities relax to the
values given by the gradient expansion within the time scales $\tau_\pi$ and $\tau_n$. In general,
$\tau_\pi$ and $\tau_n$ are scales of the underlying microscopic theory~\cite{Denicol:2011fa}.
The various coefficients appearing in the equations can be obtained by matching to the underlying microscopic theory.

If one takes the Boltzmann equation as the microscopic theory, it is possible to explicitly calculate all transport coefficients
appearing in Eqs.\ (\ref{eq:IS_equations_diffusion}) and (\ref{eq:IS_equations_shear}).
The Boltzmann equation is an evolution equation for the single-particle momentum-distribution function $f_{\mathbf{k}}(x)$,
\begin{equation}
 k^\mu\partial_{\mu}f_{\mathbf{k}}(x) = C\left[f\right]\;,
\end{equation}
where $k^\mu= (k^0,\mathbf{k})$
is the (on-shell) 4-momentum of the particle, and $C\left[f\right]$ is the collision integral.
In the following, we consider a gas of massless classical particles undergoing elastic binary collisions only, which conserve
particle number, energy, and momentum.
In this case, conservation of charge is synonymous with conservation of particle number. Moreover,
the bulk viscosity is zero and the isotropic pressure is equal to the thermodynamic equilibrium
pressure $P = p_0 = e_0/3$. 

In thermal equilibrium the distribution function is given by
\begin{equation}
 f_{0\mathbf{k}} = \exp(\alpha_0 - \beta_0 E_\mathbf{k})\;,
 \label{eq:LTE_distribution}
\end{equation}
where $\beta_0 = 1/T_0$ is the inverse temperature, and $E_\mathbf{k} = k^{\mu}u_{\mu}$ is the energy of the
particle in the LRF.
Fluid dynamics can be obtained by considering deviations $\delta f_{\mathbf{k}}(x)$ from
$f_{0\mathbf{k}}$,
\begin{equation}
 f_{\mathbf{k}} = f_{0\mathbf{k}} + \delta f_{\mathbf{k}} \;,
\label{eq:fk}
\end{equation}
and expanding $\delta f_{\mathbf{k}}$ in a complete basis of irreducible tensors $k^{\left\langle \mu_1\right.} \cdots
k^{\left. \mu_\ell \right\rangle}$ [where the angular brackets denote a generalization of the rank-4 projection
tensor in Eq.\ (\ref{eq:transverse_traceless_projector}) to rank $2 \ell$, for more
details, see Ref.\ \cite{Denicol:2012cn}]. The coefficients of this expansion are the so-called
irreducible moments of $\delta f_{\mathbf{k}}$,
\begin{equation}
 \rho_i^{\mu_1 \cdots \mu_\ell} = \int \frac{\dd^3\mathbf{k}}{\left( 2\pi \right)^3 k^0} E_{\mathbf{k}}^i
 k^{\langle \mu_1}\cdots k^{\mu_\ell\rangle}\delta f_{\mathbf{k}}\;,
\end{equation}
which can be taken as dynamical variables instead of the distribution function $f_{\mathbf{k}}$ itself.
Note that $\pi^{\mu\nu} = \rho_0^{\mu\nu}$ and $n^{\mu}= \rho_0^{\mu}$. In order to derive the
fluid-dynamical equations of motion in terms of quantities that solely appear in $T^{\mu\nu}$ and $N^{\mu}$, the infinite
set of irreducible moments needs to be
truncated and reduced to the fluid-dynamical degrees of freedom. An implicit assumption here is that the system is
sufficiently close to local thermal equilibrium so that the deviations from equilibrium can be expressed by only a
few dissipative quantities.

A simple way to do this is the so-called 14-moment approximation. Here, the expansion of $\delta f_{\mathbf{k}}$
is truncated after the first 14 lowest-rank irreducible tensors~\cite{Israel:1979wp}. Then, by matching this expansion
to $T^{\mu \nu}$ and $N^\mu$, the irreducible tensors are expressed in terms of the dissipative
quantities appearing in $T^{\mu \nu}$ and $N^\mu$, in this case the shear-stress tensor and the particle-diffusion current.
Hence, $\delta f_{\mathbf{k}}$ can be written as
\begin{equation}
\begin{split}
 \delta f_{\mathbf{k}} = & f_{0\mathbf{k}}\left(\frac{1}{8p_0 T^2}k_{\langle\mu}k_{\nu\rangle} \pi^{\mu\nu}\right. \\
 & \left. - \frac{5}{p_0} k_{\mu} n^\mu + \frac{1}{p_0 T}E_{\mathbf{k}}  k_{\mu} n^\mu \right)\;.
\label{eq:deltaf}
\end{split}
\end{equation}
Inserting this ansatz into the Boltzmann equation leads to the fluid-dynamical equations of motion,
Eqs.~(\ref{eq:IS_equations_diffusion}) and (\ref{eq:IS_equations_shear}). It should be noted that this approach only
gives approximate values for the transport coefficients. The reason for this is that directly truncating the moment
expansion does not correspond to a truncation of given order in the gradient expansion. In order to account for all 
terms of first and second order in the gradient expansion we need to resum an infinite number of moments,
see Ref.~\cite{Denicol:2012cn} for details of
this procedure. The resummation also produces new terms that can violate the
causality of the theory. In order to overcome such problems it would be necessary to extend the basis of
dynamical variables beyond those appearing in the energy-momentum tensor and
particle 4-current~\cite{Denicol:2012vq}. The investigation of these extensions of the fluid-dynamical theory is left as
future work. Here, we take the Israel-Stewart 14-moment approximation as fluid-dynamical theory
[however, with the prescription of Ref.\ \cite{Denicol:2010xn} to compute the transport coefficients], which is also the
most widely used theory to describe the dynamics of ultrarelativistic heavy-ion collisions. The coefficients in this
approximation for the massless, classical gas with constant cross section $\sigma$ are shown in
Tabs.~\ref{tab:diffusion} and \ref{tab:shear}, respectively \cite{Denicol:2010xn, Denicol:2012es}.
\begin{table}[h]
\begin{center}
\begin{tabular}{|c|c|c|c|c|c|c|}
\hline
$\kappa _{n}$ & $\tau _{n}/\lambda _{\mathrm{mfp}}$ & $\delta _{nn}/\tau_{n}$
& ${\lambda }_{nn}/\tau _{n}$ & ${\lambda }_{n\pi }/\tau _{n}$ & $\ell _{n\pi }/\tau _{n}$
& $\tau _{n\pi }/\tau _{n}$ \\ \hline
${3}/\left( 16{\sigma }\right) $ & $9/4$ & $1$ & $3/5$ & $\beta _{0}/{20}$ &
${\beta _{0}}/{20}$ & ${\beta _{0}}/{(80p_0)}$ \\ \hline
\end{tabular}
\end{center}
\caption{{\protect\small The coefficients of the particle-diffusion current in the 14-moment approximation.}}
\label{tab:diffusion}
\end{table}
\begin{table}[h]
\begin{center}
\begin{tabular}{|c|c|c|c|c|c|c|}
\hline
$\eta $ & $\tau _{\pi }/\lambda _{\mathrm{mfp}}$ & ${\tau }_{\pi \pi }/\tau_{\pi}$
& ${\lambda }_{\pi n}/\tau _{\pi}$ & $\delta _{\pi \pi }/\tau_{\pi} $ & $\ell _{\pi n}/\tau _{\pi}$
& $\tau _{\pi n}/\tau _{\pi}$ \\
\hline
${4}/({3\sigma \beta _{0}})$ & $5/3$ & $10/7$ & $0$ & $4/3$ & $0$ & $0$ \\
\hline
\end{tabular}
\end{center}
\caption{{\protect\small The coefficients of the shear-stress tensor in the 14-moment approximation.}}
\label{tab:shear}
\end{table}

The requirements that the dynamics of the system is dominated by long-wavelength excitations and that it is
sufficiently close to local equilibrium manifest in Eqs.~(\ref{eq:IS_equations_diffusion}) and (\ref{eq:IS_equations_shear})
in such a way that they can be considered as a power series in two types of quantities, Knudsen number and inverse
Reynolds number. The Knudsen number can in general be defined as a ratio of a microscopic and a macroscopic time or
length scale, ${\rm Kn} = \ell_{\rm micr}/L_{\rm macr}$, and thus quantifies the separation between these two types of
scales. In the Boltzmann equation the microscopic scales are proportional and of the same order as the mean free path,
and the slowest of these scales appear as relaxation times in transient fluid dynamics, $\tau_n$ and $\tau_\pi$ in
Eqs.~(\ref{eq:IS_equations_diffusion}) and (\ref{eq:IS_equations_shear}), respectively. The local macroscopic time scales
can be calculated e.g.~from the expansion rate $L_{\rm macr}^{-1} = \theta$, or from the energy density
$L_{\rm macr}^{-1} = \sqrt{|\nabla^{\mu}e\nabla_{\mu}e|}/e$.
The inverse Reynolds number can be defined as a ratio of a dissipative and
a thermodynamic quantity, e.g.~$\mathrm{R}^{-1}=|n^{\mu}|/n_0$ or $\mathrm{R}^{-1}=|\pi^{\mu\nu}|/p_0$.
Both definitions measure the
local deviation from equilibrium. Powers of Knudsen and inverse Reynolds numbers higher than two
are neglected in the derivation of the fluid-dynamical
equations of motion (\ref{eq:IS_equations_diffusion}), (\ref{eq:IS_equations_shear}).
It is then natural to expect that the applicability of fluid dynamics requires that these two types of
quantities are sufficiently small, i.e., ${\rm Kn}\lesssim 1$ and $\mathrm{R}^{-1}\lesssim 1$.

\section{Numerical methods}
\label{sec:numerical}

\subsection{Fluid dynamics: SHASTA}

In the following the fluid-dynamical equations of motion in the longitudinally boost-invariant case \cite{Bjorken:1982qr}
with arbitrary transverse expansion are solved. In this case the solutions depend only on the transverse coordinates
$\mathbf{x}=(x,y)$ and the longitudinal proper time $\tau = \sqrt{t^2-z^2}$, i.e., they do not depend on the
space-time rapidity $\eta_s = \frac{1}{2} \ln[ (t + z)/(t - z)]$. Thus it is sufficient to solve the system of equations numerically
in $(2+1)$ dimensions.

The conservation laws \eqref{eq:econs} and \eqref{eq:ncons} together with Eqs.~\eqref{eq:IS_equations_diffusion}
and \eqref{eq:IS_equations_shear} are solved numerically using the time-honored SHASTA
algorithm~\cite{Boris73, Zalesak79}. The details of this algorithm are presented e.g.\ in Ref.~\cite{Molnar:2009tx}.
The additional ingredient is that the time step must be adapted in order to resolve the two fast time scales
$\tau_n$ and $\tau_\pi$ in Eqs.~\eqref{eq:IS_equations_diffusion} and \eqref{eq:IS_equations_shear},
which become small when the cross section is large. Another fast time scale is given
by the initially large longitudinal expansion rate $\sim 1/\tau$. All of these
time scales need to be resolved by the numerical algorithm.

\begin{figure}
\includegraphics[width = 0.45\textwidth]{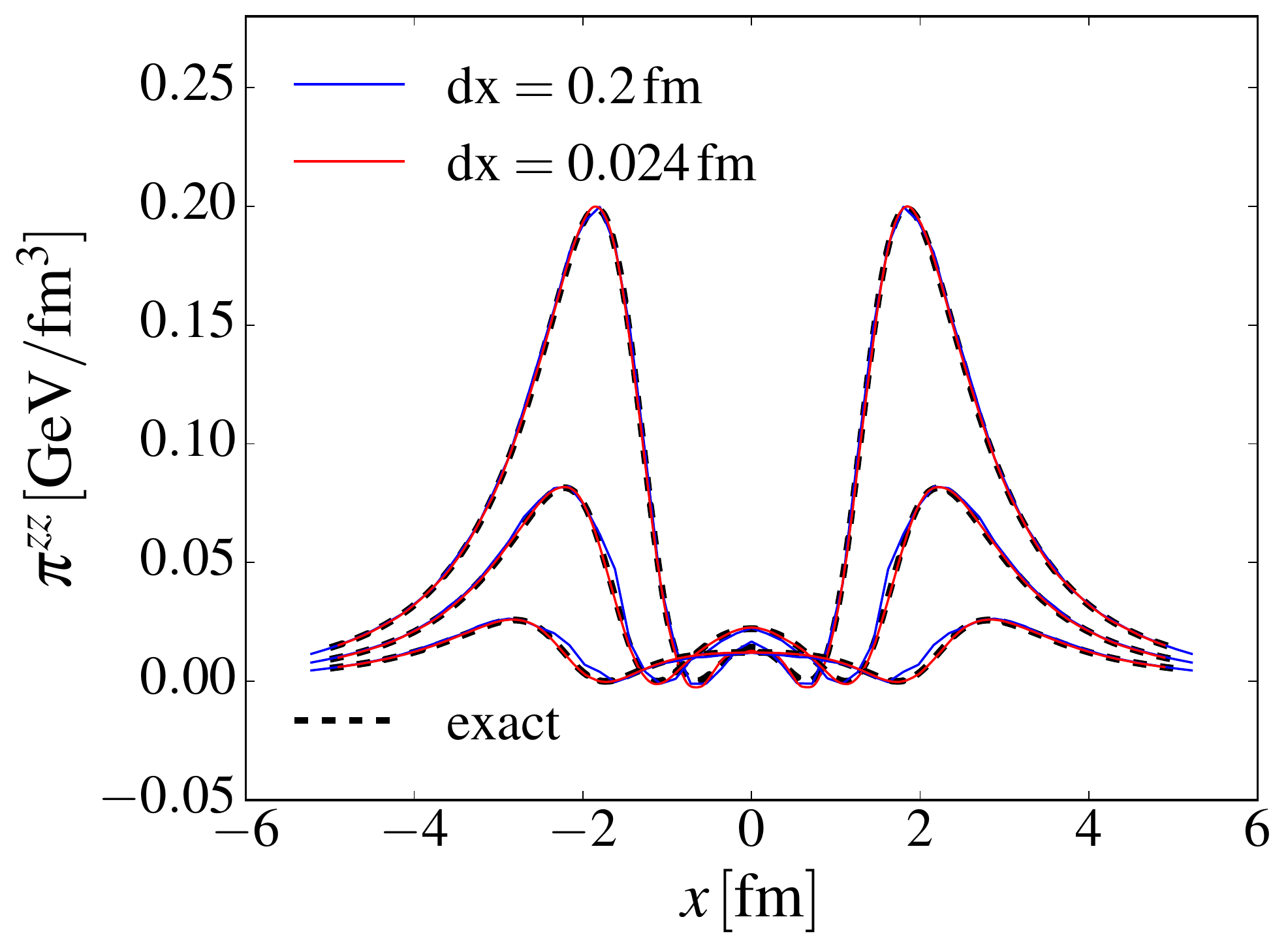}
\includegraphics[width = 0.45\textwidth]{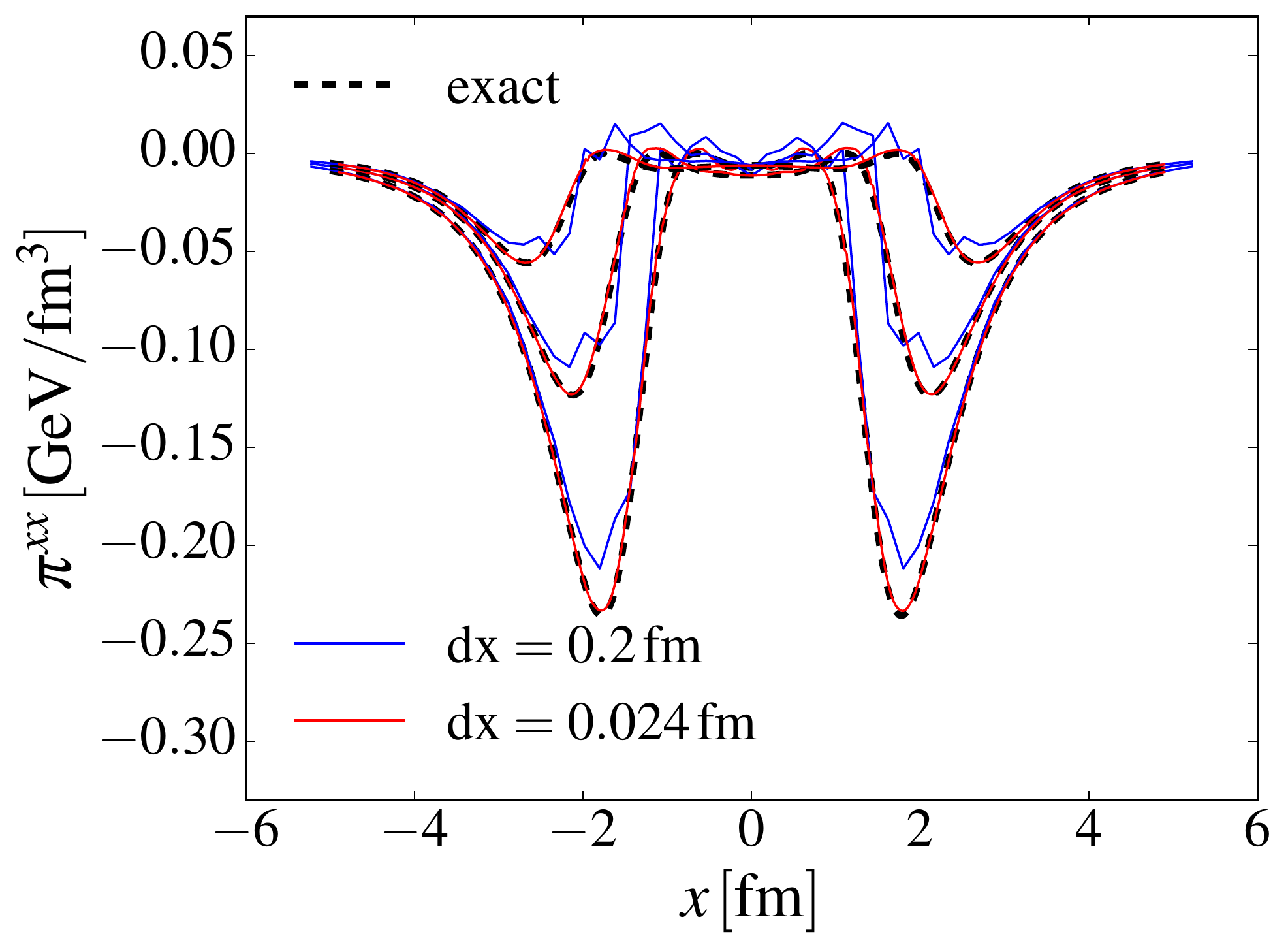}
 \caption{(Color online)
   Comparing the semi-analytic ``Gubser'' solution of the Israel-Stewart equations \cite{Marrochio:2013wla} 
   to the numerical solution using SHASTA
   with two different numerical resolutions as indicated in the plots. Figure (a) shows the $\pi^{zz}$ profile and
   figure (b) shows the $\pi^{xx}$ profile at different times $\tau = 1.2,\,1.5,\,2.0\fm$.}
 \label{fig:gubser_pixxzz}
\end{figure}

The effectiveness of the algorithm can be tested by comparing the numerical solutions to a semi-analytic solution,
the so-called ``Gubser'' solution, which describes a radially expanding system which is boost-invariant in the
longitudinal direction \cite{Marrochio:2013wla}. In this case simplified Israel-Stewart equations
with a constant $\eta/s=0.08$ and without particle number and diffusion can be solved semi-analytically, i.e., the solution can
be obtained by solving an ordinary differential equation. This solution can then be compared to a corresponding
numerical solution using SHASTA, without explicitly accounting for the symmetries in the Gubser solution, except for
boost invariance.
A comparison of the resulting shear-stress tensor components $\pi^{zz}$ and $\pi^{xx}$ is shown in
Fig.~\ref{fig:gubser_pixxzz} at several fixed values of $\tau$ and two different numerical resolutions.
As can be seen in Fig.\  \ref{fig:gubser_pixxzz}, the numerical solution reproduces the semi-analytic solution very well
for both resolution scales, only $\pi^{xx}$ shows clear deviations for the lower resolution, but converges rapidly
to the actual solution when the resolution is increased.

\subsection{Boltzmann equation: BAMPS}

As microscopic transport model we choose the Boltzmann equation, which is solved via the BAMPS
algorithm \cite{Xu:2004mz}.
Besides many other studies, this model has already been applied very successfully to study elliptic and triangular flow in
ultrarelativistic p+p collisions \cite{Deng:2011at}.
BAMPS was also already applied to a comparison with
fluid-dynamical calculations in the case of
a (0+1) dimensional expansion scenario \cite{El:2009vj} and in the framework of the relativistic Riemann
problem \cite{Bouras:2009nn,Bouras:2010hm}.

Many of the complex features implemented in BAMPS are not used in the present study. But a very important aspect is
the correct realization of the collision rate. This is guaranteed by the implementation of the collision criterion not by
a geometrical, but a stochastic interpretation of the underlying cross section. The transport equations are solved via the
testparticle method. Scaling the
cross section guarantees the correct solution of the underlying Boltzmann equation in the limit
of infinitely many testparticles.

A drawback of the stochastic implementation of the cross section is the discretization of the spatial coordinates.
Particles are grouped into spatial cells according to their $x$, $y$, and $\eta_s$ coordinates and only interactions 
between particles in each of these cells are considered. Since
the number of testparticles ($N_{\rm test}=r N_{\rm phys}$, where $r$ is 
1000 for A+A and 7000 for p+p and p+A collisions) 
is limited by the computational resources, also the number of testparticles in a cell is finite. If
a cell contains only one particle, no interactions occur in this cell. On the other hand, if one
increases the overall number of testparticles by e.g.~a factor of ten, then also approximately ten testparticles would be
found in the specific cell. Thus, interactions are possible. In the actual solution, only interactions in cells with
at least $N_{\rm test} \geq 4$ testparticles are allowed. In physical terms, the particle density
is then below $1\fm^{-3}$ for proper times larger than $1\fm$ [for $r=1000\, (7000)$, the transverse cell size is
$\Delta x= \Delta y= 0.25 (0.08)\fm$, while $\Delta\eta_s\sim0.06$].
As we shall see below, in the comparison with fluid-dynamical calculations this has to be kept in mind.

One of the main challenges for this study is the fact that BAMPS uses a Cartesian coordinate system for
time $t$ and (longitudinal) space coordinate $z$, while the boost-invariant Bjorken expansion scenario
is formulated in proper time $\tau$ and space-time rapidity $\eta_s$.
This introduces the following three problems:
\begin{enumerate}
\item[(i)] Because of boost invariance,
we can restrict our considerations to a rapidity interval $I_{\Delta \eta_s} \equiv [-\Delta \eta_s/2, +\Delta \eta_s/2]$
around $\eta_s=0$. In the calculations, we choose $\Delta \eta_s = 1$ (see below).
In a boost-invariant system, the rapidity range extends formally from $-\infty$ to $\infty$, but computational
resources (computer memory and CPU time) limit the rapidity range realizable in the actual calculations,
$I_{\eta_{s,\mathrm{max}}} = [- \eta_{s,\mathrm{max}},+\eta_{s,\mathrm{max}}]$.
The value for $\eta_{s,\mathrm{max}}$ has to be chosen sufficiently large, such that there is no information
propagating from the boundaries $\pm \eta_{s,\mathrm{max}}$ to the mid-rapidity interval $I_{\Delta \eta_s}$, where
the actual comparison between fluid dynamics and BAMPS is performed. Otherwise, the vacuum at
$|\eta_s| > \eta_{s,\mathrm{max}}$ will start to influence the solution in the rapidity interval $I_{\Delta \eta_s}$, leading
to artefacts like a decrease in the particle density that is stronger than permitted in a boost-invariant scenario.
\item[(ii)]
A related problem is that the boost-invariant fluid-dynamical calculation starts at proper time $\tau_0$, while the
BAMPS calculation is initialized at a fixed Cartesian time $t_0=\tau_0$. One therefore has to make sure that
the initial condition for BAMPS at $t_0$ gives the same initial condition as for fluid dynamics at $\tau_0$. At
$z=\eta_s=0$, this is formally guaranteed [see, however, (iii) below],
but at any nonzero value for $\eta_s$, one has to propagate the information
about the initial condition from the $\tau_0=const.$ hypersurface back to the $t_0= const.$ hypersurface.
In practice, this is done by propagating the particles without collisions
from $t_0$ to a time $t_p$ when the particle wordline crosses the $\tau_0$ hypersurface.
\item[(iii)] While the initial conditions for BAMPS and fluid dynamics are formally the same at $z = \eta_s =0$ (see
above), in practice one has a cell of non-zero longitudinal extension at this point. This cell is initially filled with
particles, but the ones with positions $z_p \neq 0$ are at a longitudinal proper time $\tau_p < \tau_0$.
One therefore has to wait until these particles have also reached the $\tau_0= const.$ hypersurface. 
For a typical cell size used in
BAMPS this means that the BAMPS calculation starts at a time slightly later than $t_0$ (typically $t_0+0.01\fm$).
\end{enumerate}
A boost-invariant (2+1) dimensional formulation of
the BAMPS algorithm would avoid all three problems, but is currently not available.

\section{Initial Conditions}
\label{sec:setup}

In all test cases presented here, the evolution is started in local equilibrium, i.e., the initial local particle
distributions are given by Eq.~(\ref{eq:LTE_distribution}).
Three different initializations for the initial particle density in the transverse plane are considered.
The first two scenarios are given by a radially symmetric Gaussian profile with a width parameter
$w = 1$ or $3\fm$,
\begin{equation}
 n(\tau_0, \mathbf{x}) \sim \exp\left(-\frac{\mathbf{x}^2}{2w^2}\right)\;.
\end{equation}
The smaller width parameter corresponds to a system created in p+p or p+A collisions, while the larger
width parameter corresponds to the case of event-averaged A+A collisions.

In the third initialization scenario, labeled ``nBC'' (for ``particle density, binary collisions'') in the following, the  initial transverse density profile is
proportional to the density of binary collisions computed within the optical Glauber model,
\begin{equation}
 n(\tau_0, \mathbf{x}) \propto T_A(\mathbf{x} - \mathbf{b}/2) T_A(\mathbf{x} + \mathbf{b}/2)\;,
\end{equation}
with impact parameter $b = 7.5\fm$, taken along the $x$-direction, and where $T_A$ is the standard nuclear
thickness function,
\begin{equation}
 T_A = \int_{-\infty}^{\infty} \dd z\, \rho(\mathbf{x}, z)\;,
\end{equation}
with $\rho(\mathbf{x}, z)$ being the Woods-Saxon parametrization of the nucleon density with radius
$R = 6.38\fm$ and surface thickness $d = 0.55\fm$.

\begin{figure}
 \includegraphics[width = 0.45\textwidth]{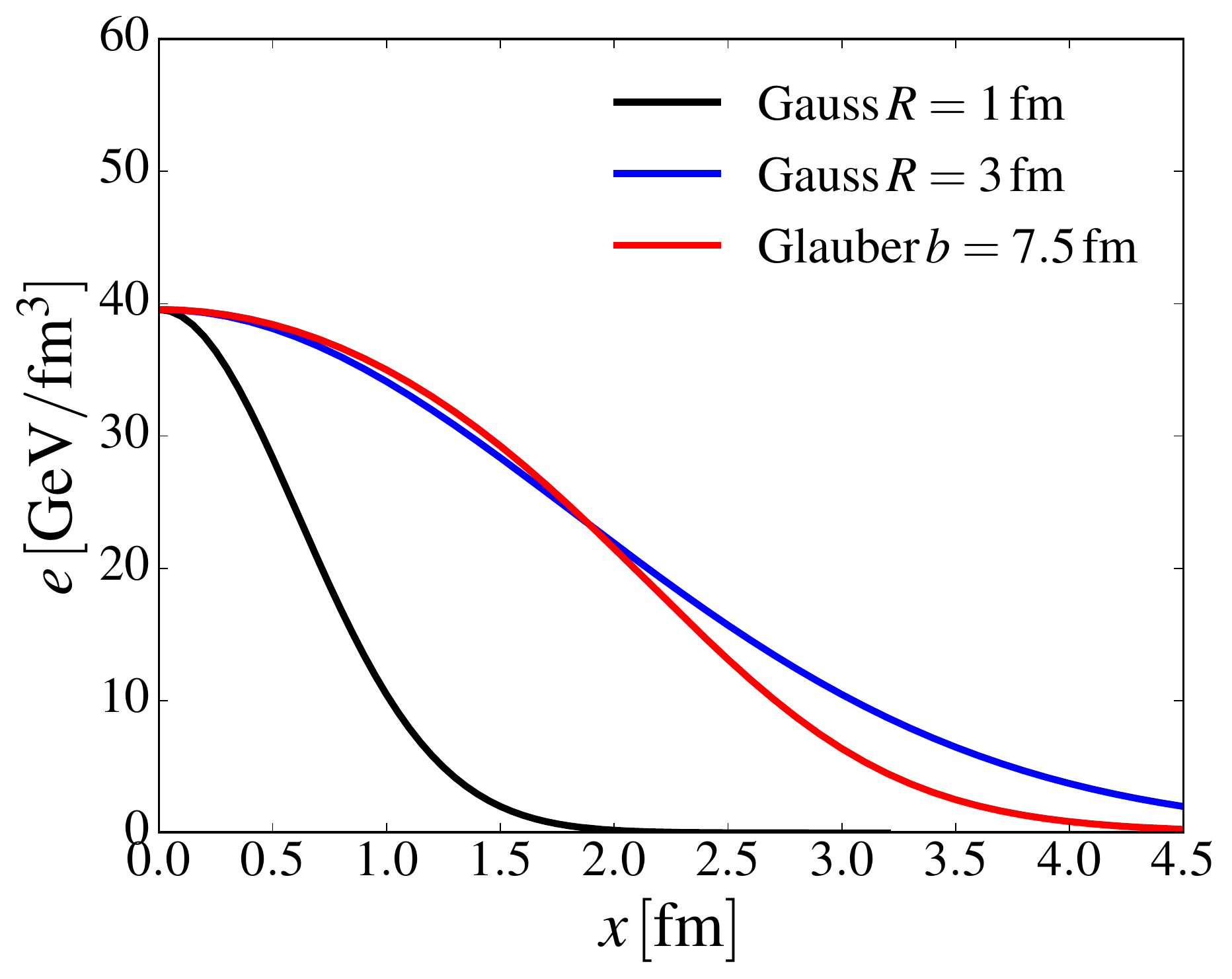}
 \caption{(Color online) Initial energy-density profiles.}
 \label{fig:iniconditions}
\end{figure}

The initialization time is always set to $\tau_0 = 0.2\fm$, and the maximum temperature at
$x = y = 0$ is $T_0 = 500\MeV$, with the temperature dropping towards larger radii guaranteeing a
constant fugacity  $e^{\alpha_0}=1$ for all radii at initial time $\tau_0$. The constant isotropic cross section is varied from
$\sigma = 1\mb$ to $\sigma = 20\mb$. The three different energy-density profiles are shown in Fig.~\ref{fig:iniconditions}.

\begin{figure}
 \includegraphics[width = 0.45\textwidth]{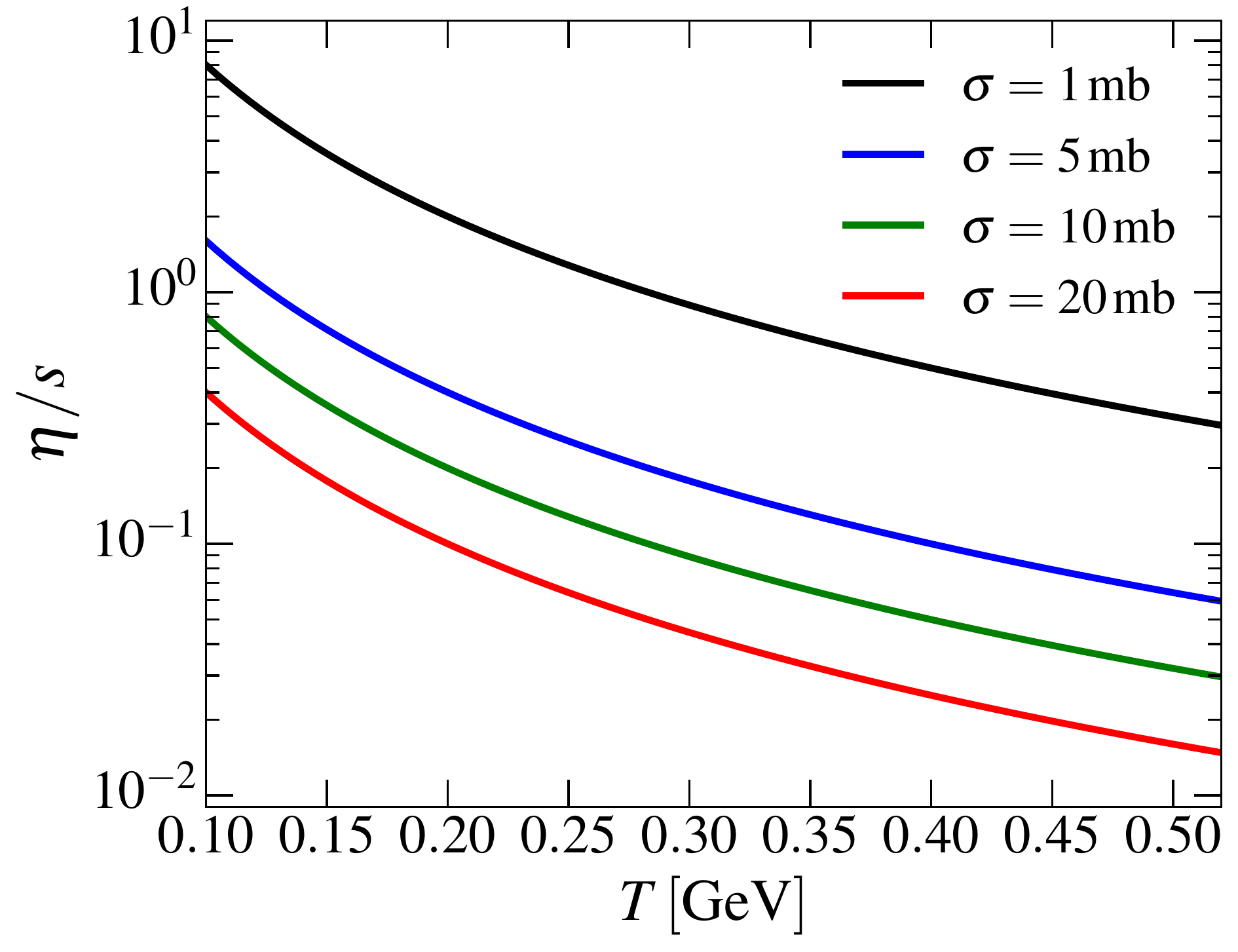}
 \caption{(Color online) Shear viscosity-to-entropy density ratio as a function of temperature for a constant isotropic
 cross section.}
 \label{fig:etapers}
\end{figure}

For the case of an ultrarelativistic Maxwell-Boltzmann gas (with degeneracy factor $g$), the entropy density $s$ and the
shear viscosity $\eta$ are given by \cite{Denicol:2010xn}
\begin{equation}
s=e^{\alpha_0}(4 -\alpha_0)\frac{g}{\pi^2}T^3 \;,\qquad
\eta=\frac{4}{3}\frac{T}{\sigma}\; .
\end{equation}
Here the degeneracy factor is taken as $g = 16$. In Fig.~\ref{fig:etapers}, the ratio $\eta/s$ as a function of temperature
corresponding to different constant isotropic cross sections and $\alpha_0 =0$ is shown. One observes
that quite a wide range of $\eta/s$ values is covered.

\section{Space-time evolution}
\label{sec:Comparisons}

\subsection{Rapidity-averaging procedure}
\label{subsec:Setup}

The numerical algorithm to solve the fluid-dynamical equations of motion is explicitly boost-invariant, i.e.,
the solution is independent of the space-time rapidity $\eta_s$. It thus uses the
coordinates $(\tau, x, y)$, where $\tau$ is the proper time.
However, the BAMPS code solves the Boltzmann equation in Cartesian $(t,x,y,z)$-coordinates. In practice the
components of $T^{\mu\nu}$ and $N^{\mu}$ need to be obtained by averaging over a finite space-time rapidity
range $\Delta \eta_s$ at fixed Cartesian time. However, if we did that in the
BAMPS calculational frame, the averaging would result in e.g.\ a non-zero shear-stress tensor even if the fluid is
locally isotropic. In order to avoid this problem,
every particle is boosted by $- \eta_{s,p}$, where $\eta_{s,p}$ is the space-time rapidity of
the particle, before performing the average over $\Delta \eta_s$. This makes sure that for the averaging procedure
the particle is considered in a rest frame moving with $+\eta_{s,p}$ with respect to the
BAMPS calculational frame. The decomposition \eqref{eq:Tmunu_decomposition} and
\eqref{eq:Nmu_decomposition} of $T^{\mu\nu}$ and $N^{\mu}$, respectively, is then done after boosting
and averaging. This
procedure minimizes an artificial creation of dissipative quantities due to the averaging procedure.

In order to have a one-to-one comparison of the space-time evolution
of different quantities, the fluid-dynamical solutions need to be averaged in the same way as in the BAMPS calculation,
i.e.,
\begin{equation}
 \langle T^{\mu\nu} \rangle_{\Delta \eta_s, t} = \frac{1}{\Delta z}\int_{-\Delta z/2}^{\Delta z/2} \dd z\,
 T^{\mu\nu}(\tau = \sqrt{t^2-z^2}, x, y)\;,
\label{eq:Tzaverage}
\end{equation}
where $\Delta z/2 = t \tanh(\Delta \eta_s/2)$. We choose $\Delta \eta_{s}=1$ in all our calculations.

\subsection{\texorpdfstring{Gaussian profile $w=3\fm$}{Gaussian profile w=3fm}}
\label{subsec:Gauss3fm}

\begin{figure*}
\includegraphics[width = 0.32\textwidth]{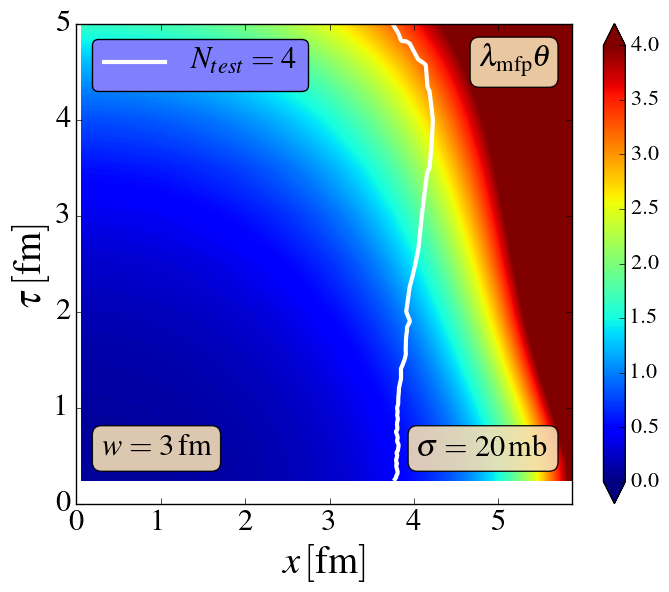}
\includegraphics[width = 0.32\textwidth]{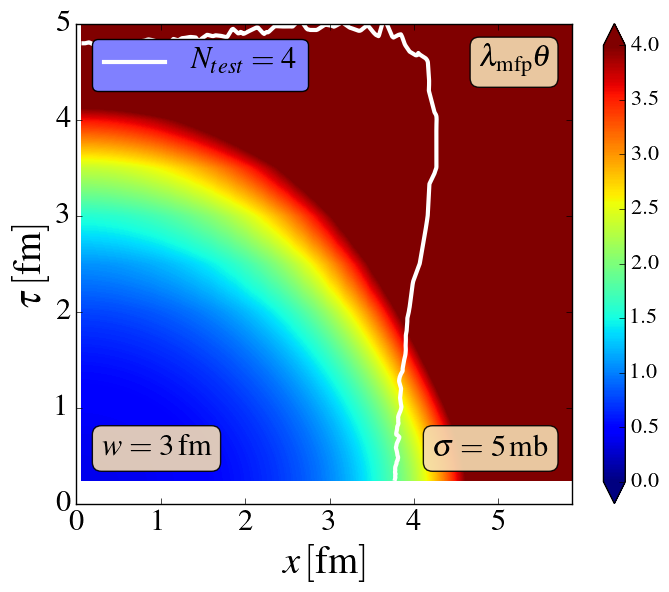}
\includegraphics[width = 0.32\textwidth]{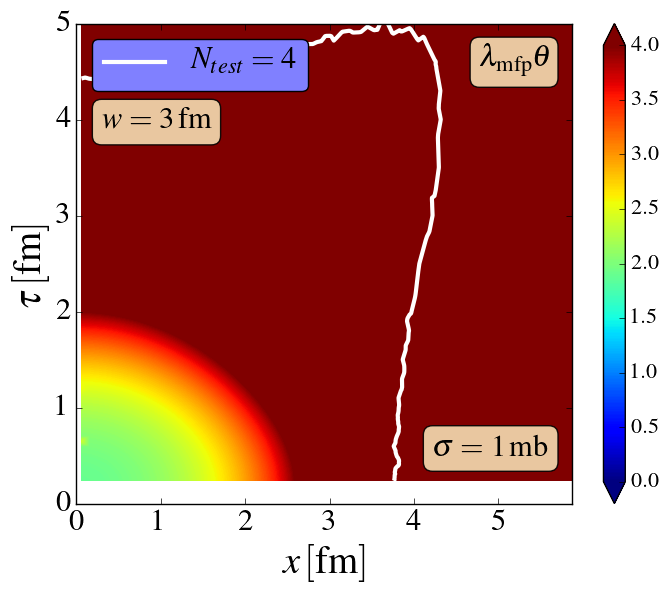}

\includegraphics[width = 0.32\textwidth]{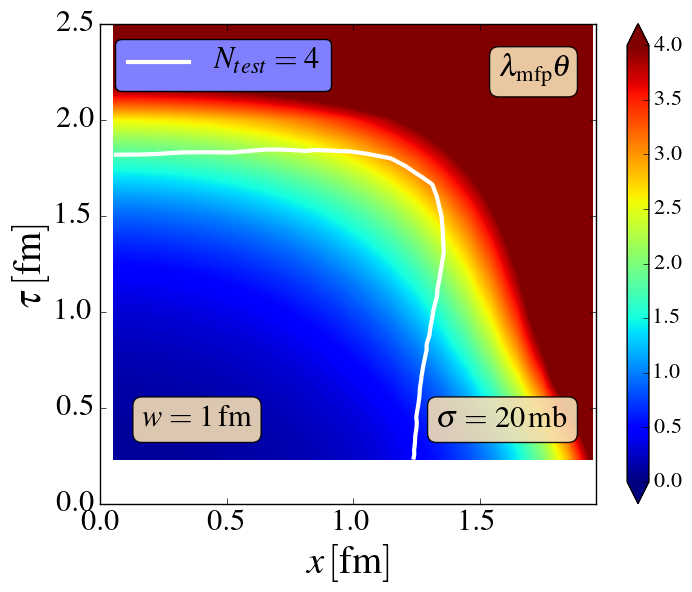}
\includegraphics[width = 0.32\textwidth]{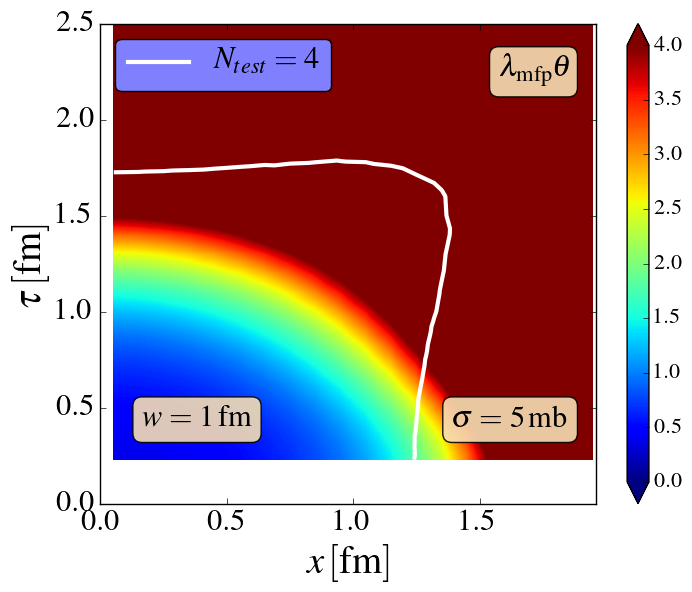}
\includegraphics[width = 0.32\textwidth]{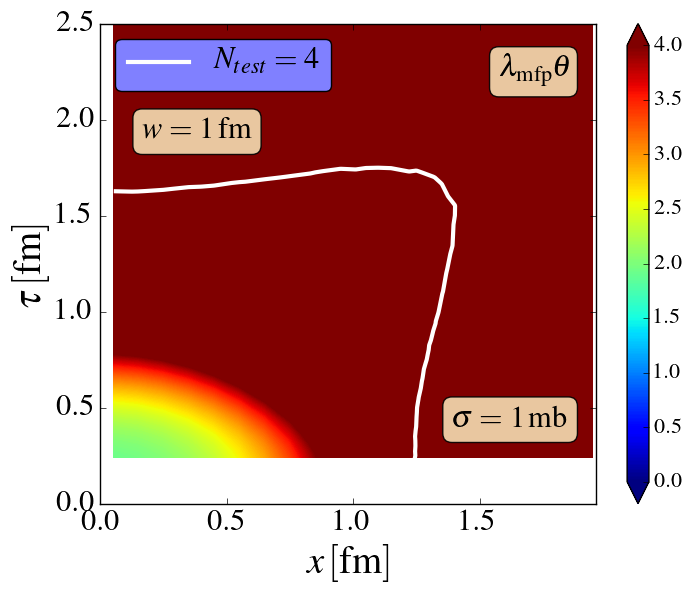}
\caption{(Color online) Knudsen-number contours for a Gaussian $w=3\fm$ (top) and $w=1\fm$ (bottom) profile,
with $\sigma = 20\mb, 5\mb$, and $1\mb$ (from left to right). The contour line shows where the testparticle number
per computational cell in the BAMPS calculation assumes the value four.}
\label{fig:knudsen_contour_gauss3fm}
\end{figure*}

We first consider a Gaussian profile with $w=3\fm$ and vary the cross section from $\sigma = 20\mb $ down to
$1\mb$. In the upper row of Fig.~\ref{fig:knudsen_contour_gauss3fm}, the space-time evolution of the Knudsen number
${\rm Kn} = \lambda_{\rm mfp}\theta$ extracted from the fluid-dynamical calculations is shown
for this case. The Knudsen number is
smallest at earliest times and in the center of the system. Each figure shows also the contour where the testparticle
number per computational cell in the BAMPS calculation is four. As discussed above, outside this region the particles are
not interacting anymore, and therefore the comparison between the BAMPS solution of the Boltzmann equation and fluid
dynamics is meaningful only inside this region. As can be seen from the figures, if the cross section is large, the Knudsen
number stays small ${\rm Kn} \lesssim 1$ within almost the whole $N_{\rm test} > 4$ region, but as the cross section
decreases, the Knudsen number increases, and for $\sigma = 1\mb$, one finds ${\rm Kn} > 1$ in the whole region.

\begin{figure*}
\includegraphics[width = 0.4\textwidth]{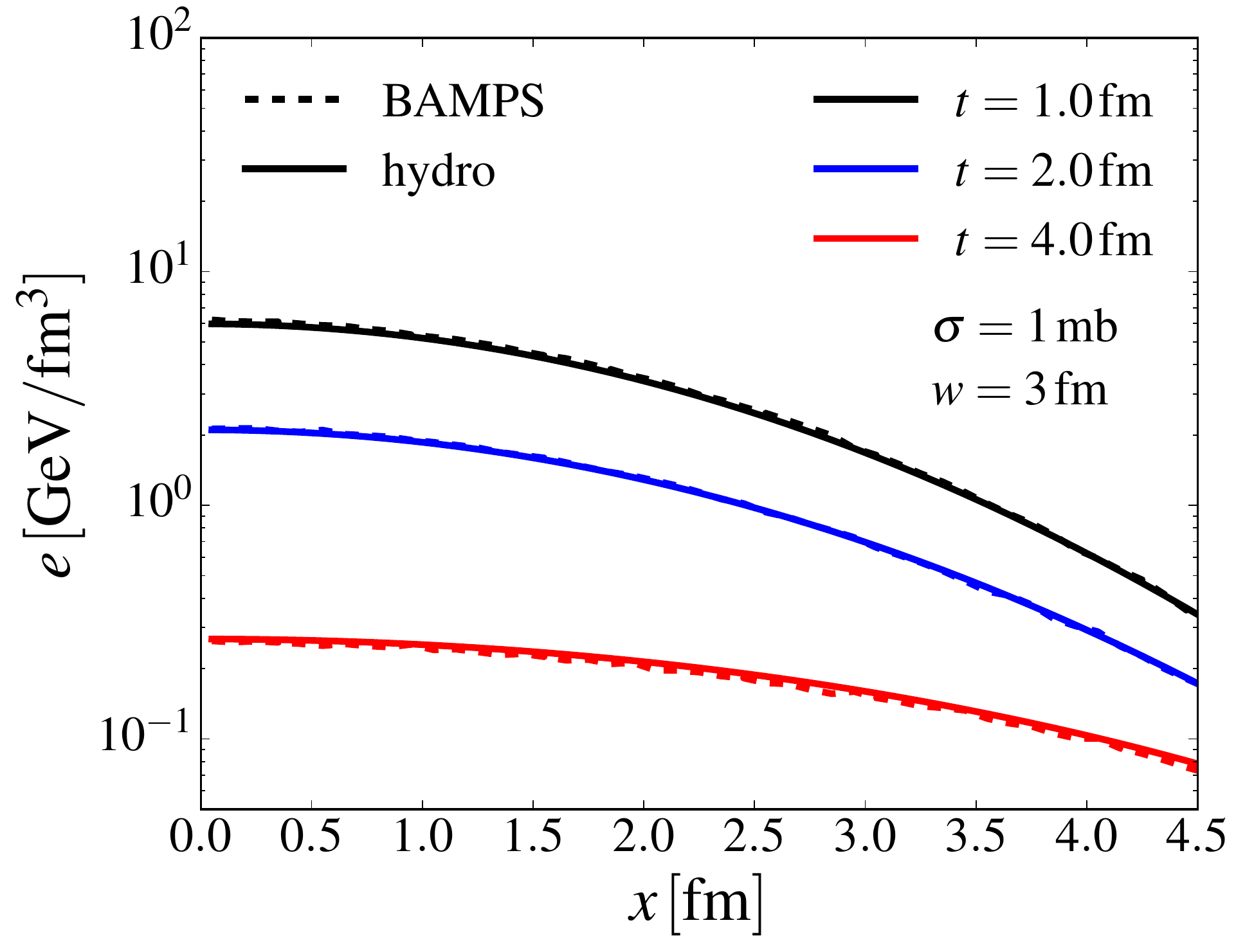}
\includegraphics[width = 0.4\textwidth]{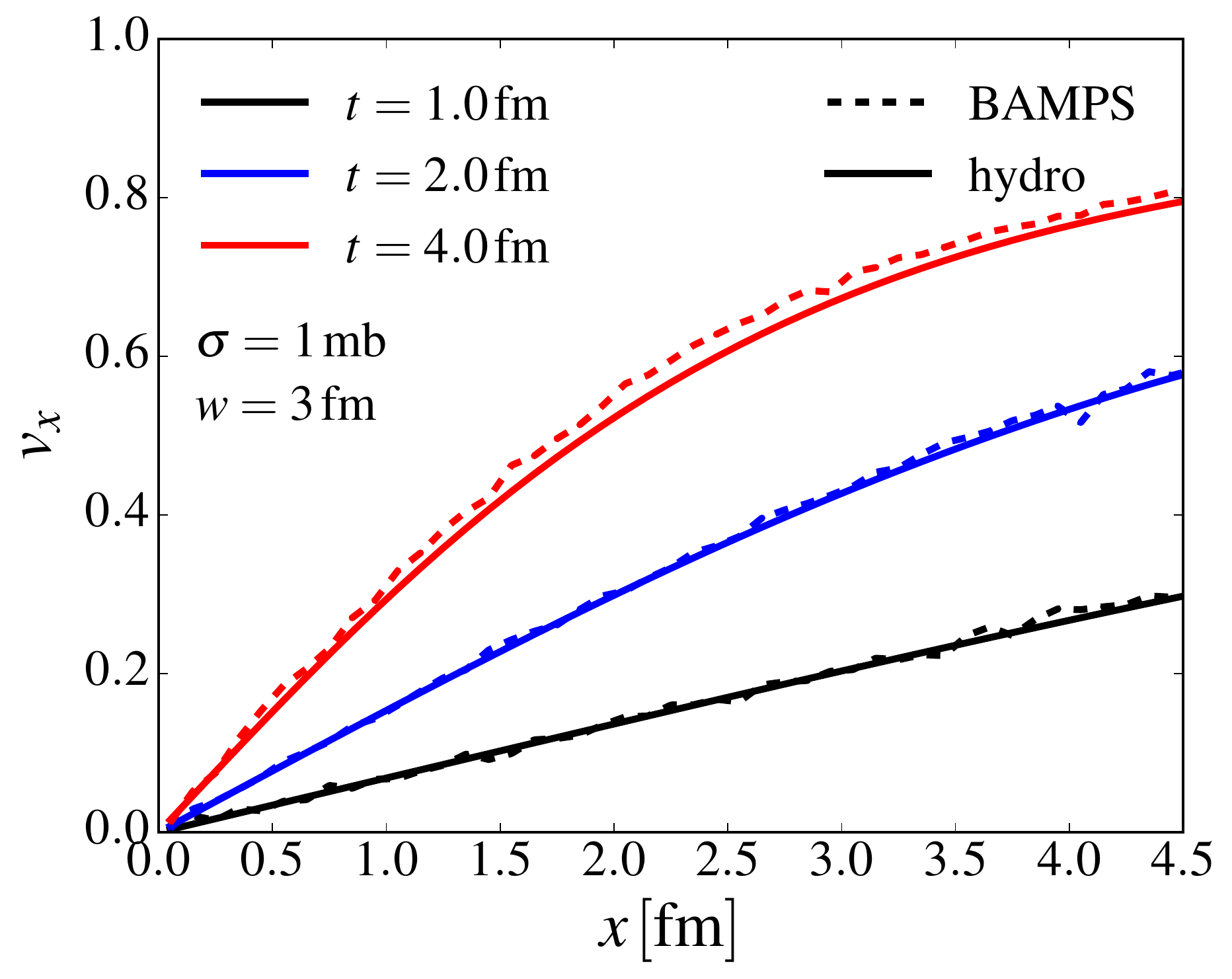}
\caption{(Color online) Energy-density (left) and $v_x$ (right) profiles for $\sigma=1\mb$ and a
$w=3\fm$ Gaussian initial density profile for different times $t= 1, \, 2$, and $4$ fm. Solid lines represent
fluid-dynamical results, while dashed lines show the BAMPS solutions.}
\label{fig:edvx_gauss3fm}
\end{figure*}

The energy-density and transverse-velocity profiles at different times are shown in Fig.~\ref{fig:edvx_gauss3fm}
for $\sigma = 1\mb$.
As can be seen from the figures, the agreement between the Boltzmann equation and fluid dynamics remains extremely
good over the whole space-time evolution even up to quite high Knudsen numbers. Naturally, the agreement remains
good for larger values of the cross section.

\begin{figure*}
 \includegraphics[width = 0.32\textwidth]{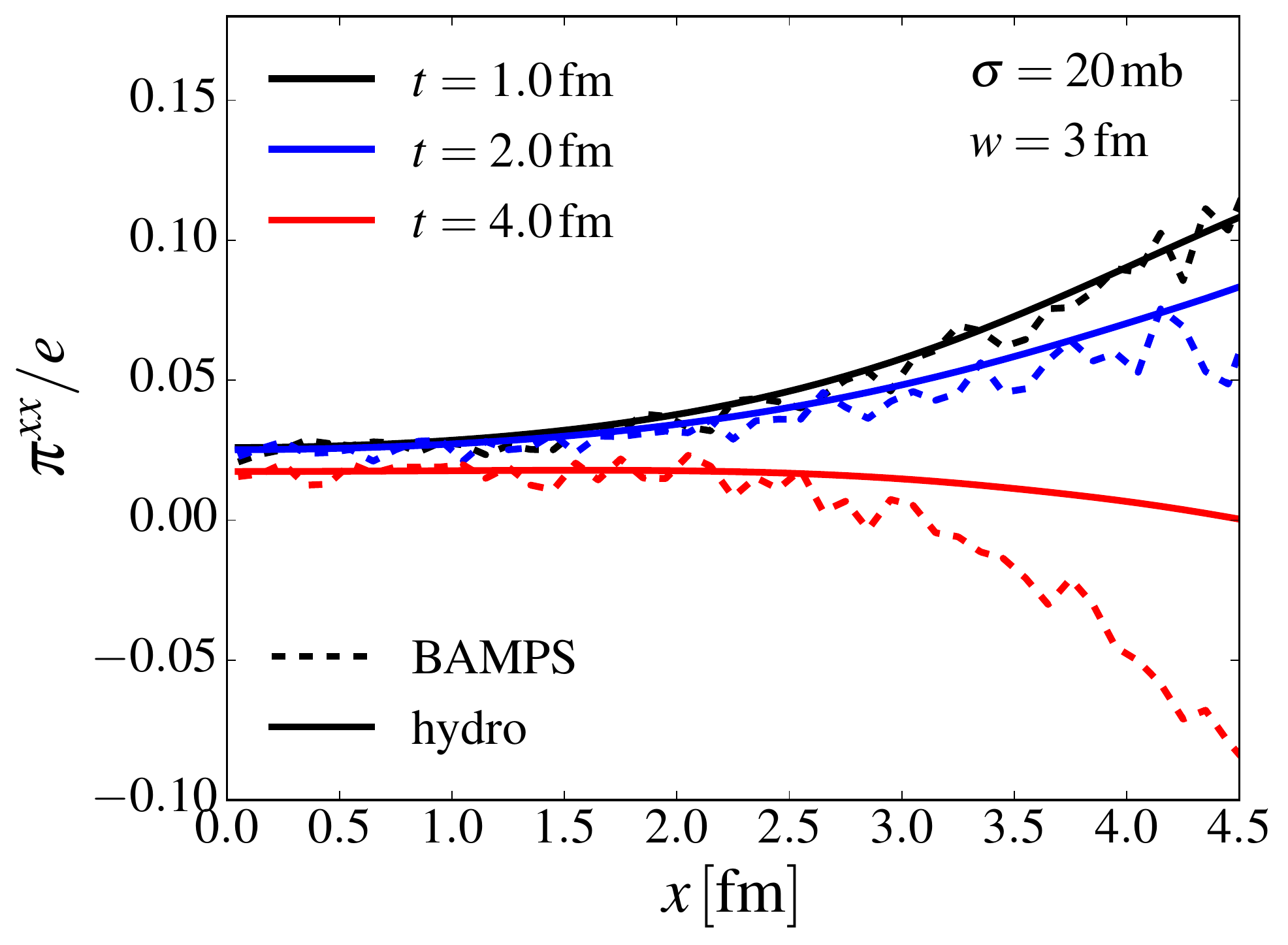}
 \includegraphics[width = 0.32\textwidth]{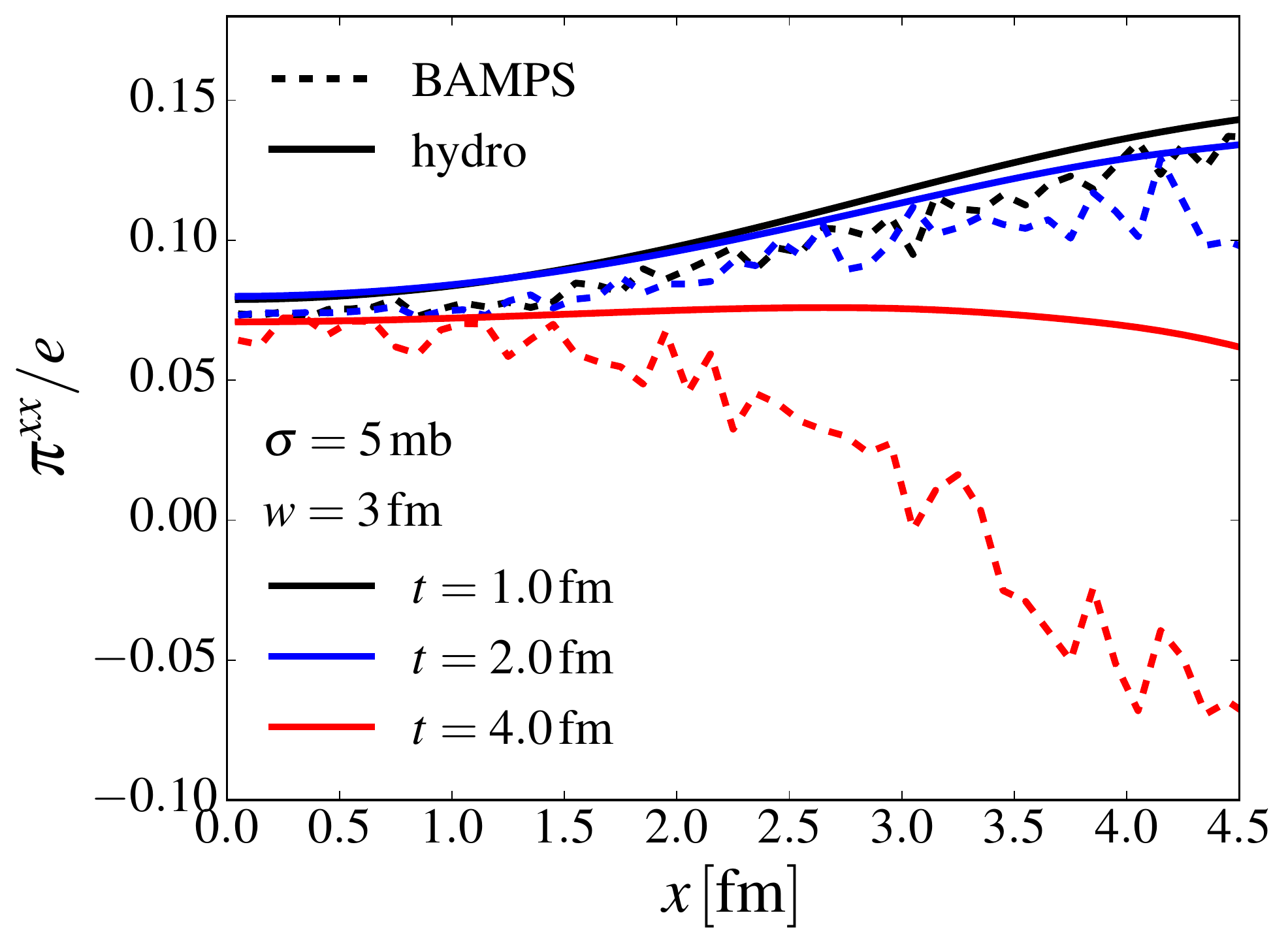}
 \includegraphics[width = 0.32\textwidth]{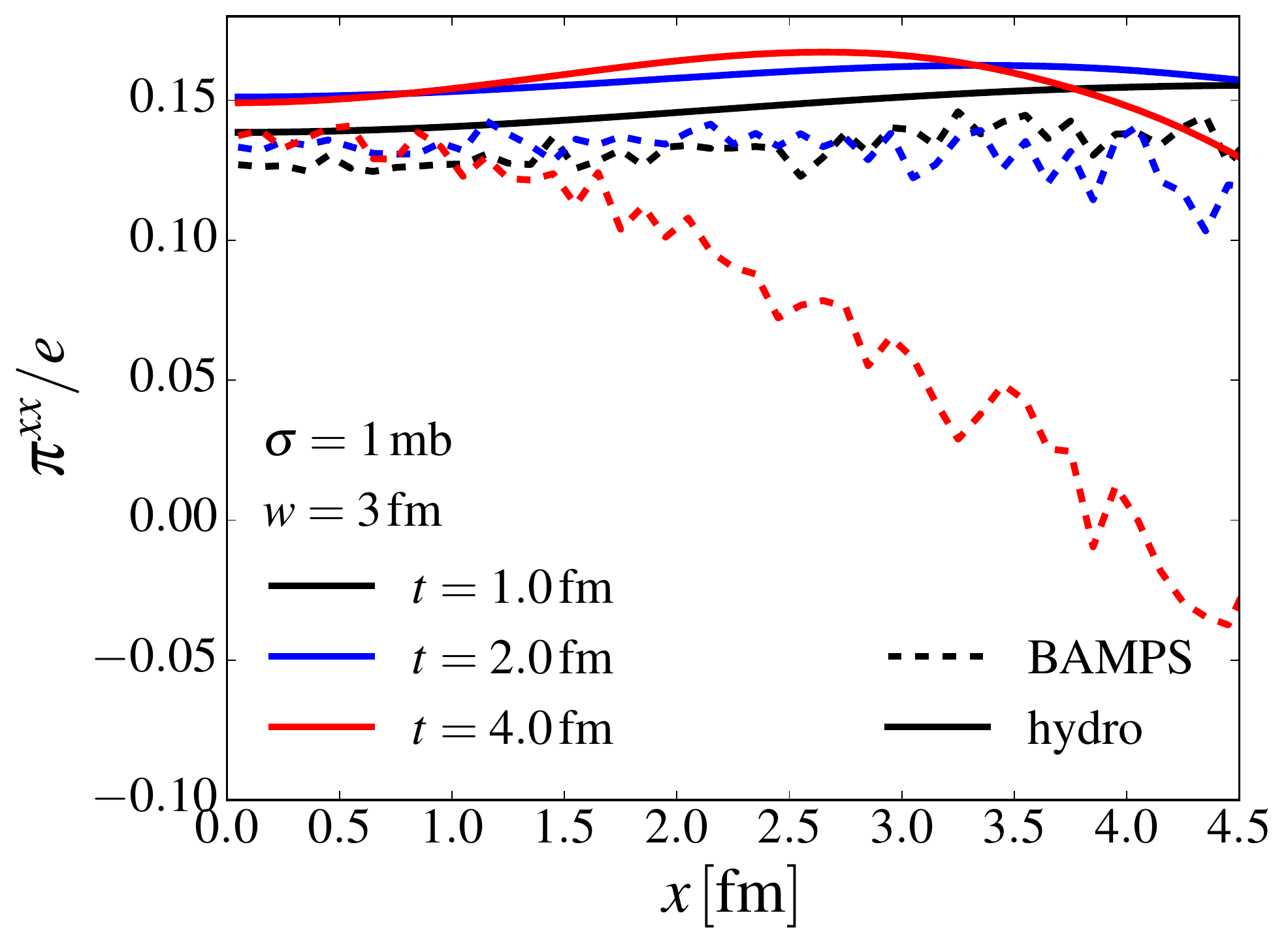}
 \caption{(Color online) The $\pi^{xx}$ profiles scaled by energy density $e$ for $\sigma=20\mb, 5\mb$, and $1\mb$
 (from left to right) and a $w=3\fm$ Gaussian initial density profile.}
 \label{fig:pizzxx_gauss3fm}
\end{figure*}

The components of the shear-stress tensor $\pi^{\mu\nu}$ provide a more sensitive probe of the applicability of fluid
dynamics. Figure \ref{fig:pizzxx_gauss3fm} shows as an example the transverse profile of $\pi^{xx}/e$ at different
times for $\sigma = 20$, $5$, and $1$ mb. Here already at early times small deviations are visible, especially in the case
of small cross sections. At later times, $t=4\fm$, differences become large.
This occurs in
regions where the Knudsen number becomes quite large, ${\rm Kn} \sim 2 - 4$.
As expected, for even larger cross sections,
the deviations become smaller (not shown here).

\begin{figure*}
 \includegraphics[width = 0.32\textwidth]{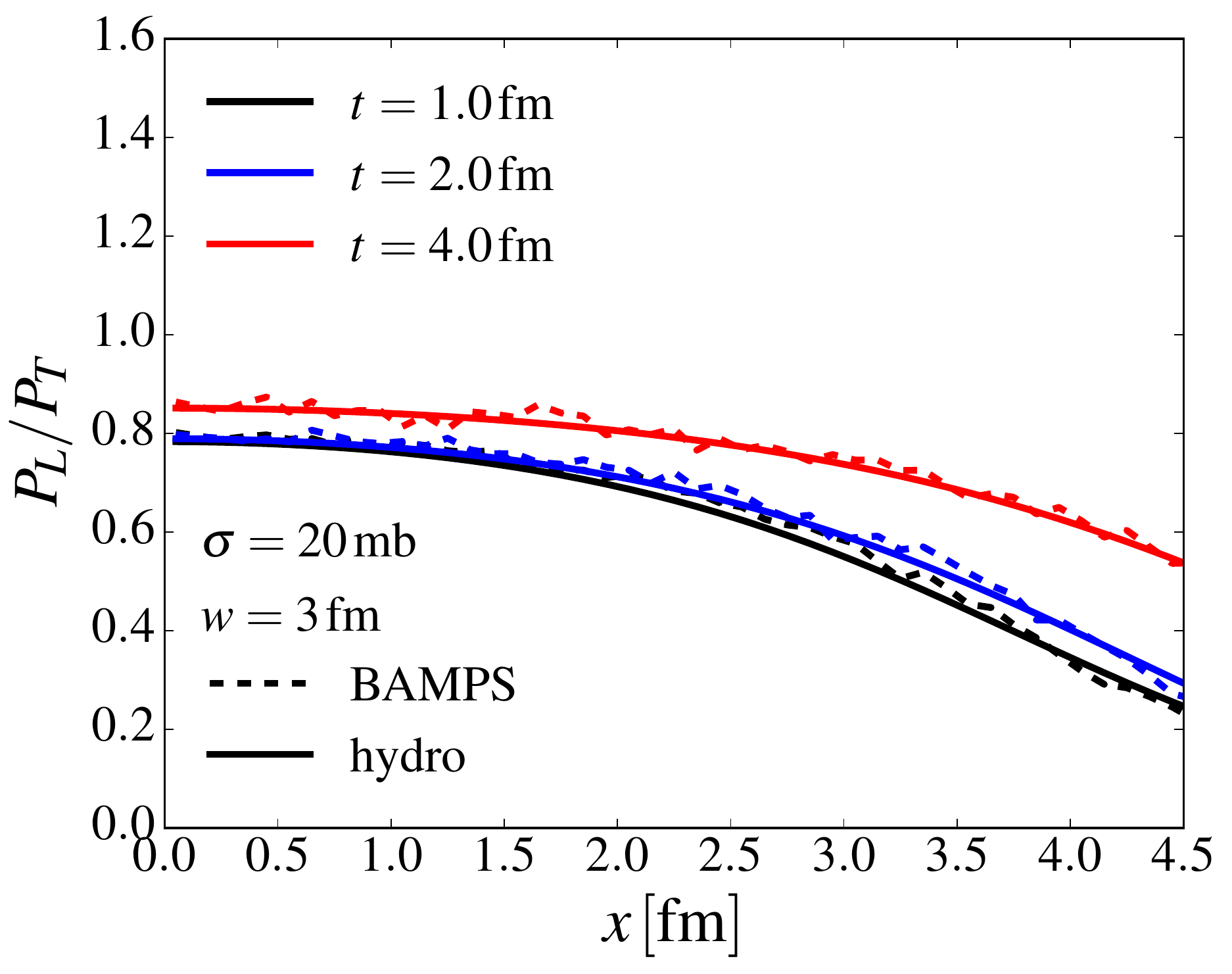}
 \includegraphics[width = 0.32\textwidth]{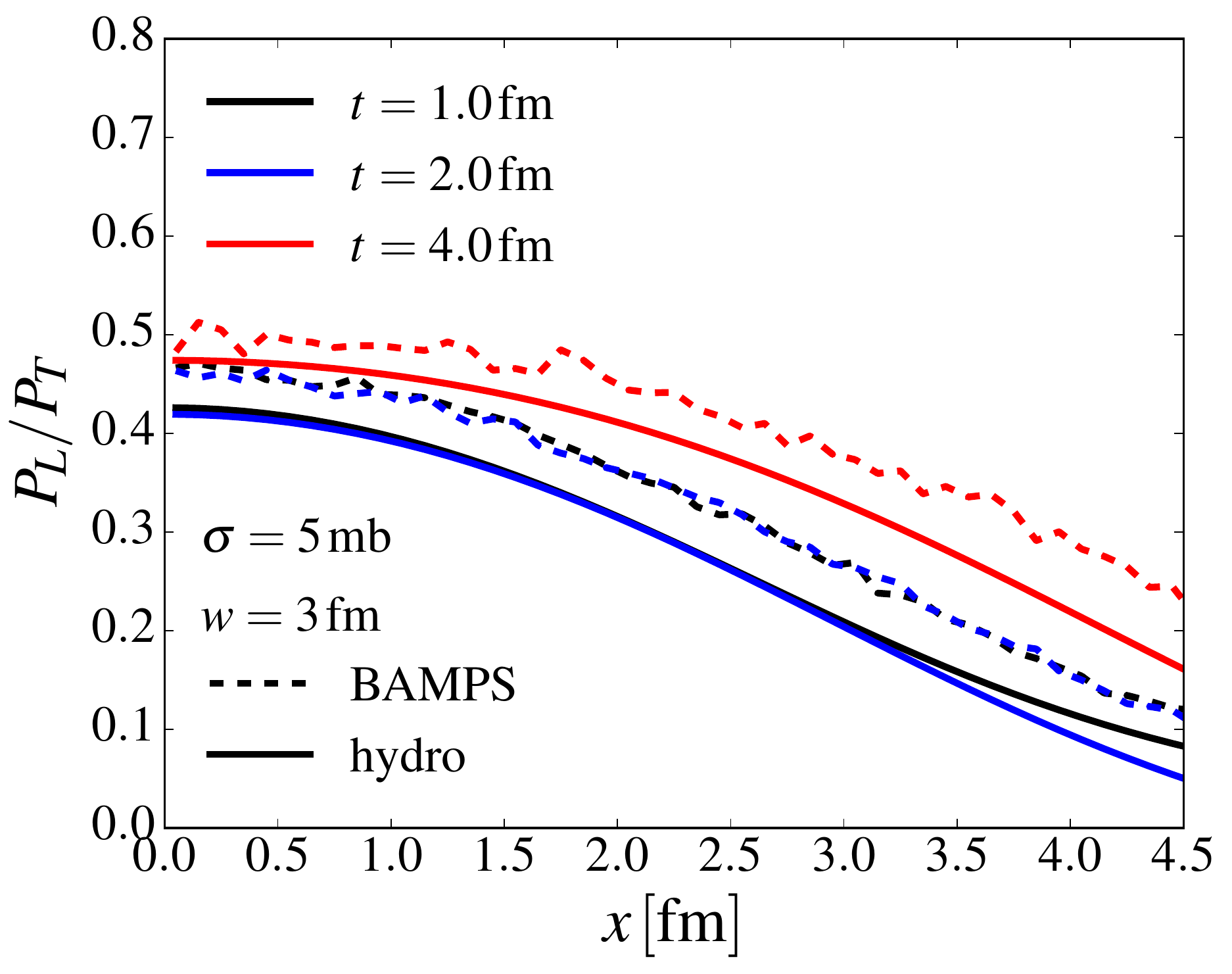}
 \includegraphics[width = 0.32\textwidth]{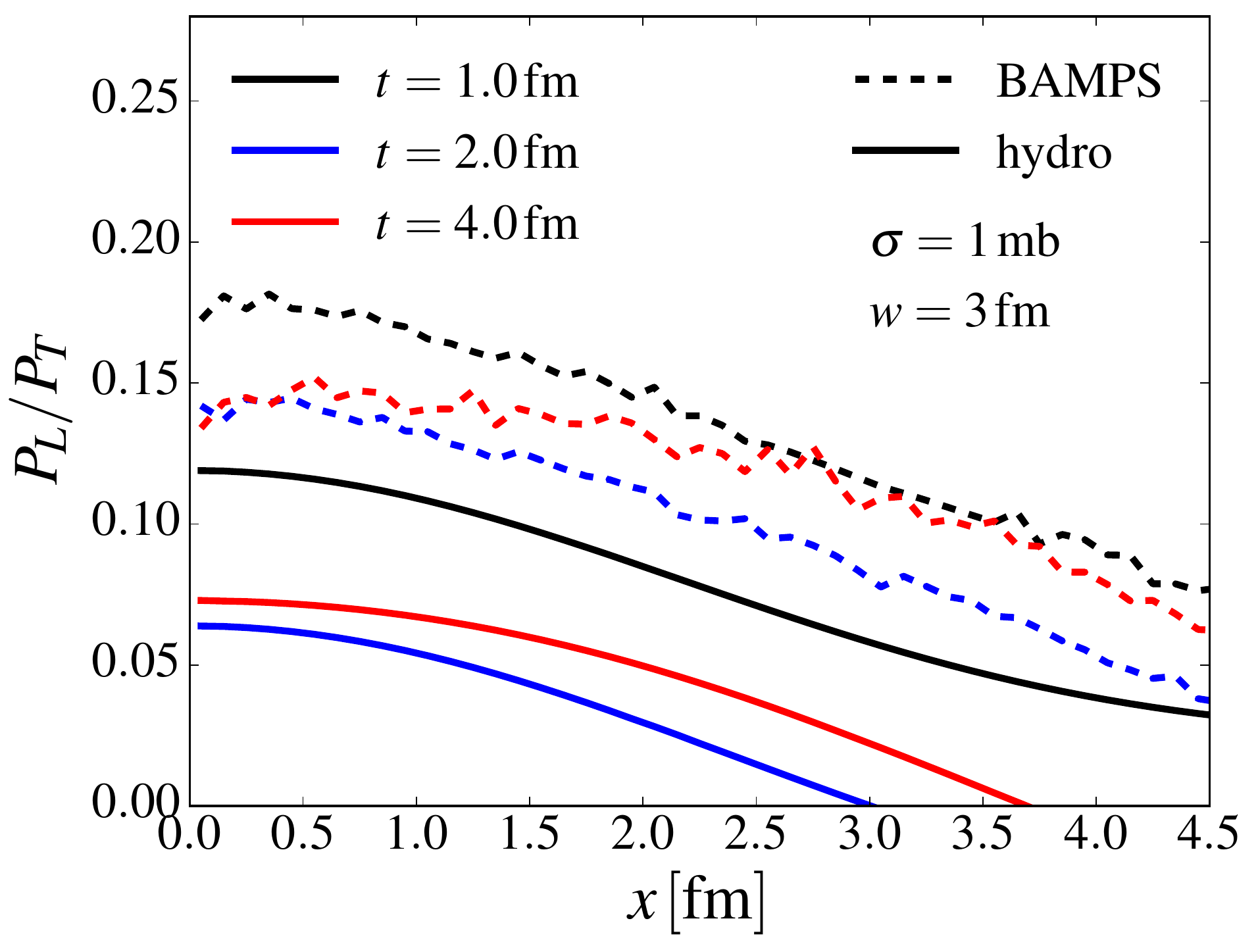}
 \caption{(Color online) The ratio of longitudinal over transverse pressure $P_L/P_T$ for $\sigma=20\mb, 5\mb$, and
  $1\mb$ (from left to right) and a $w=3\fm$ Gaussian initial density profile.}
 \label{fig:pl_per_pt_gauss3fm}
\end{figure*}

In order to illustrate the degree of isotropisation, the ratio of longitudinal over transverse pressure as
function of the radial coordinate is shown in Fig.~\ref{fig:pl_per_pt_gauss3fm}. The longitudinal and transverse
pressures can be calculated as
\begin{align}
\label{eq:pl}
 P_L &= T^{\mu\nu} \ell_\mu \ell_\nu = P - \pi^{zz}\;, \\
\label{eq:pt}
 P_T &= \frac{1}{2} (3P - P_L) = P + \frac{1}{2}\pi^{zz}\;,
\end{align}
where $\ell^\mu = \gamma_z(v_z, 0, 0, 1)$ is a space-like unit vector pointing into the longitudinal direction, with
$\gamma_z = (1 - v_z^2)^{-1/2}$. The last equalities in Eqs.~\eqref{eq:pl} and \eqref{eq:pt} hold in our case of
a boost-invariant expansion.
The comparison of fluid dynamics and BAMPS shows a similar pattern as $\pi^{xx}/e$, but deviations for
small cross sections are larger. As expected, the ratio stays always positive in the kinetic description,
while for small cross sections, the fluid-dynamical results assume negative values.

\subsection{\texorpdfstring{Gaussian profile $w=1\fm$}{Gaussian profile w=1fm}}
\label{subsec:Gauss1fm}

Now we consider the smaller value $w=1\fm$ for the width of the Gaussian number-density profile.
The cross section is varied from $\sigma = 20 \mb$ down to $1\mb$, as before.
The space-time evolution of the Knudsen number
${\rm Kn} = \lambda_{\rm mfp}\theta$ is shown for each case in the lower row of
Fig.~\ref{fig:knudsen_contour_gauss3fm}. As before, the $N_{\rm test}=4$ contour shows the region where the
comparison between fluid dynamics and BAMPS is meaningful. The plots in Fig.~\ref{fig:knudsen_contour_gauss3fm}
for the small width $w=1\fm$ look very similar to that for the larger one, but notice the difference in
the space and time scales.

\begin{figure*}
\includegraphics[width = 0.4\textwidth]{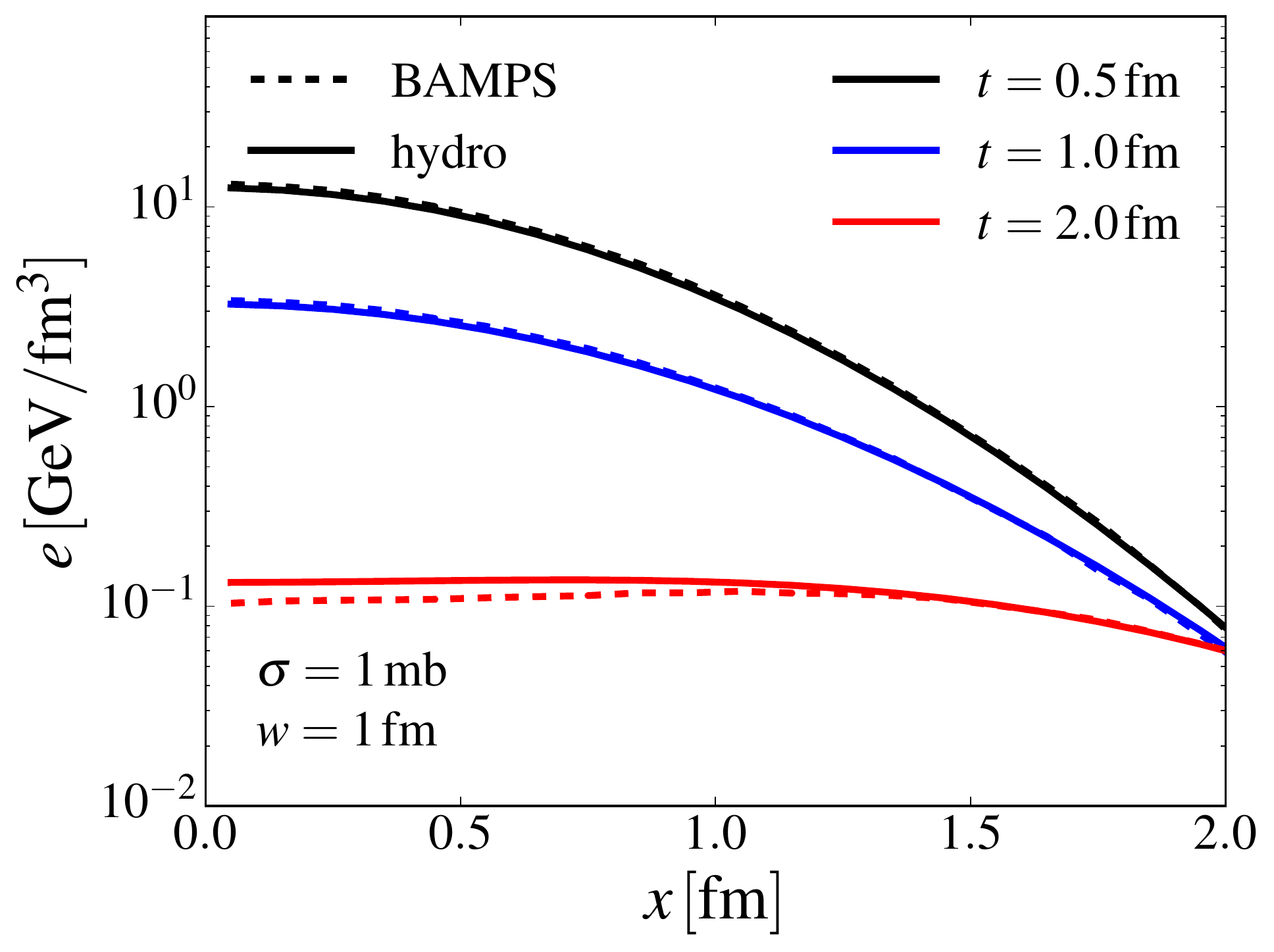}
\includegraphics[width = 0.4\textwidth]{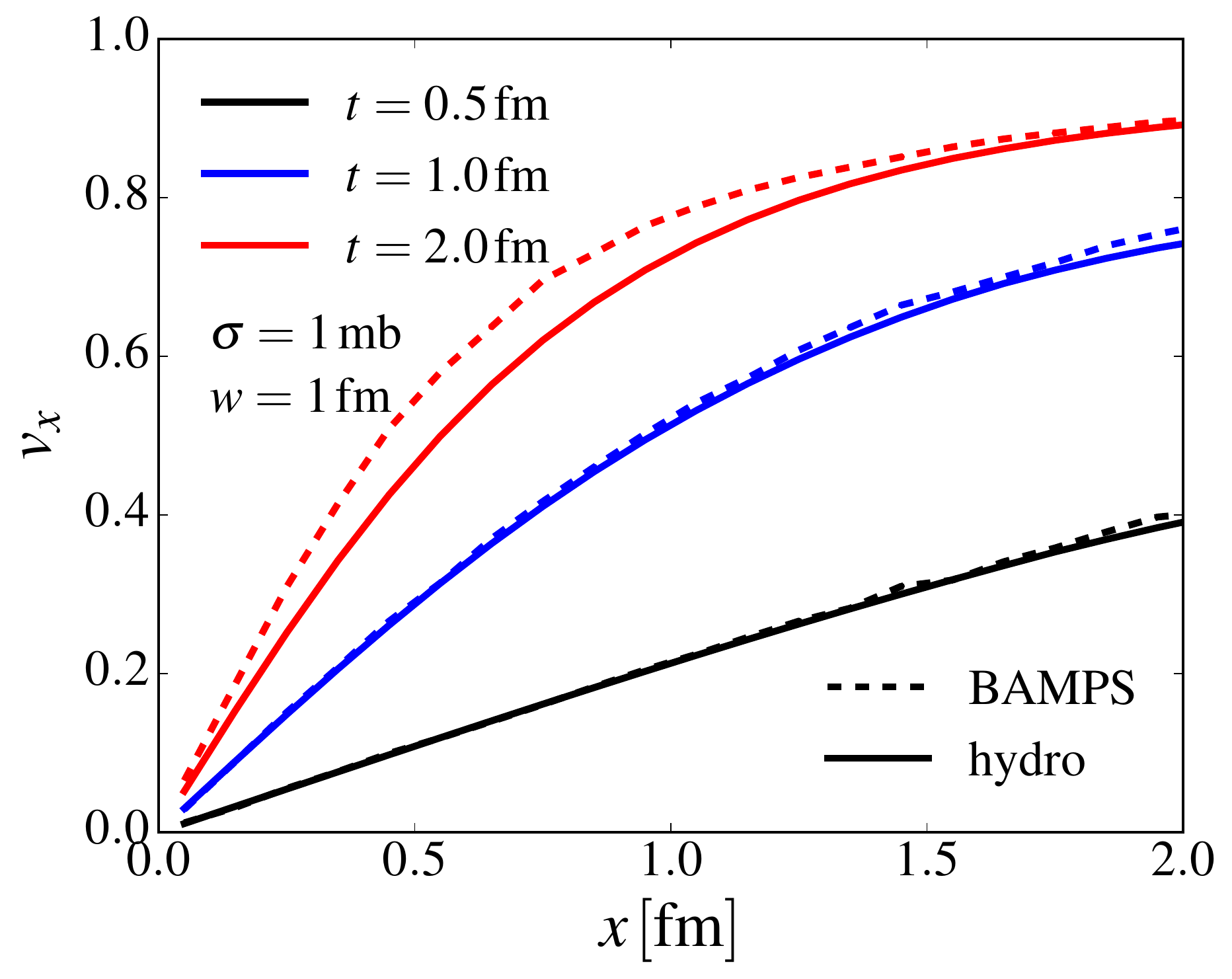}
 \caption{(Color online) Energy-density (left) and $v_x$ (right) profiles for $\sigma=1\mb$ and a $w=1\fm$
 Gaussian initial density profile for different times $t=0.5, \, 1$, and 2 fm. Solid lines represent the fluid-dynamical results,
 while dashed lines show the BAMPS solutions, as in Fig.~\ref{fig:edvx_gauss3fm}.}
 \label{fig:edvx_gauss1fm}
\end{figure*}

The spatial profiles of energy density and velocity with $\sigma = 1$ mb are shown in Fig.~\ref{fig:edvx_gauss1fm} for
different times. As for the case $w=3\fm$, the agreement between fluid dynamics and the Boltzmann equation is
very good for all cross sections. Notable differences appear only in the region where $N_{\rm test} < 4$, i.e., where the
particles in the BAMPS calculation are essentially free-streaming.

\begin{figure*}
 \includegraphics[width = 0.32\textwidth]{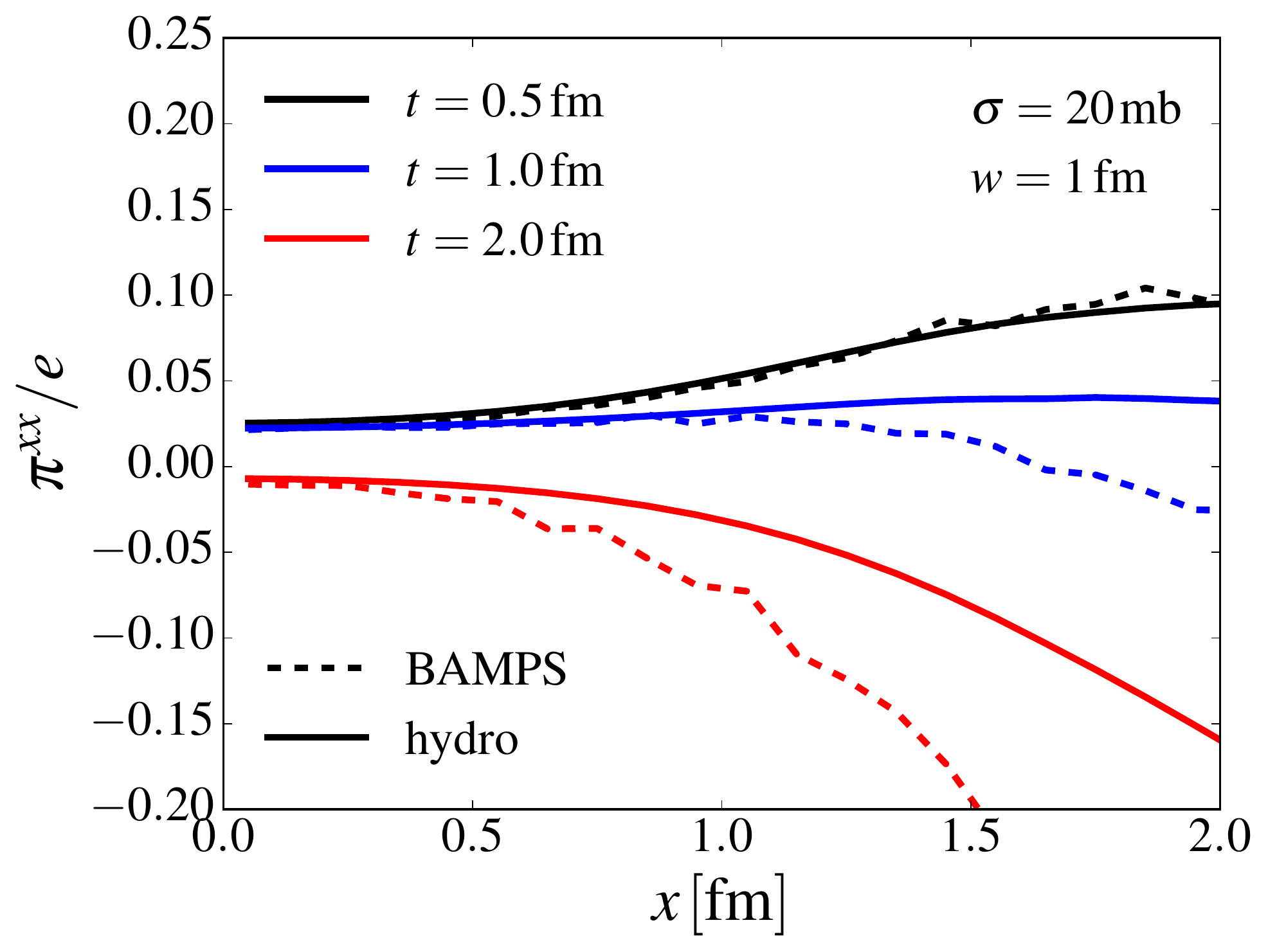}
 \includegraphics[width = 0.32\textwidth]{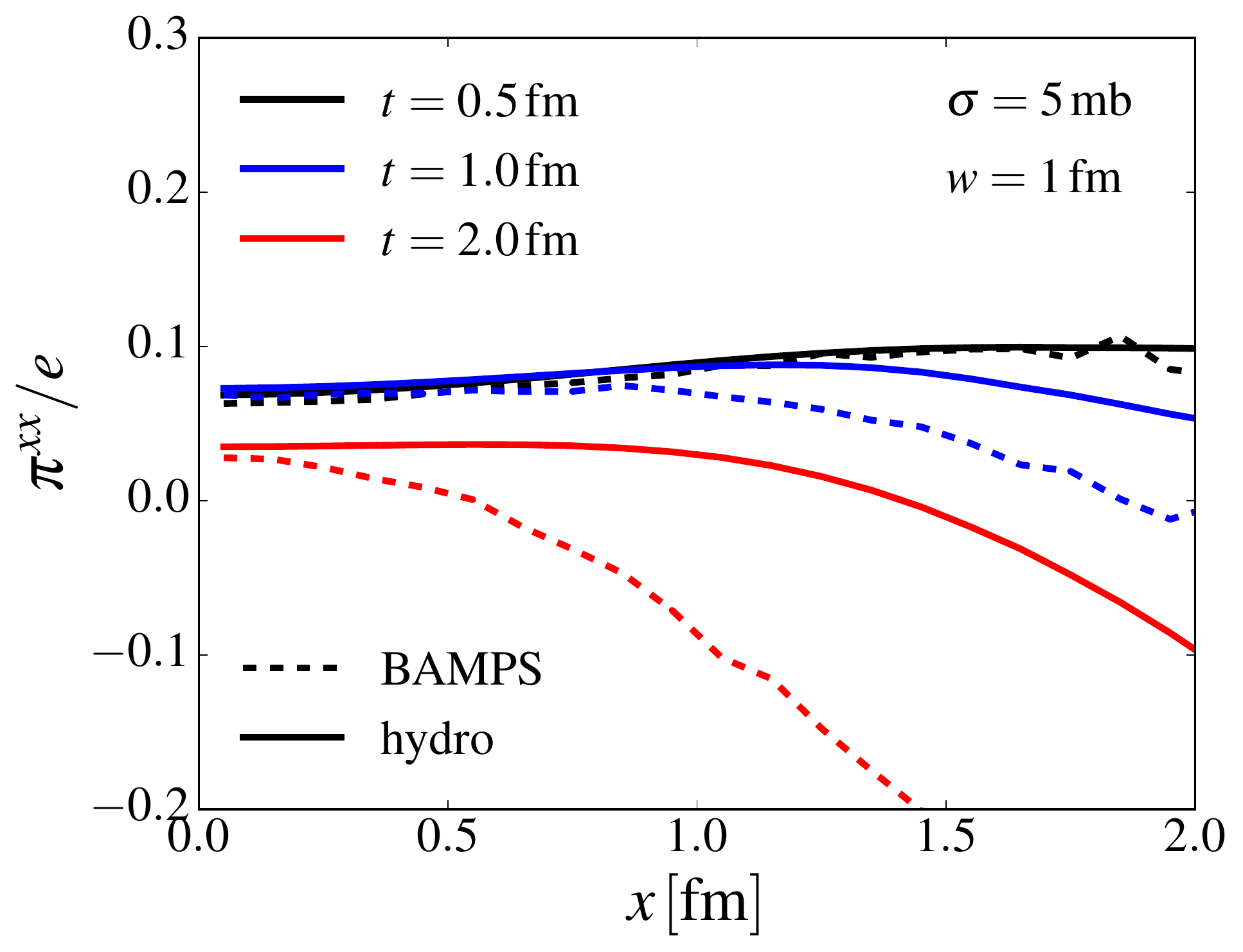}
 \includegraphics[width = 0.32\textwidth]{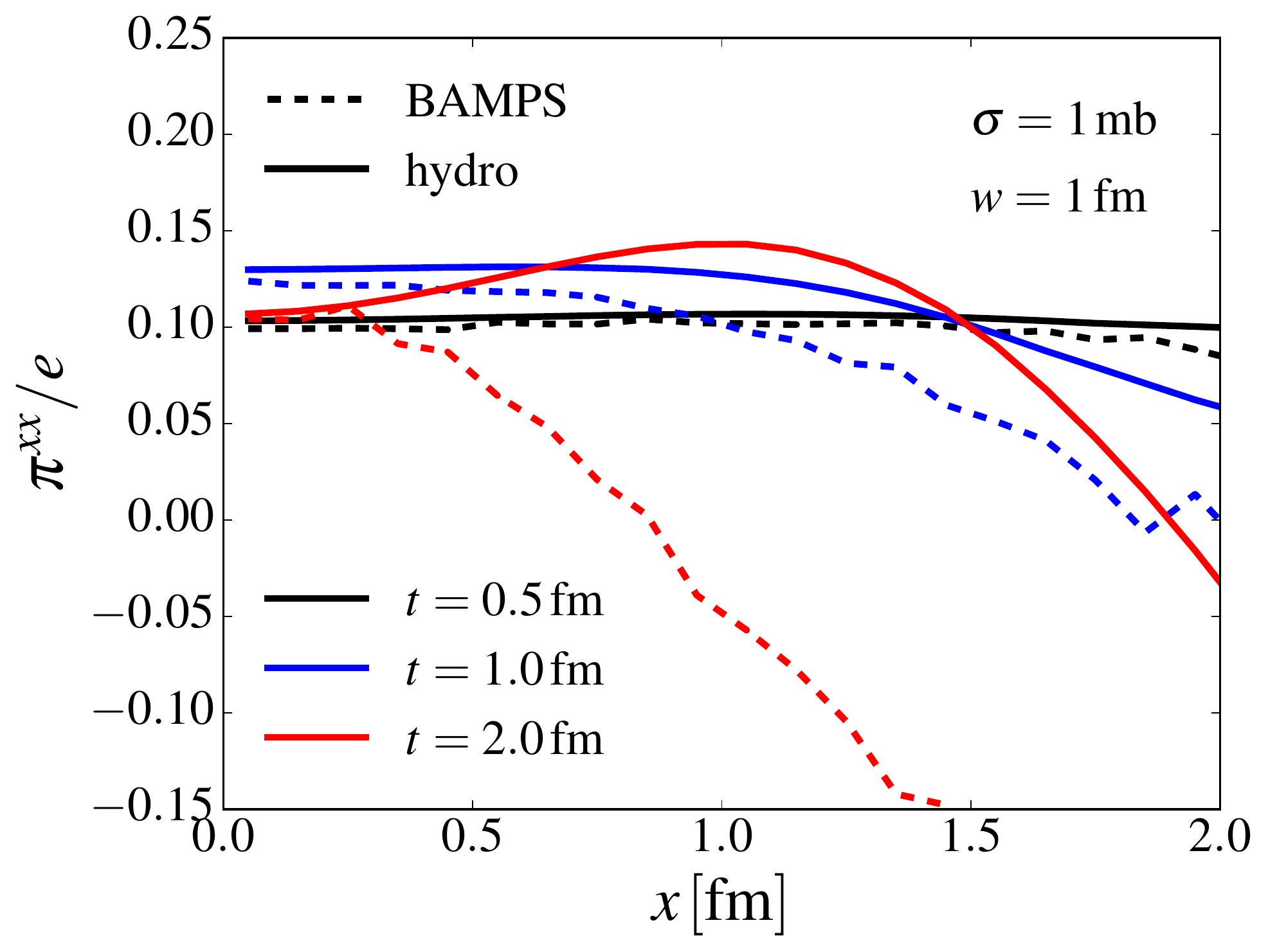}
 \caption{(Color online) The $\pi^{xx}$ profiles scaled by energy density for $\sigma=20\mb, 5\mb$, and $1\mb$
 (from left to right) and a $w=1\fm$ Gaussian initial density profile, as in Fig.~\ref{fig:pizzxx_gauss3fm}.}
 \label{fig:pizzxx_gauss1fm}

 \includegraphics[width = 0.32\textwidth]{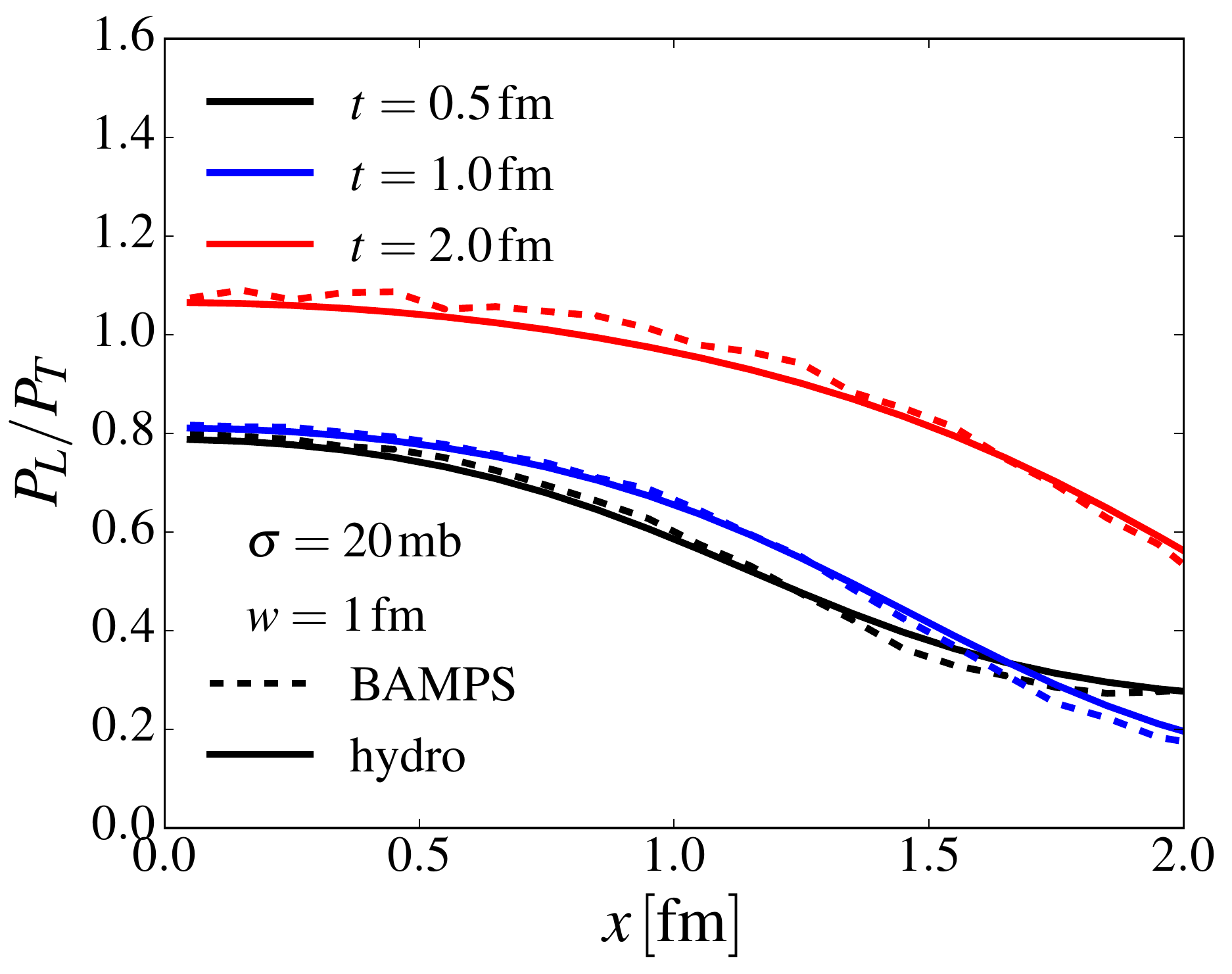}
 \includegraphics[width = 0.32\textwidth]{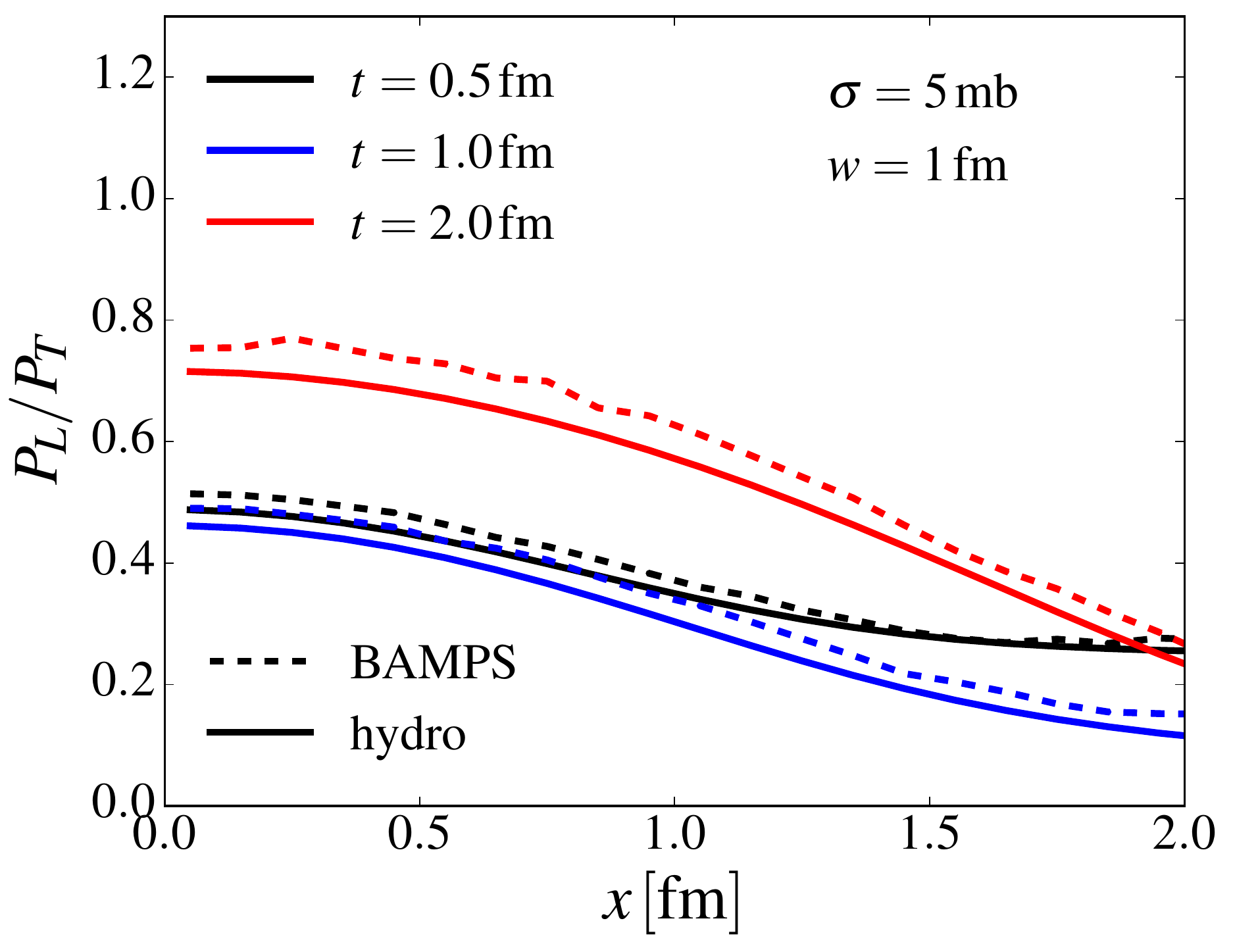}
 \includegraphics[width = 0.32\textwidth]{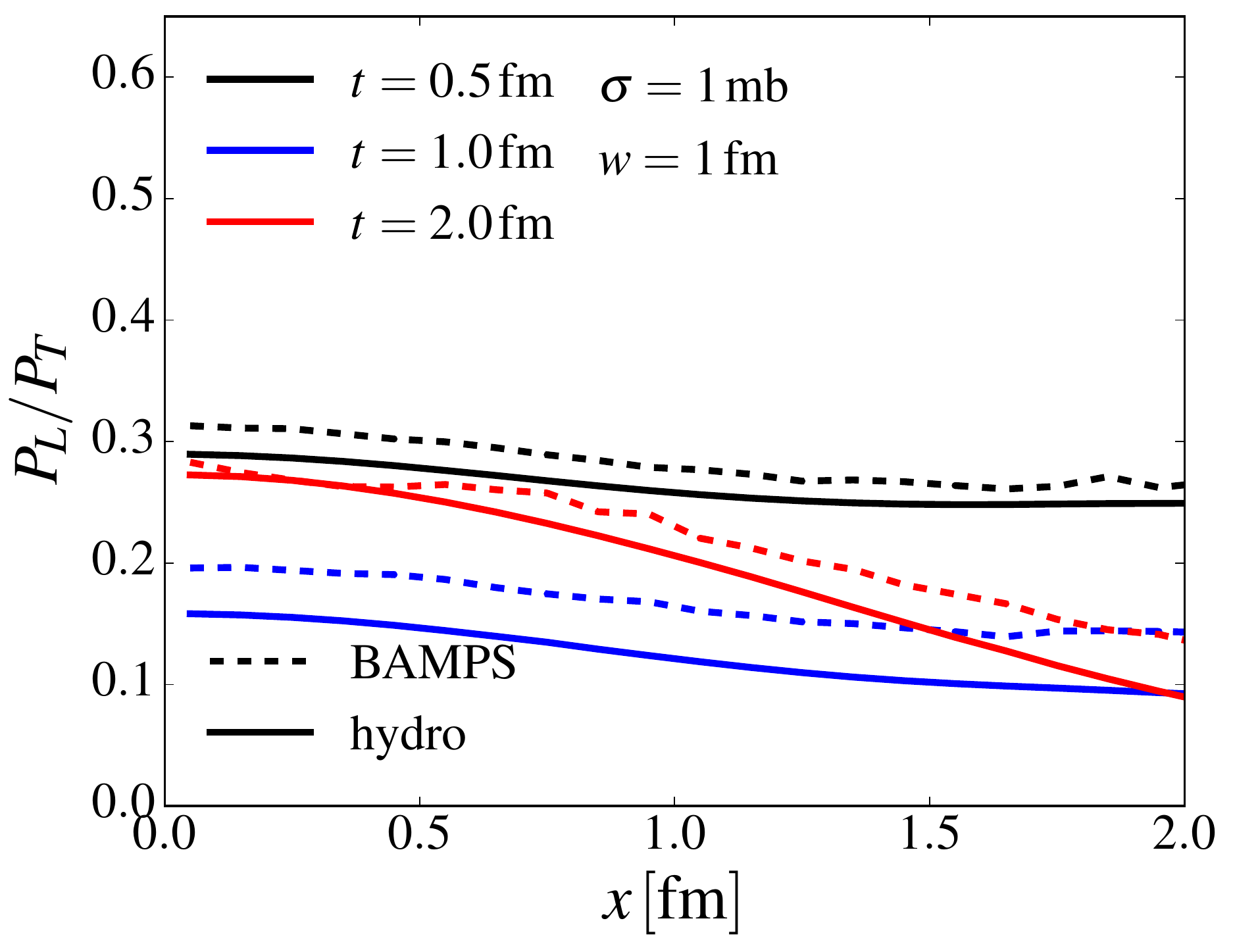}
 \caption{(Color online) The ratio of longitudinal over transversal pressure $P_L/P_T$ for $\sigma=20\mb, 5\mb$,
 and $1\mb$ (from left to right) and a $w=1\fm$ Gaussian initial density profile, as in Fig.~\ref{fig:pl_per_pt_gauss3fm}.}
 \label{fig:pl_per_pt_gauss1fm}
\end{figure*}

As mentioned and already observed above, the shear-stress tensor exhibits a much larger
sensitivity to the Knudsen number. The $\pi^{xx}/e$ components for the smaller system are shown for different
cross sections in Fig.~\ref{fig:pizzxx_gauss1fm}.
For the large $\sigma = 20\mb$ cross section the agreement between the two approaches remains good until
$N_{\rm test}=4$ is reached, but for smaller cross sections there are larger deviations at the end of the
evolution, $t \sim 1 - 2\fm$. We note, however, that even with the smallest cross section $\sigma = 1\mb$, 
the profiles are
quite well described up to $t=1\fm$, even if $\rm Kn \sim 2 - 4$ throughout the evolution.

Again, as for the larger system, also the ratio of longitudinal over transverse pressure is examined, see
Fig.~\ref{fig:pl_per_pt_gauss1fm}. The discrepancies between the kinetic and the fluid-dynamical calculations are not
as big as in the case of the large system with $w=3\fm$. We note that in the smaller system also the momentum-space
distributions deviate less from an isotropic distribution (characterized by $P_L/P_T = 1$). The reason for this is the
stronger transverse expansion in the smaller system that counteracts the initial strong asymmetry in the longitudinal
versus the transverse expansion, which is the main driving force in reducing the
$P_L/P_T$ ratio.

Overall, the agreement between kinetic theory and fluid dynamics is very good for $\sigma = 20\mb $ and
$\sigma = 5\mb$. Only for $\sigma = 1\mb$ there are significant deviations between the two approaches.
This is also expected, as for $\sigma = 1\mb$ one finds ${\rm Kn} \gtrsim 2$ already at the beginning of the evolution.
It is notable, however, that the energy-density and velocity profiles are well described by fluid dynamics in all cases.

\subsection{\texorpdfstring{Binary Glauber profile $b=7.5\fm$}{Binary Glauber profile b=7.5fm}}
\label{subsec:Binary7fm}

\begin{figure*}
 \includegraphics[width = 0.32\textwidth]{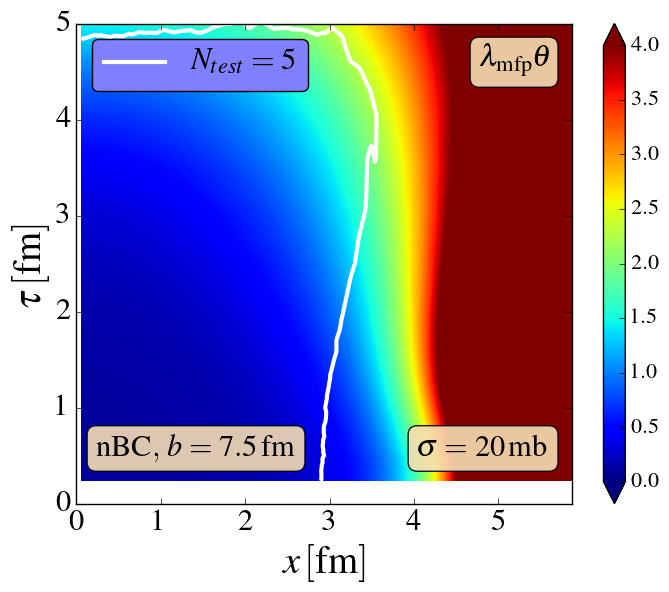}
 \includegraphics[width = 0.32\textwidth]{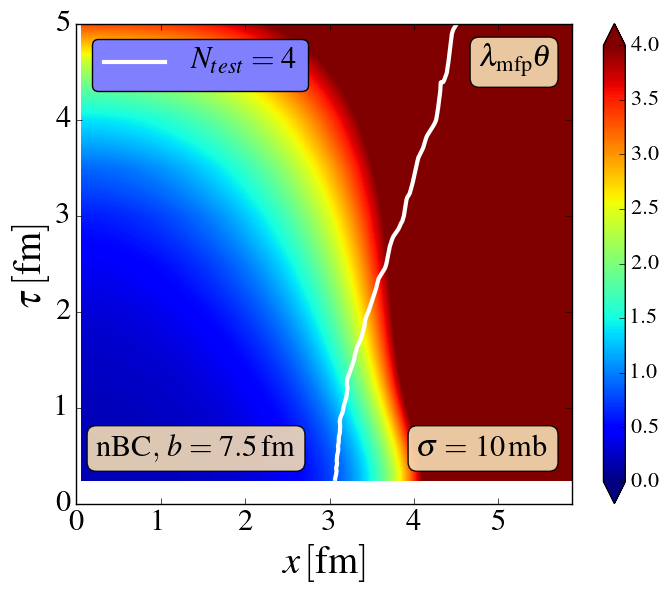}
 \includegraphics[width = 0.32\textwidth]{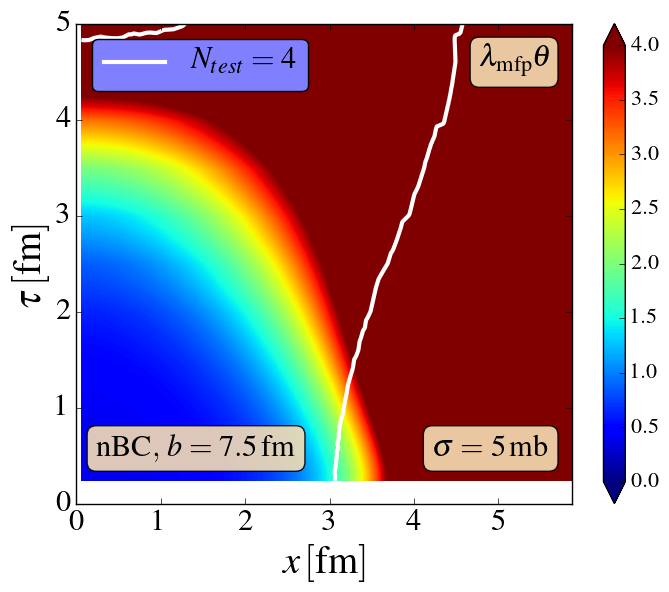}
 \caption{(Color online) Knudsen-number contours for an nBC Glauber initial profile for $\sigma = 20$, $10$, and
 $5\mb$ (from left to right). The white line corresponds to the testparticle number per computational cell
 $N_{\rm test}=4$ contour in the BAMPS calculation.}
 \label{fig:knudsen_contour_binary}
\end{figure*}

\begin{figure*}
 \includegraphics[width = 0.45\textwidth]{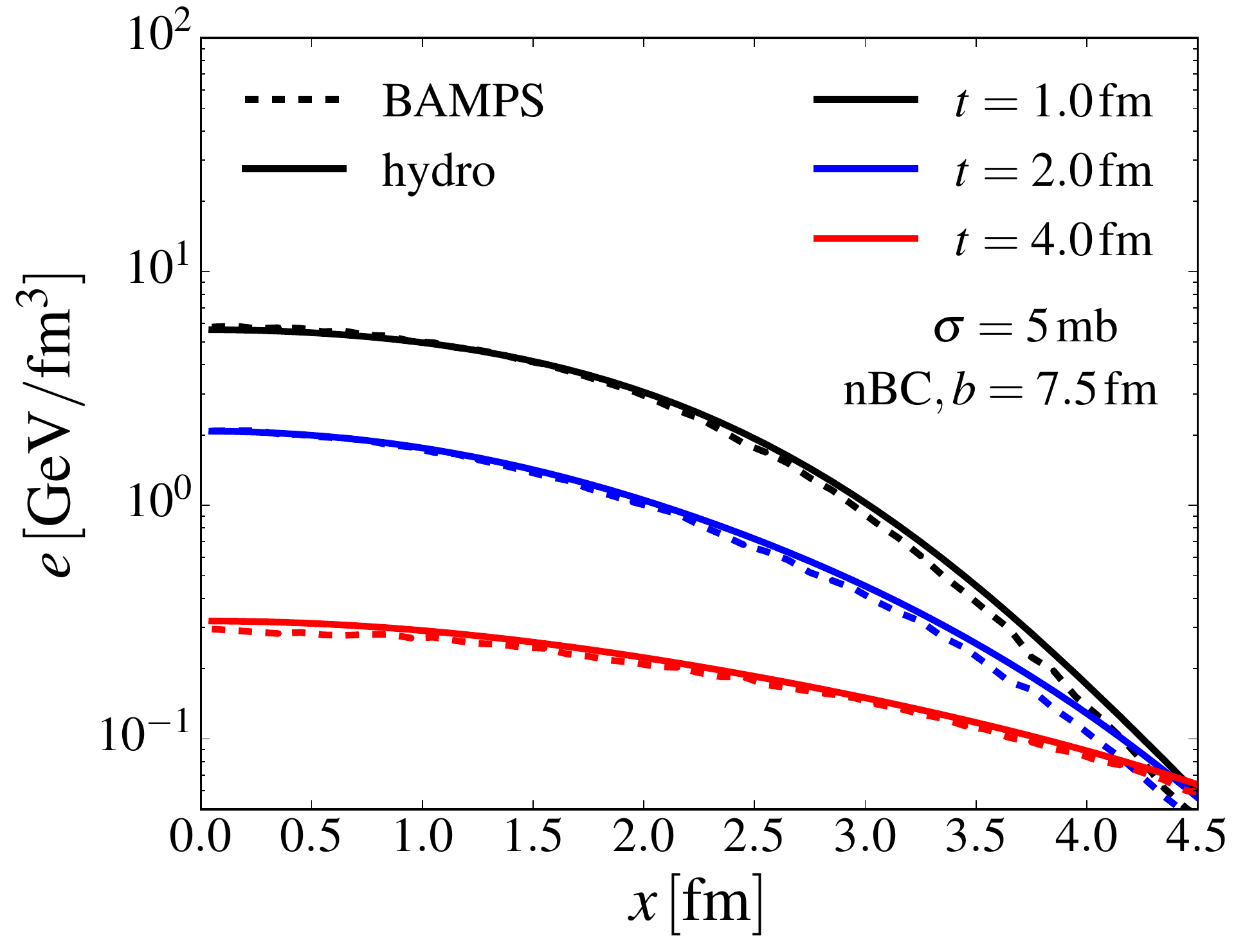}
 \includegraphics[width = 0.45\textwidth]{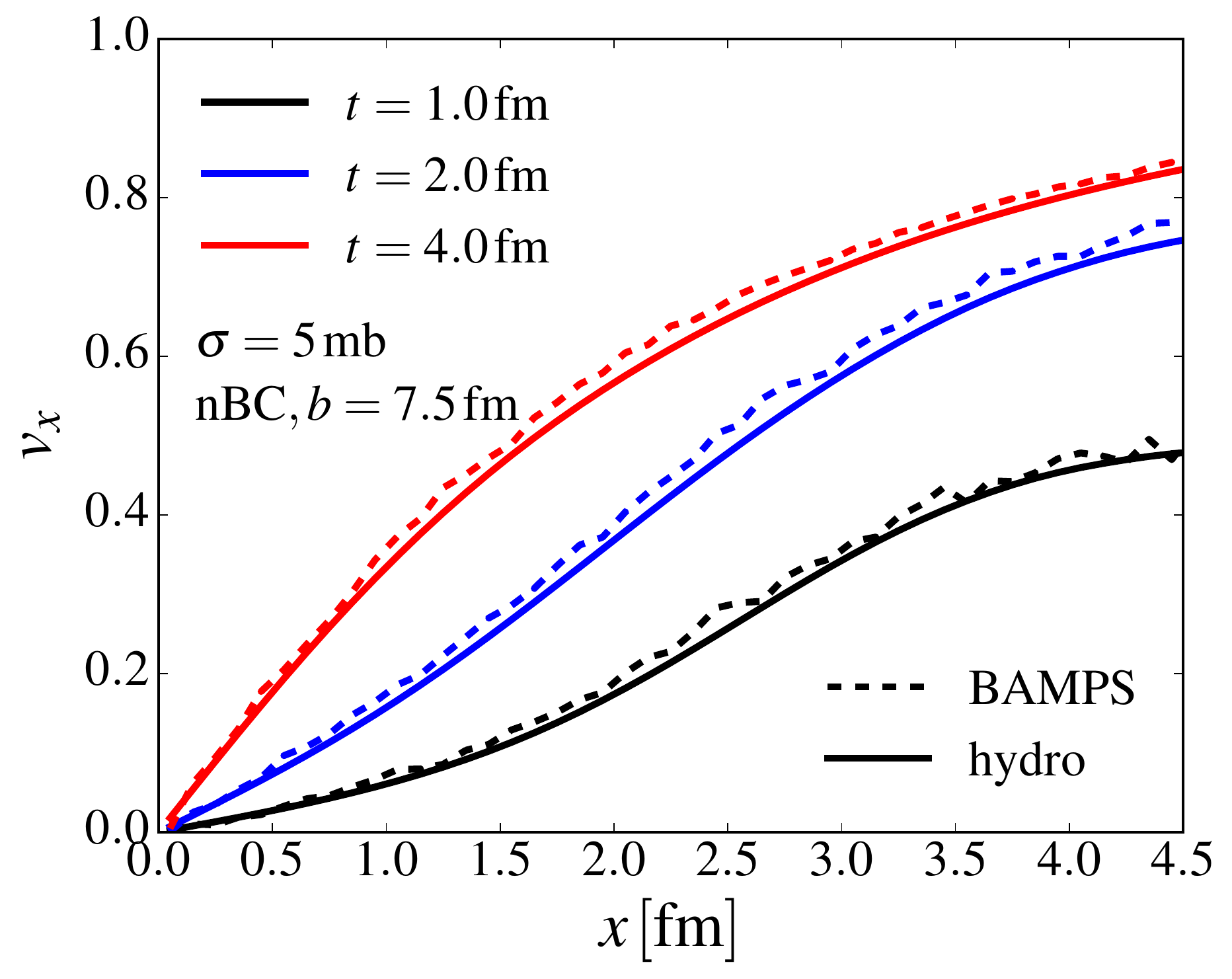}
 \caption{(Color online) Energy-density (left) and $v_x$ (right) profiles for $\sigma=5\mb$ and an nBC
 Glauber initial density profile for times $t=1,\, 2,$ and 4 fm. Solid lines represent fluid-dynamical
 results, while dashed lines show the BAMPS solutions.}
 \label{fig:edvx_binary}
\end{figure*}

\begin{figure*}
 \includegraphics[width = 0.32\textwidth]{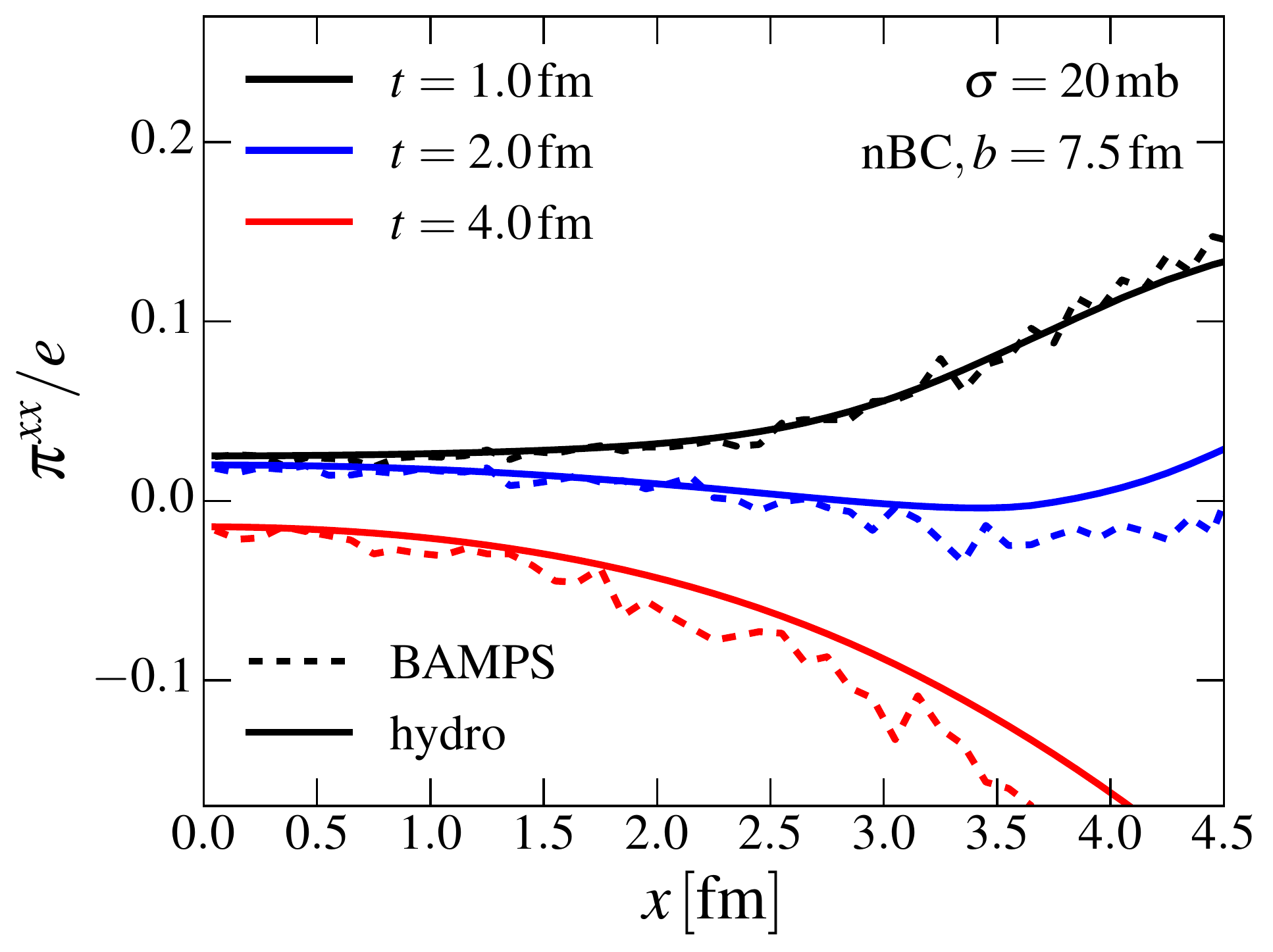}
 \includegraphics[width = 0.32\textwidth]{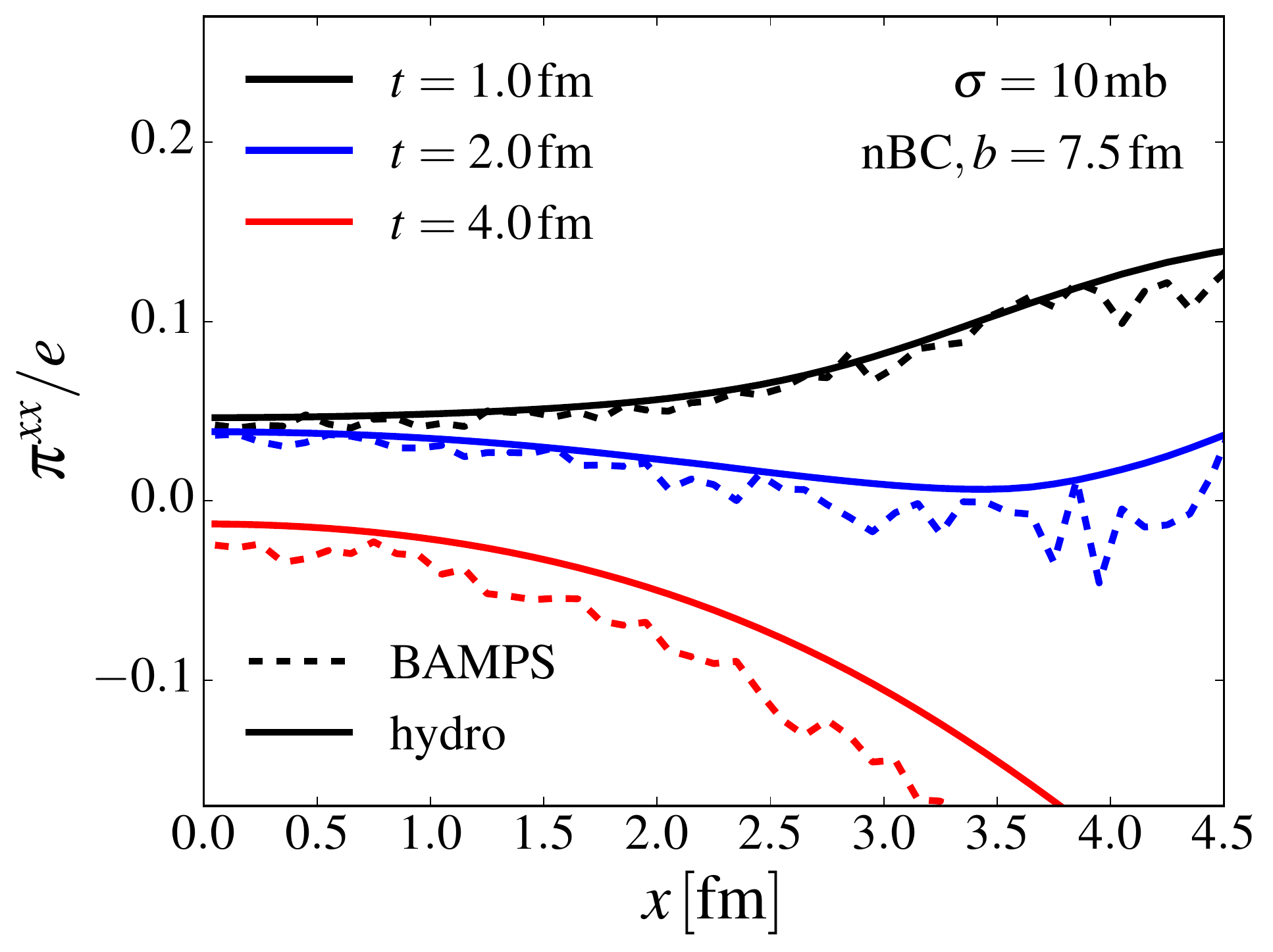}
 \includegraphics[width = 0.32\textwidth]{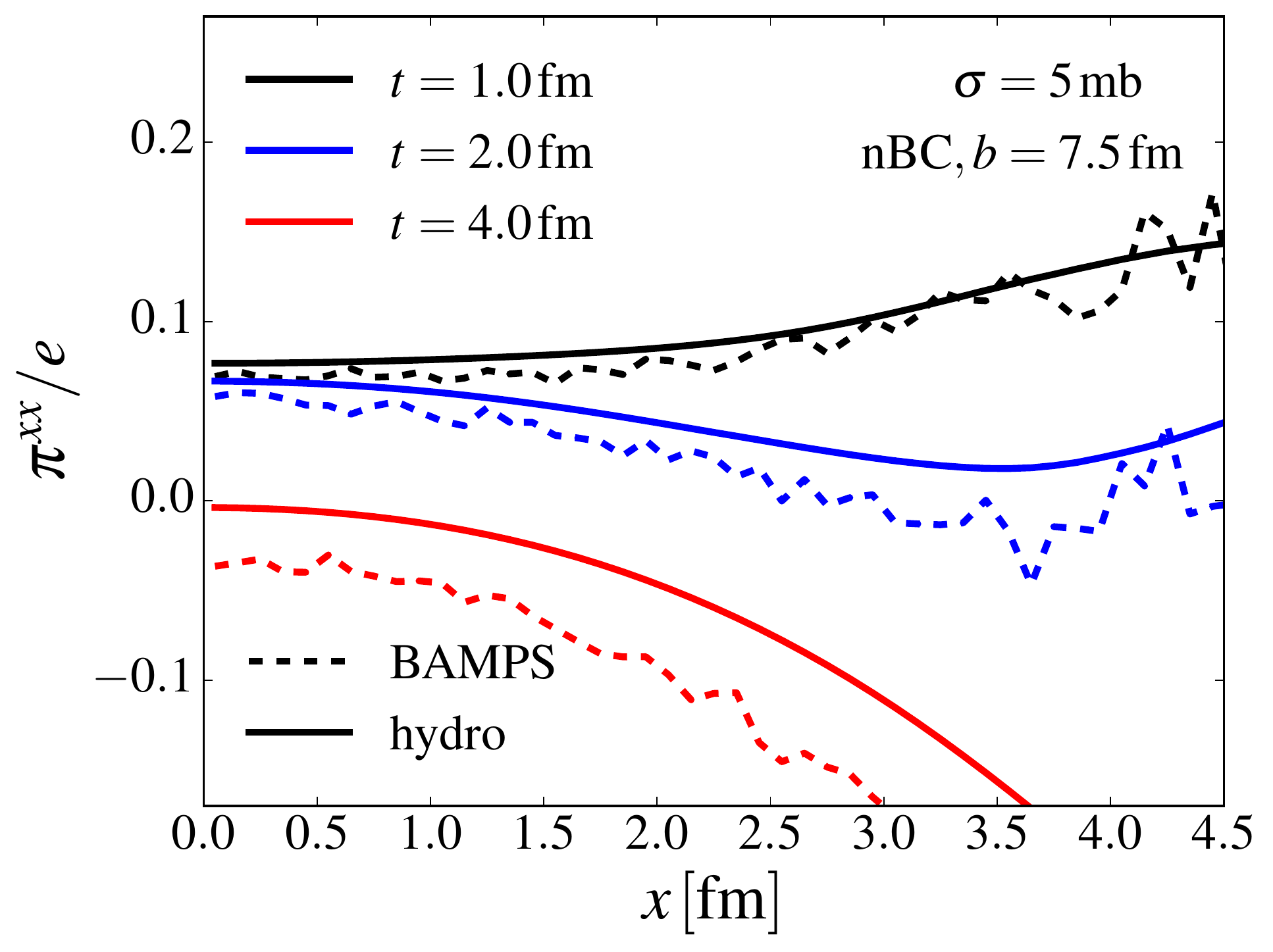}
 \caption{(Color online) The $\pi^{xx}$ profiles scaled by energy density $e$ for $\sigma = 20$, $10$, and $5\mb$ and an
 nBC Glauber initial density profile.}
 \label{fig:pizzxx_binary}
 \includegraphics[width = 0.32\textwidth]{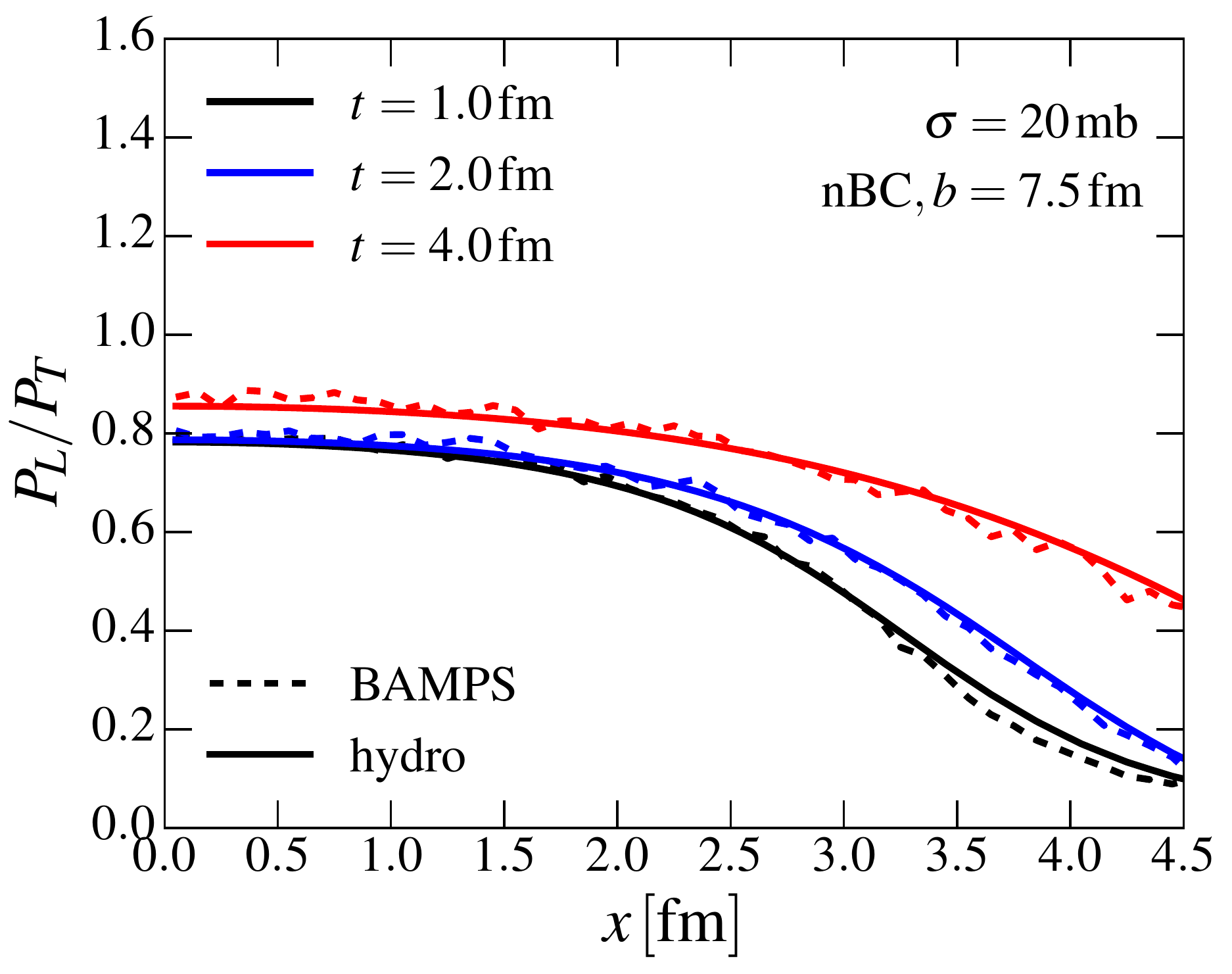}
 \includegraphics[width = 0.32\textwidth]{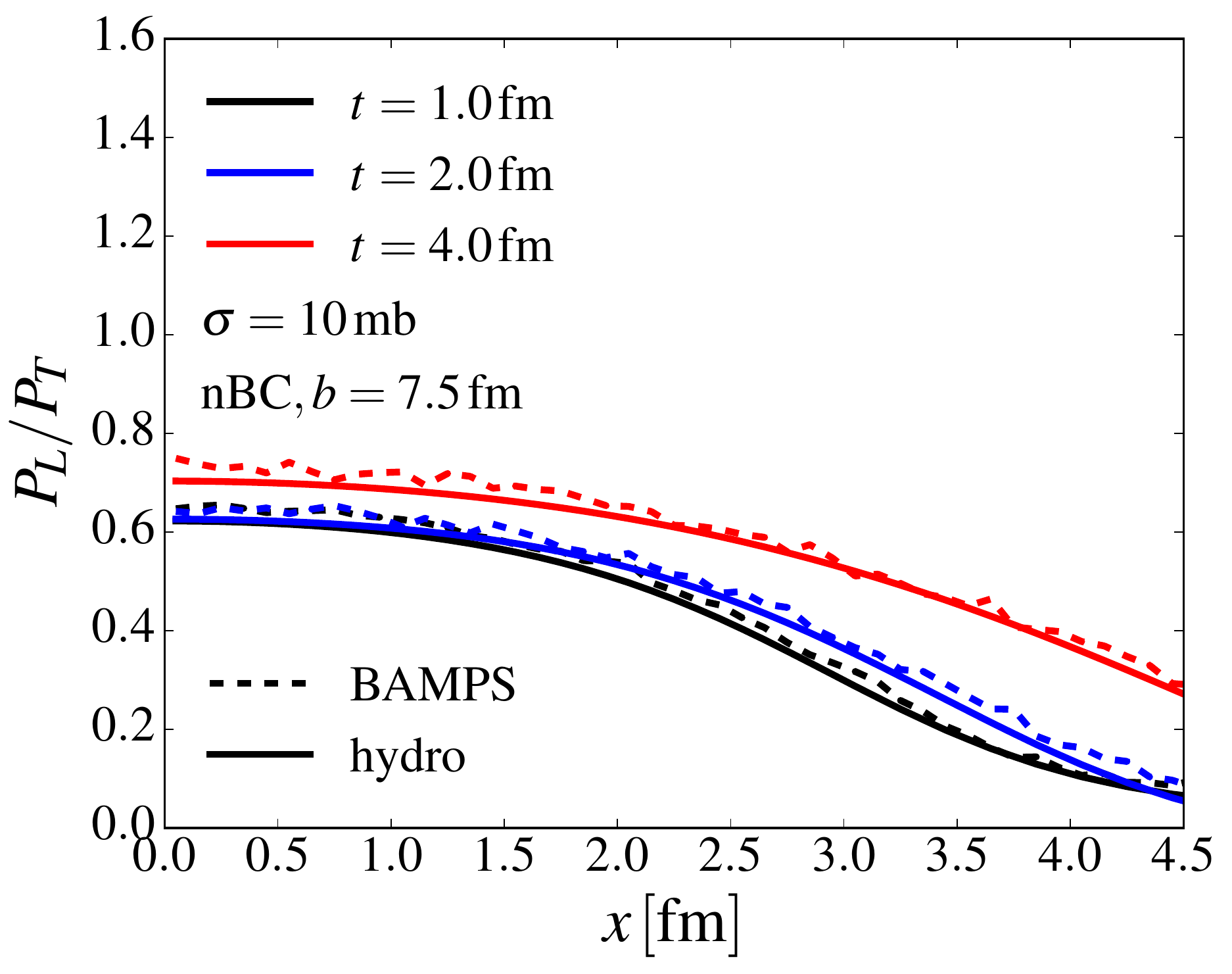}
 \includegraphics[width = 0.32\textwidth]{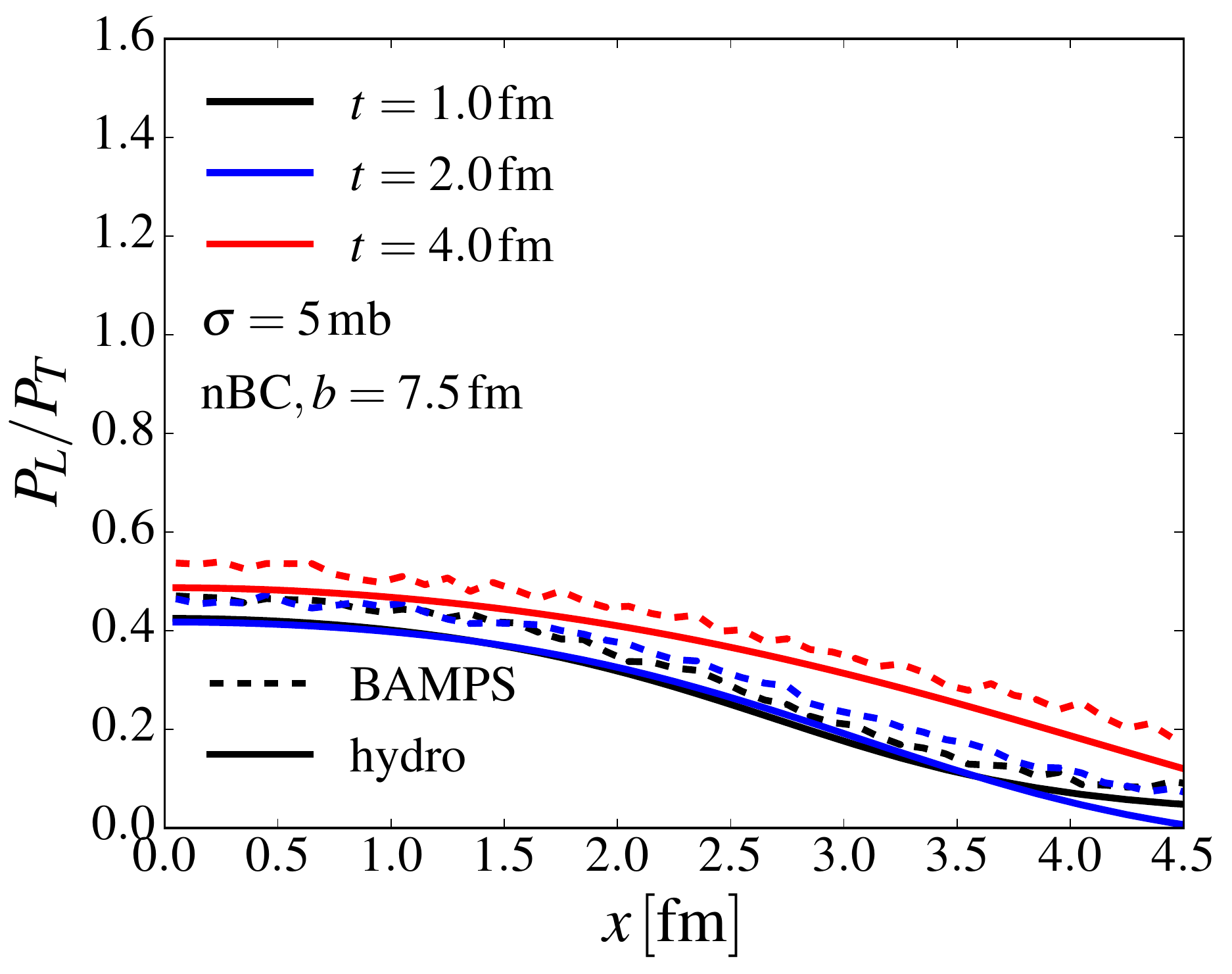}
 \caption{(Color online) The ratio of longitudinal over transverse pressure $P_L/P_T$ for $\sigma = 20$, $10$,
 and $5\mb$ and an nBC Glauber initial density profile.}
 \label{fig:pl_per_pt_binary}
\end{figure*}

The last case considered is an initial nBC profile from the Glauber model
with an impact parameter $b = 7.5\fm$. The Knudsen-number contours are shown in
Fig.~\ref{fig:knudsen_contour_binary}, the energy-density and
velocity profiles are shown in Fig.~\ref{fig:edvx_binary}, and the $\pi^{xx}/e$ profiles in Fig.~\ref{fig:pizzxx_binary}.
Note that in this case the system is not azimuthally symmetric anymore, and the profiles are shown along the $x$-axis,
which is the direction of the impact parameter. In this case, calculations for three values of the cross section,
$\sigma = 20$, $10$, and $5\mb$, are shown. The agreement between BAMPS and fluid dynamics is quite
similar to the Gaussian profile with $w = 3\fm$, which is not surprising as the initial profiles are quite similar as seen from
Fig.~\ref{fig:iniconditions}. The comparison looks very similar also along the $y$-axis.

As before, also in this case the agreement of the energy-density and velocity profiles remains good even in space-time
regions where the Knudsen number becomes large. The good agreement holds for all values of the cross section. The
same is true for $\pi^{xx}/e$, shown in Fig.~\ref{fig:pizzxx_binary}, and for the $P_L/P_T$ ratio, shown in
Fig.~\ref{fig:pl_per_pt_binary}. Only at $t=4\fm$ and with $\sigma = 5\mb$ one can observe more significant
deviations in $\pi^{xx}/e$. This is also the region where the Knudsen number becomes clearly larger than one.

\begin{figure*}
 \includegraphics[width = 0.32\textwidth]{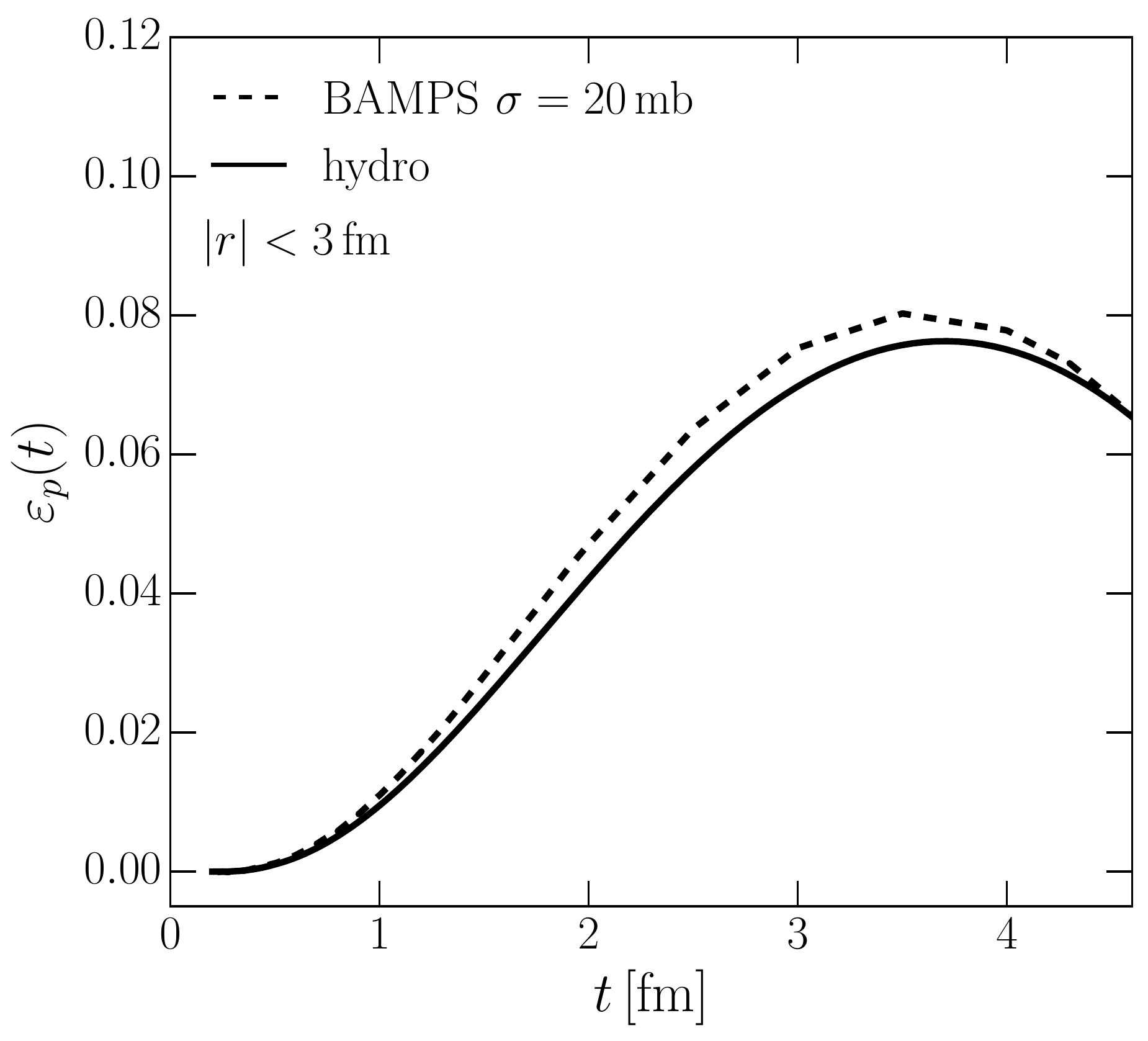}
 \includegraphics[width = 0.32\textwidth]{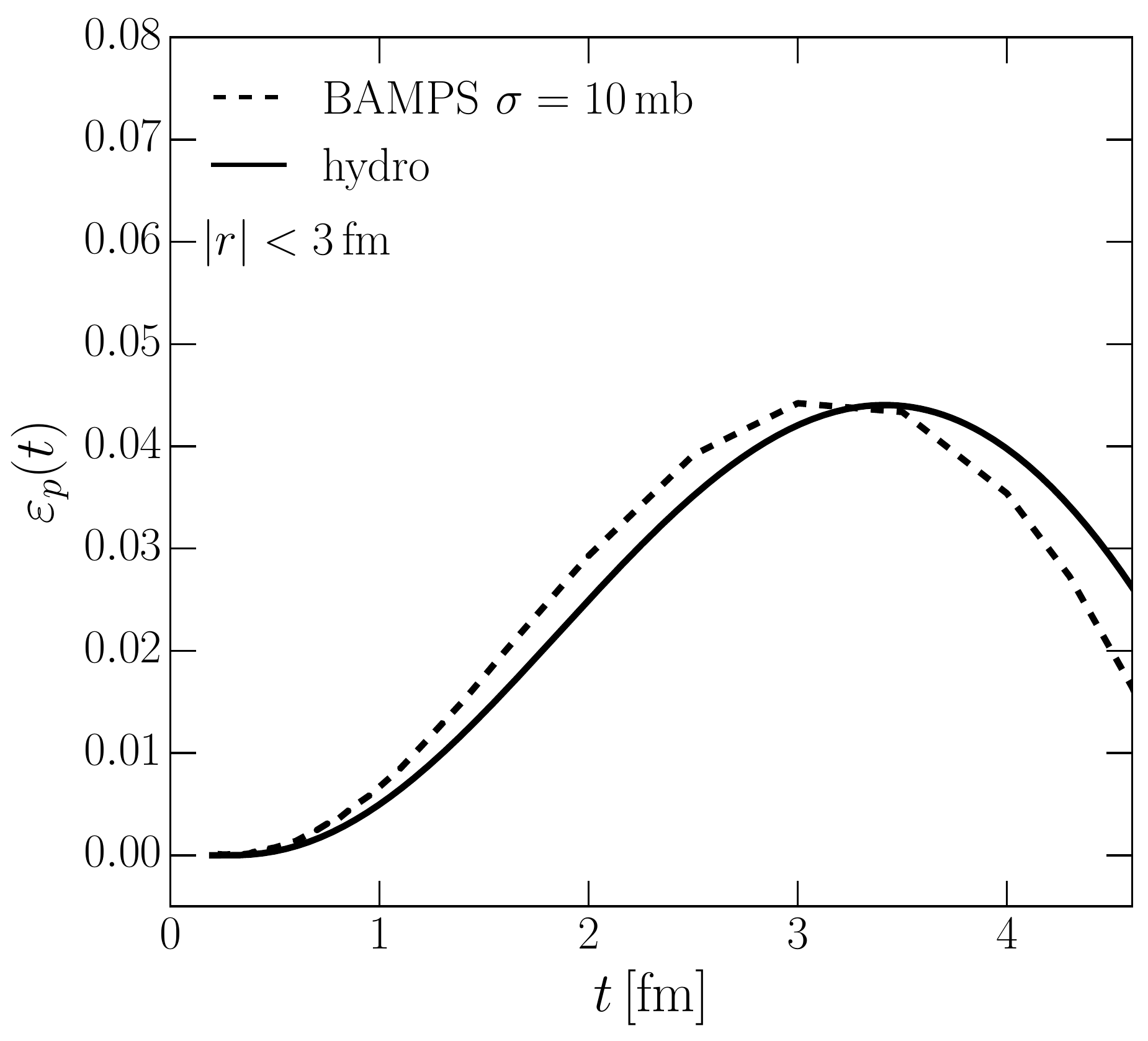}
 \includegraphics[width = 0.32\textwidth]{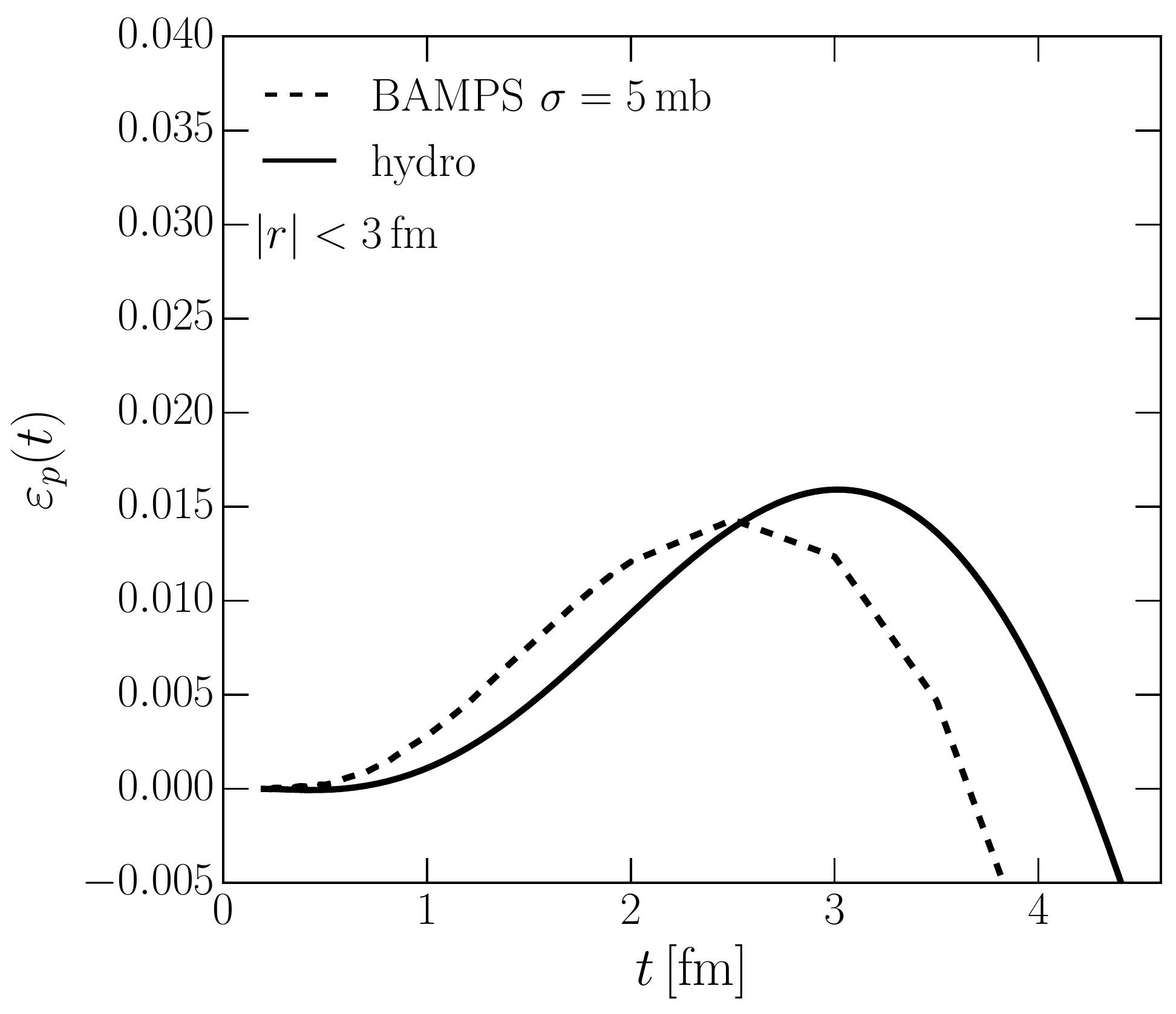}
 \caption{(Color online) Momentum-space eccentricity as function of time, for $\sigma = 20$,  $10$, and $5\mb$ and
 nBC initial conditions.}
 \label{fig:epsilon_p}
\end{figure*}

The momentum-space asymmetry of the solutions can be quantified by calculating the so-called momentum-space
eccentricity,
\begin{equation}
 \varepsilon_p = \frac{\langle T^{xx} - T^{yy}\rangle}{\langle T^{xx} + T^{yy}\rangle}\; ,
\end{equation}
where the angular brackets denote the integral over the transverse plane at fixed time,
\begin{equation}
 \langle \cdots \rangle = \frac{1}{\Delta z}\int \dd x \dd y \dd z (\cdots)\; ,
\end{equation}
with the average over the $z$-coordinate performed as in Eq.~\eqref{eq:Tzaverage}. In order to avoid space-time
regions where the BAMPS particles are free-streaming, we restrict the transverse integrals to radial
distances $r\leq 3\fm$. The
time evolution of the momentum eccentricities is shown in Fig.~\ref{fig:epsilon_p} for all three values of the cross section.
As seen in the figures, in the $\sigma = 20$ and $10\mb$ cases the agreement between fluid dynamics and the
Boltzmann equation is good throughout the evolution, but for the $5\mb$ cross section the differences, especially
towards the end of the evolution, are more pronounced. However, the overall magnitude of the momentum
eccentricity is still well described even for $\sigma = 5\mb$.

\section{Transverse-momentum spectra and elliptic flow}
\label{sec:Freezeout}

In the solution of the Boltzmann equation the particles automatically decouple when the mean free path
becomes sufficiently large, and the transverse-momentum spectra can be calculated simply by counting
particles at sufficiently late time. However, in fluid dynamics this freeze-out must be modeled separately,
as the applicability of the fluid-dynamical equations of motion does not extend all the way to free-streaming.
The standard way to model decoupling is the so-called Cooper-Frye freeze-out procedure~\cite{Cooper:1974mv},
where the final transverse-momentum spectrum is calculated by integrating the particle flux through some decoupling
surface $\Sigma$,
\begin{equation}
 E\frac{\dd N}{\dd^3 \mathbf{k}} = \frac{\dd N}{\dd y_p \dd^2 \mathbf{k}_T }
 = \int_\Sigma \dd\Sigma_\mu k^\mu f_{\mathbf{k}}(x)\; ,
\end{equation}
where $\Sigma_\mu$ is the normal vector on the (suitably defined) freeze-out surface $\Sigma$,
$y_p= 1/2\,\ln[(k+k^z)/(k-k^z)]$
is the longitudinal rapidity of the (massless) particle, and
$\mathbf{k_T}$ is its transverse momentum. The single-particle distribution function
$f_{\mathbf{k}}(x)$ is given by Eqs.~(\ref{eq:fk}) and (\ref{eq:deltaf}) in the 14-moment approximation. 
In the following the standard freeze-out surface is taken as a surface of
constant Knudsen number, $\mathrm{Kn_{fr}} = \lambda_{\rm mfp}\theta =  C_{\rm fr}$, with $C_{\rm fr}$ being some
constant to be determined later. As described above, in addition to this physical freeze-out condition, in the BAMPS
calculation the particles become free-streaming when the number of testparticles in the computational cell becomes less
than four. This needs to be taken into account also in the fluid-dynamical calculation, in order to have a meaningful
comparison with BAMPS results. In practice, we use the testparticle number per cell $N_{\rm test}=5$ contour as the
freeze-out surface, if the testparticle number drops below this limit before the Knudsen-number criterion is reached. We
note that $N_{\rm test}$ fluctuates around the average in each BAMPS simulation, so that this numerical freeze-out is not
really a sharp surface. Thus, the constant $N_{\rm test}$ freeze-out in fluid dynamics should also be considered as an
effective description. In practice, $N_{\rm test}=5$ gives a good agreement with the BAMPS low-$k_T$ spectra when the
freeze-out is almost completely determined by the $N_{\rm test}$ criterion. Moreover, some of the particles decouple
immediately at the beginning of the evolution, i.e., the Knudsen-number or testparticle-number criterion is reached
already at $\tau = \tau_0$. This part needs to be included into the construction of the decoupling surface.
The contour surfaces of the solution needed for the calculation of the particle spectra are determined by the
Cornelius algorithm~\cite{Huovinen:2012is}.

The most commonly used decoupling criterion in modeling heavy-ion collisions is that of a constant temperature.
One can argue that decoupling happens when the mean free path of the particles becomes of the order of the size of
the system. If one assumes that the system size does not vary significantly between different collisions, the criterion can
also be written as $\lambda_{\rm mfp}(T,\alpha_0) = const$. If chemical potentials can be neglected this is equivalent to the
constant-temperature decoupling condition. Our standard condition above is similar to the dynamical freeze-out
condition of Refs.~\cite{Holopainen:2012id, Eskola:2007zc, Heinz:2007in, Dumitru:1999ud, Hung:1997du}. Additionally,
we also consider the constant-mean free path and constant-temperature decoupling conditions. These also include the
$N_{\rm test}=5$ surface as a part of the freeze-out surface.

The generic shape of the freeze-out surface can be read off from Fig.~\ref{fig:knudsen_contour_binary}, where both the
Knudsen-number contours and $N_{\rm test}=4$ contour are shown. The surface that corresponds to
the immediately frozen-out particles is the $\tau = 0.2\fm$ surface extending from around $x \sim 3\fm$ to the
edge of the system.

The azimuthal asymmetry of the spectrum is usually characterized by the Fourier decomposition
\begin{align}
 \frac{\dd N}{\dd y_p \dd k_T^2 \dd \phi} &= \frac{\dd N}{\dd y_p \dd k_T^2}\left[ 1 + 2v_2(k_T) \cos(2\phi) \right.\nonumber\\
&\qquad\qquad \left.+ 2v_4(k_T) \cos(4\phi) + \cdots \right]\; ,
\end{align}
where $\phi$ is the azimuthal angle. For the boost-invariant expansion one can replace the particle rapidity $y_p$ by the
space-time rapidity $\eta_s$, i.e.,
\begin{equation}
 \frac{\dd N}{\dd y_p \dd k_T^2 \dd \phi} =  \frac{\dd N}{\dd \eta_s \dd k_T^2 \dd \phi}\; .
\end{equation}

\begin{figure*}
\includegraphics[width = 0.45\textwidth]{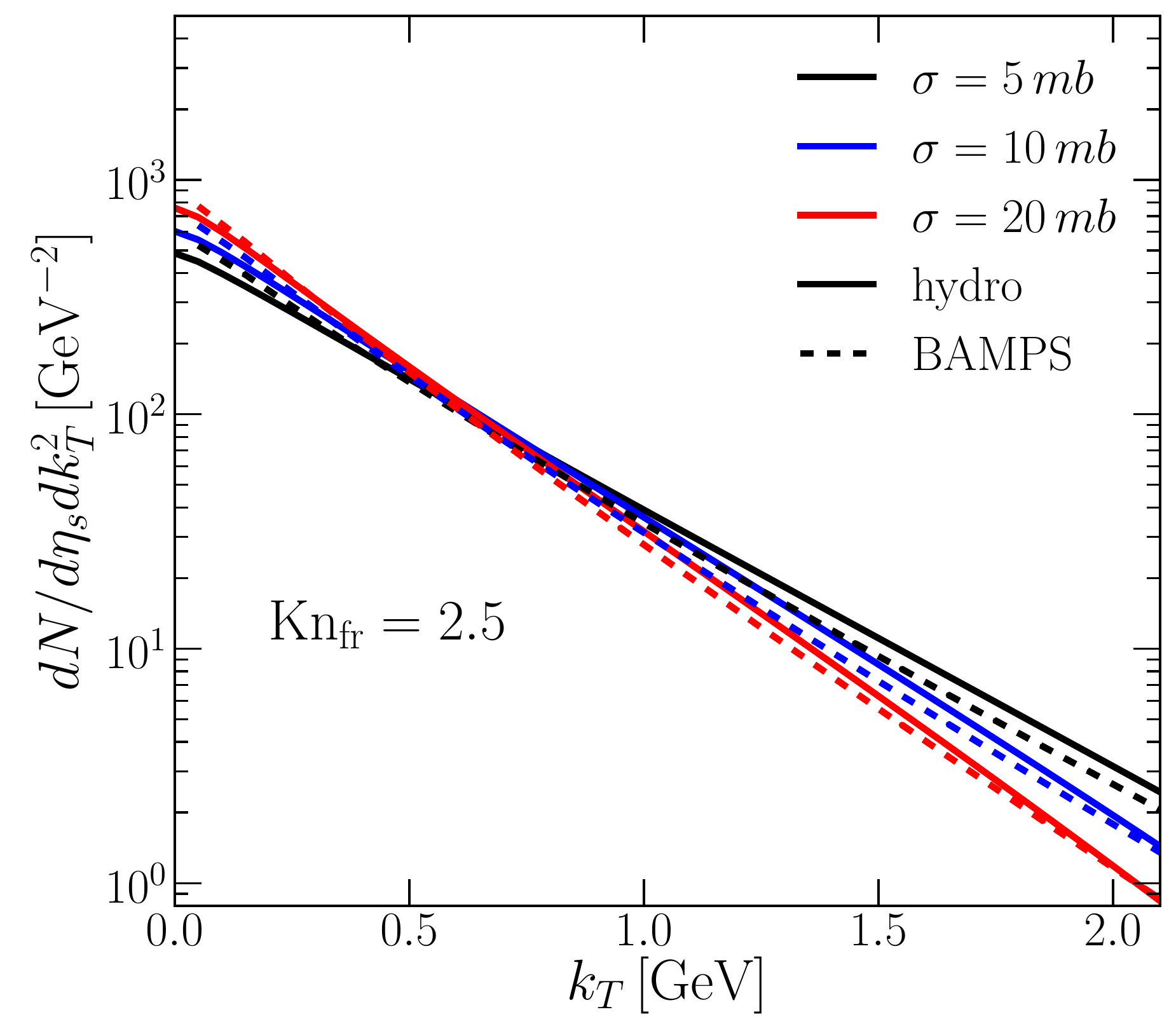}
\includegraphics[width = 0.45\textwidth]{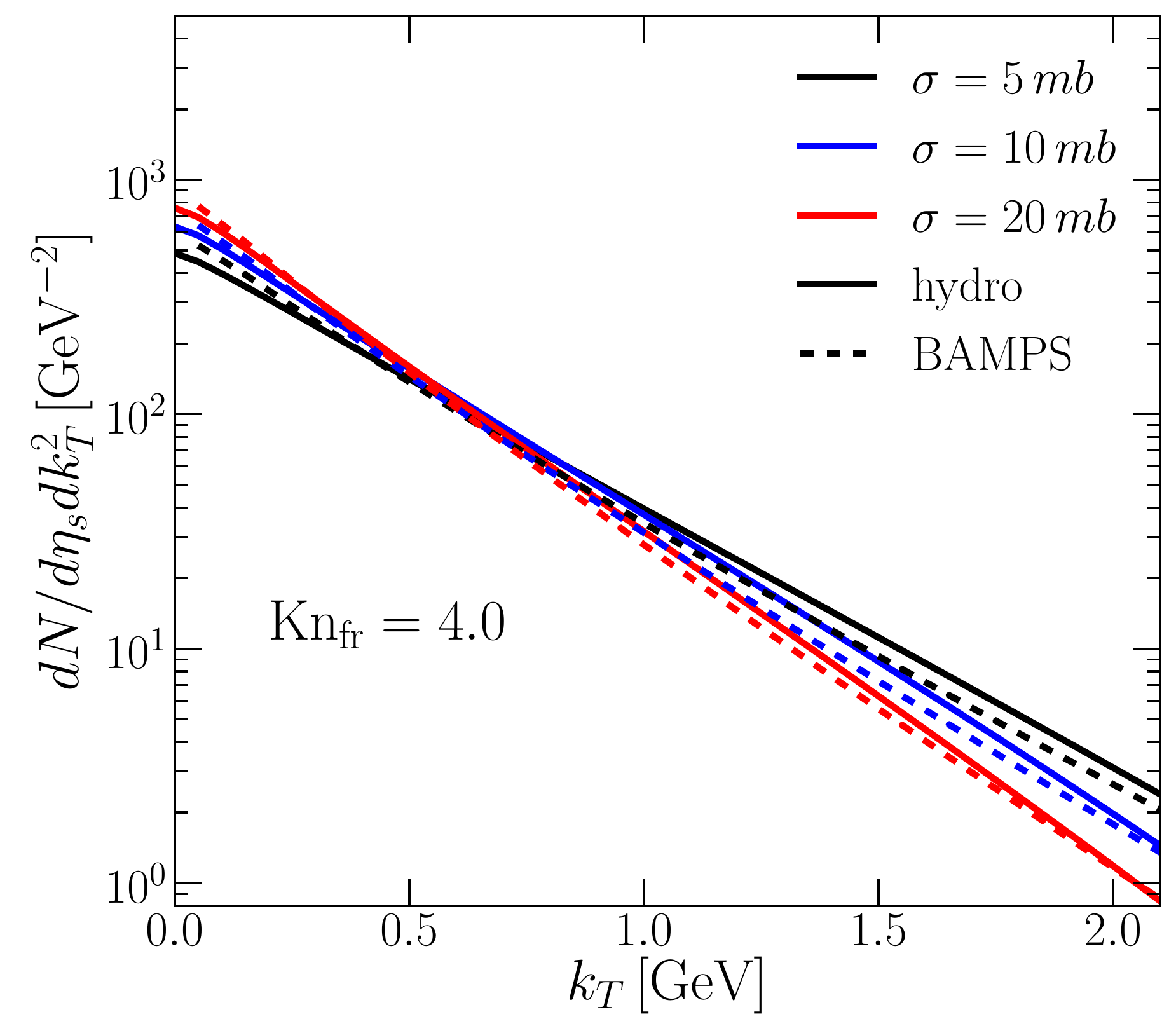}
\includegraphics[width = 0.45\textwidth]{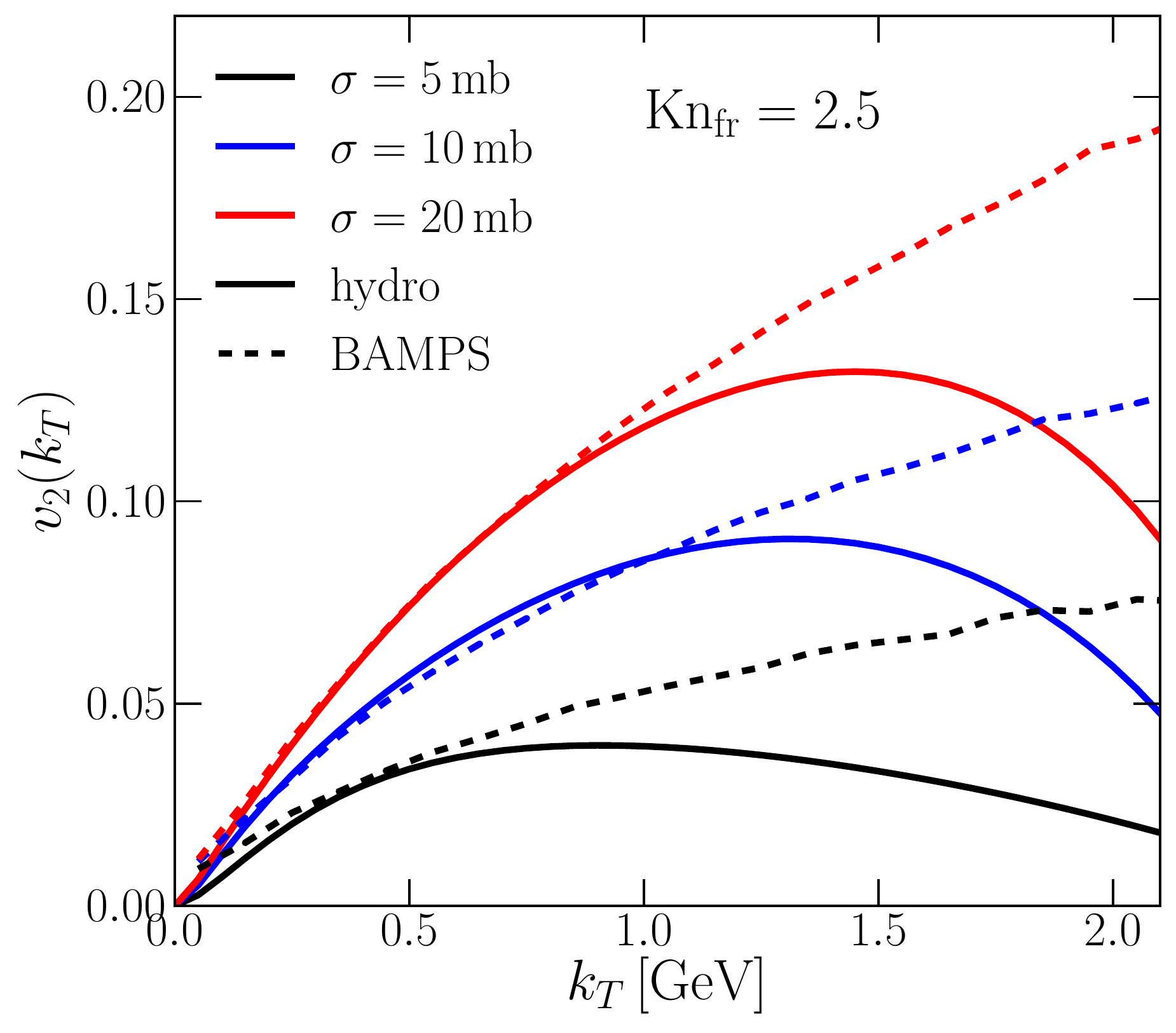}
\includegraphics[width = 0.45\textwidth]{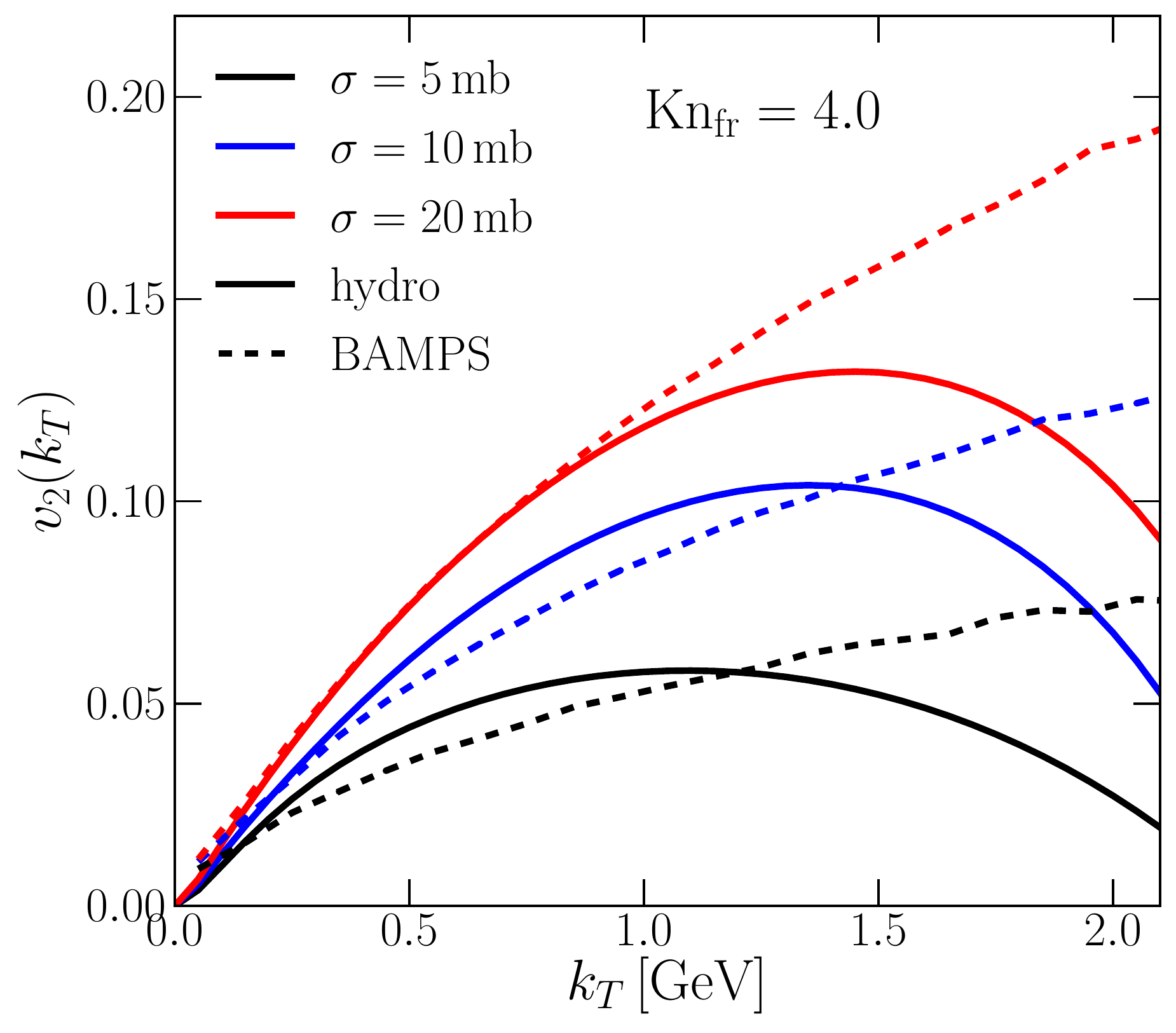}
\caption{(Color online) Transverse-momentum spectra (top panel) for a Glauber initial profile ('nBC') and $\sigma = 20$,
$10$, and $5\mb$. The freeze-out criteria $\rm{Kn_{fr}} = 2.5$ (left) and $4$ (right) are compared with each other.
The corresponding $v_2$ results are shown in the bottom panels. Solid lines represent fluid-dynamics results, while
the dashed lines show BAMPS solutions.}
\label{fig:spectra}
\end{figure*}

\begin{figure*}
\includegraphics[width = 0.32\textwidth]{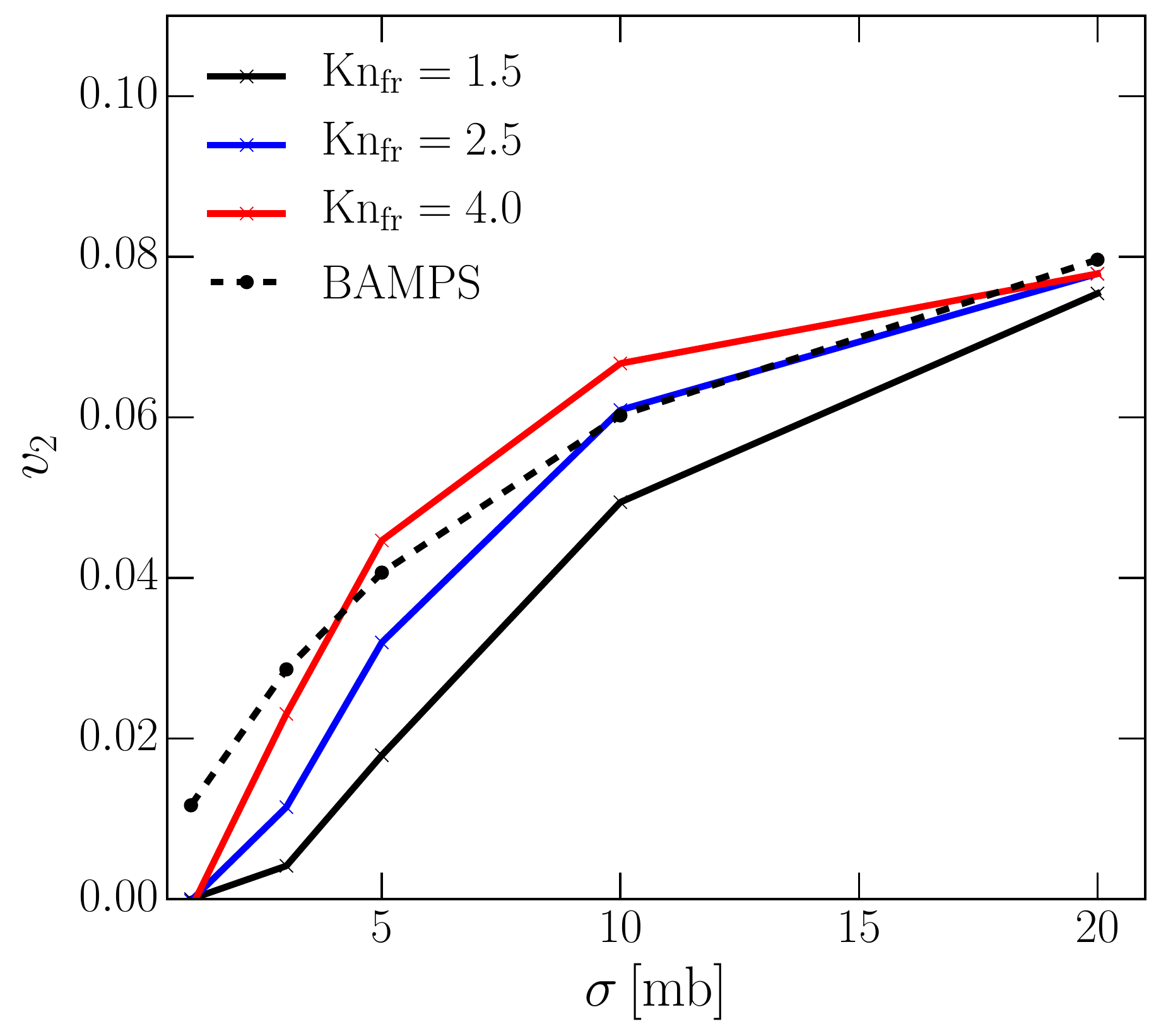}
\includegraphics[width = 0.32\textwidth]{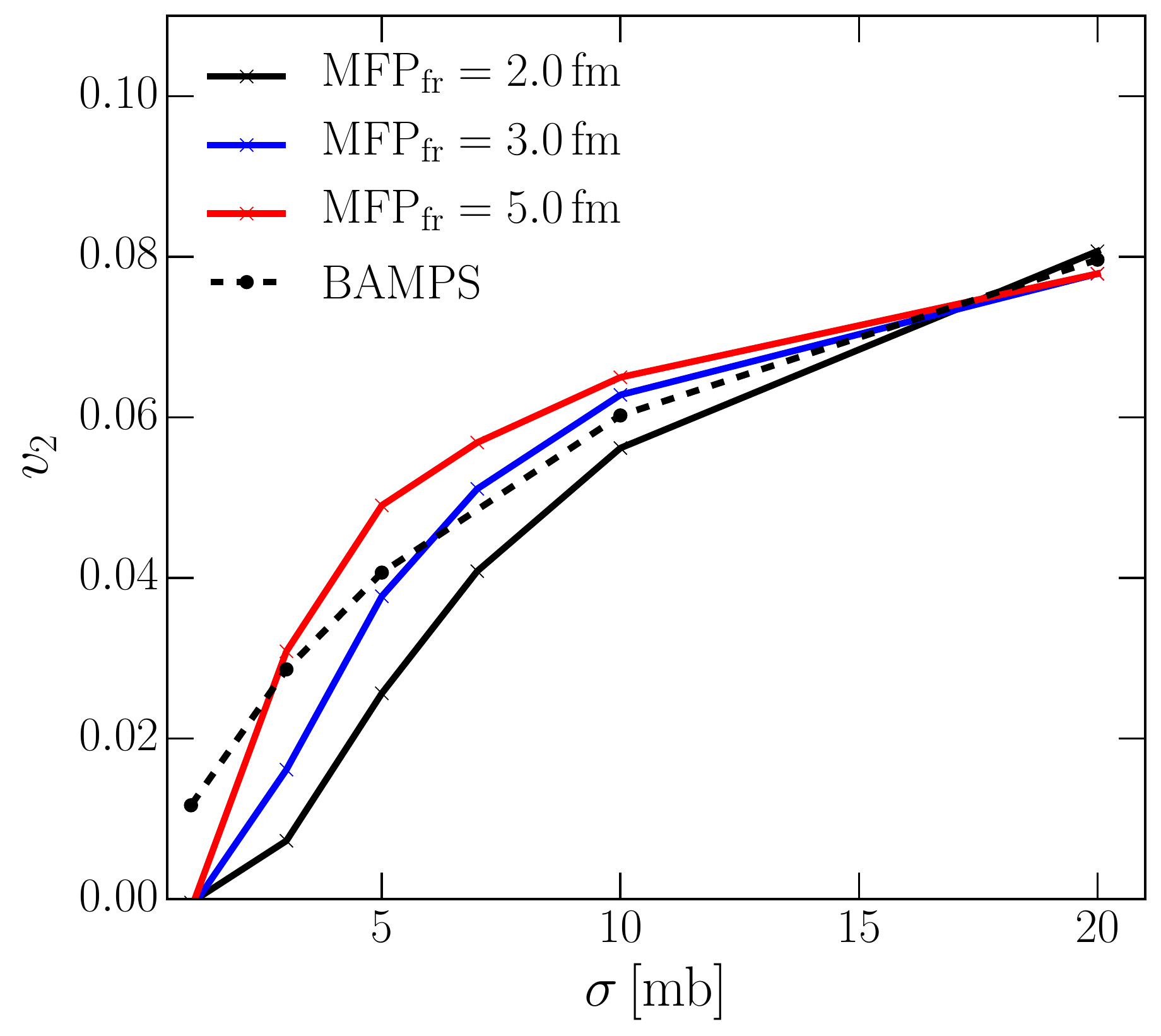}
\includegraphics[width = 0.32\textwidth]{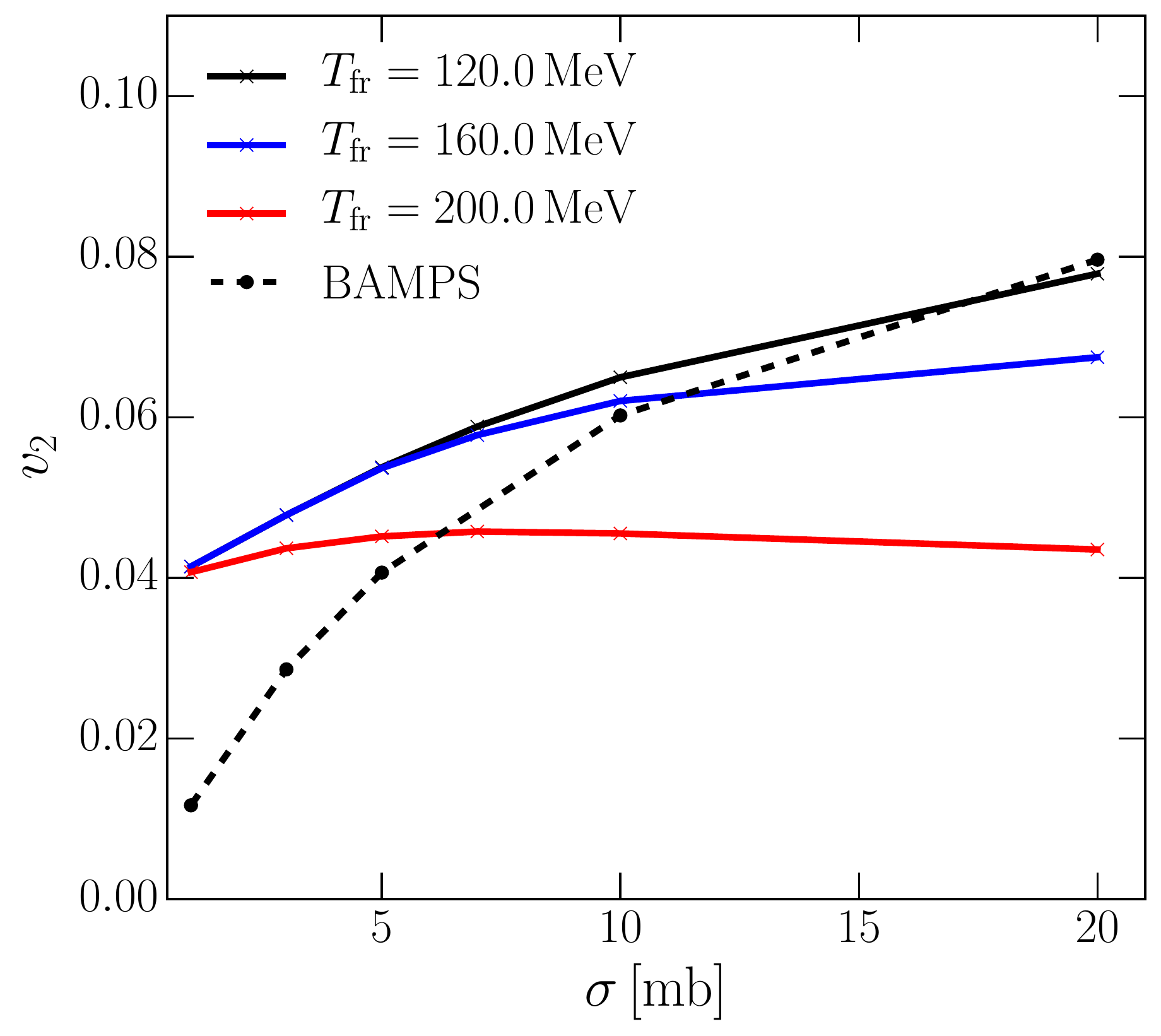}
\caption{(Color online) $k_T$-integrated elliptic flow $v_2$ for nBC initial conditions as a function of  the
cross section varying from $\sigma = 20$ to $1\mb$. The fluid-dynamical results are calculated with different decoupling
conditions: constant Knudsen number (left), constant mean free path (middle), and constant temperature (right). The
corresponding constants are shown in the legend. Solid lines represent fluid-dynamical results, while the dashed lines
show BAMPS solutions.}
\label{fig:v2integrated}
\end{figure*}

We have calculated both the azimuthally averaged transverse-momentum spectra, as well as the elliptic flow $v_2(k_T)$
with the nBC initialization using $\sigma = 20$, $10$, and $5\mb$. The resulting transverse-momentum spectra and
elliptic-flow coefficients are shown in Fig.~\ref{fig:spectra} for two different freeze-out criteria, $\rm{Kn_{fr}} = 2.5$ and
$4$. If the cross section is large, i.e., $\sigma = 20\mb$, a large part of the freeze-out is determined by the numerical
condition $N_{\rm test}=5$. For this reason, there is practically no effect of $\rm{Kn_{fr}}$ on the final results in this case.
The elliptic flow is well described up to $k_T \sim 1.0\GeV$, and particle spectra even up to higher values of $k_T$.
The average $k_T$ of the particles is around $0.58\GeV$, and the $k_T$-integrated $v_2$ is well described as well,
see Fig.~\ref{fig:v2integrated}.

If the cross section is reduced to $\sigma = 10\mb$, the physical freeze-out condition is starting to affect the elliptic flow,
while the spectrum is still insensitive to the choice of $\rm{Kn_{fr}}$. With a smaller cross section the final spectrum is
harder, and the average $k_T$ in this case is around $0.63\GeV$. We can still reproduce both the 
low-$k_T$ $v_2(k_T)$ and the
$k_T$-integrated $v_2$ by choosing the decoupling condition appropriately.

It turns out that if we further reduce the cross section to $\sigma = 5$ mb, we can choose the freeze-out condition to
reproduce either the low-$k_T$ part of $v_2(k_T)$, or the $k_T$-integrated $v_2$, but not both simultaneously. This is
already indicating a significant failure of the simple freeze-out description that is based on the 14-moment 
approximation Eq.(\ref{eq:deltaf}).
This situation is most prominent for larger momenta $k_T\gtrsim1.5\GeV$, where $v_2(k_T)$ shows a strong
unphysical decrease in the fluid-dynamical calculation.

We now investigate the possibility that the $v_2$ calculated from BAMPS results from the so-called escape mechanism
\cite{He:2015hfa}, where some azimuthal anisotropy is induced by particles which have not interacted at all on
their way out of the produced matter. Particles moving outwards have a smaller probability to interact than particles
which would first have to pass through the center zone. Thus a shadowing effect takes place, translating the spatial
asymmetry into an asymmetry in momentum space. In the partonic calculations, it is easy to track the number of collisions.
In the set-up given here, the overall probability that a particle leaves the medium without any interaction drops from
approx.~50\proz for $\sigma=1\mb$ down to 6\proz for $\sigma=10\mb$. For larger values of the cross section, the
probability of no interaction is negligible. In the cross-section interval used here, the average number of interactions of
the particles increases linearly from 0.7 to 7.3.

\begin{figure}
\includegraphics[width = 0.45\textwidth]{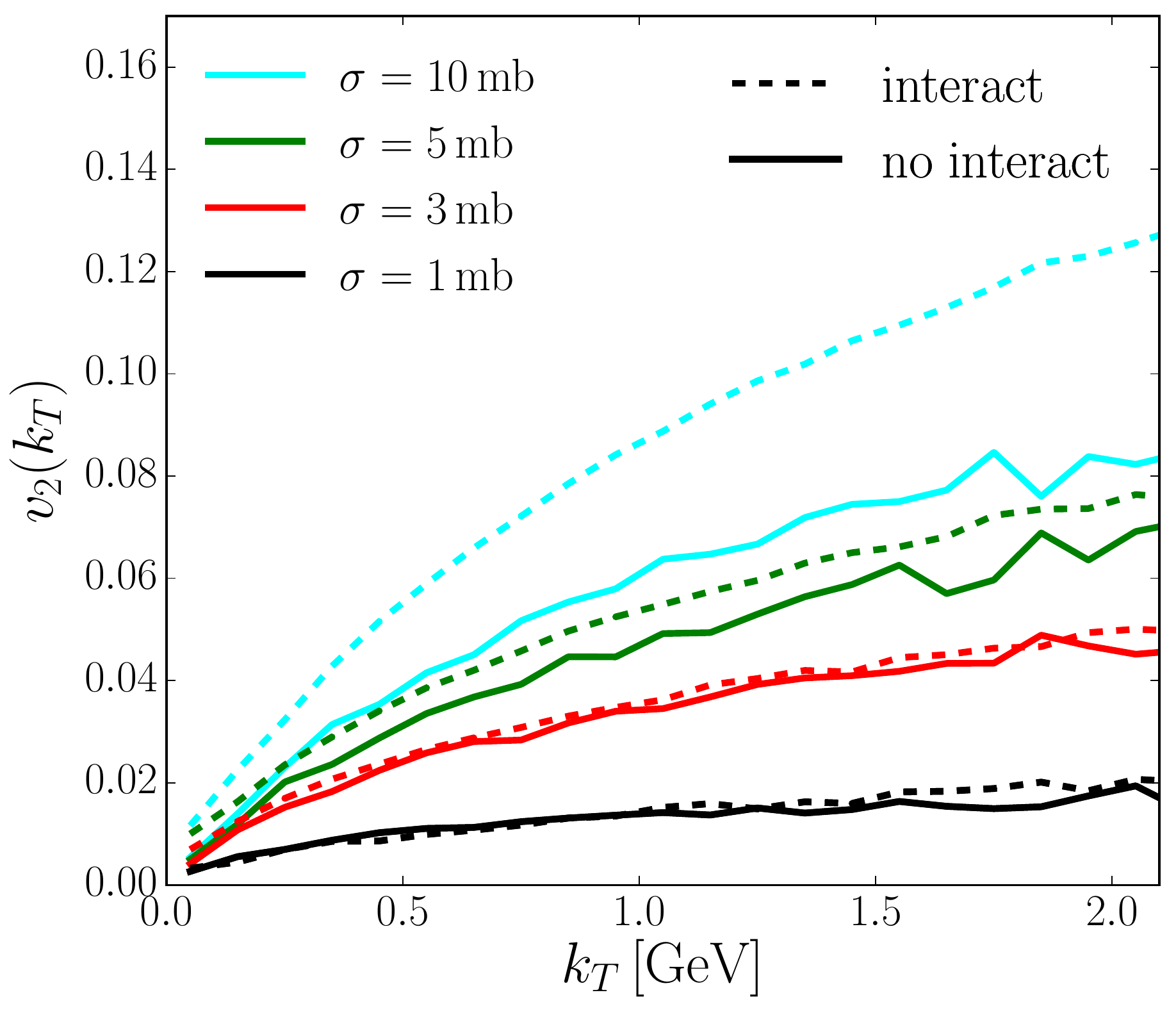}
\caption{(Color online) The $v_2$ values as function of $k_T$ for different cross sections. Solid lines show the results
for particles which have not undergone any interaction, while the dashed lines shows the values for all other particles.}
\label{fig:v2escape}
\end{figure}

In the fluid-dynamical picture the above mentioned escape mechanism is not properly accounted for, as there is no
explicit treatment of interacting particles. In order to prove that the mismatch between kinetic and fluid-dynamical results
for the flow is not induced by the escape mechanism, we show in Fig.~\ref{fig:v2escape} $v_2$ as a function of $k_T$
for different (small) values of the cross section, where we distinguish between particles which have undergone some
interactions and those which escape without any collision.
As can be seen, for $\sigma<5\mb$, these two particle classes have the same $v_2$. For larger cross sections,
the $v_2$ of particles suffering interactions becomes larger than that of the non-interacting particles. Nevertheless, the
relative importance of the latter particles decreases with increasing cross section. Thus the overall result for $v_2$ is
that of the interacting particles. Moreover, we have checked that for small cross sections there is no significant bias from
which spatial region the non-interacting particles are emitted. Only if the cross section is large, these particles are coming
mainly from the edge of the system, but as mentioned above, in this case their contribution to the total $v_2$ is also
small.

In addition, one realizes that there is no difference between the low- and high-$k_T$ ranges. Thus the non-interacting
particles cannot be the source for any mismatch between the kinetic and the fluid-dynamical results for $v_2$ at high
$k_T$. Nevertheless, for future studies of flow in smaller systems as in p+A or even p+p collisions, the escape mechanism
assumes a more prominent role, in the sense that one carefully has to study the question how the mean free path
compares to the length scales of the total system and also to the scale of fluctuations.

We can further analyze the freeze-out prescription in the fluid-dynamical calculations by considering several different
freeze-out criteria. To this end we show in Fig.~\ref{fig:v2integrated} the $k_T$-integrated $v_2$ as a function of the
cross section for constant-Knudsen number, constant-mean free path, and constant-temperature freeze-out. In principle,
if the freeze-out prescription corresponds to the real freeze-out mechanism in the BAMPS solution of the
Boltzmann equation,
we should be able to describe the BAMPS results for all values of the cross section where the fluid-dynamical limit is still
valid. As one can see from the figure, both the constant-Knudsen number and the constant-mean free path freeze-out
criteria give the correct behavior of the $\sigma$ dependence of the $k_T$-integrated $v_2$ for sufficiently
large cross sections.
Only when $\sigma$ drops below $5\mb$, we observe that fluid dynamics starts to fail to reproduce the BAMPS
results.

Here the constant-Knudsen number freeze-out can be thought of as the limit where freeze-out happens locally when fluid
dynamics ceases to be a valid approximation to the evolution. The constant-mean free path freeze-out can in turn be
thought of as a global condition: freeze-out happens when the particles become truly free, i.e., when the mean free path
becomes of the order of the overall size of the system. Both of them are physical freeze-out conditions in the sense that
they depend on the microscopic cross section. However, a constant-temperature freeze-out does not
correspond to any physical freeze-out mechanism, and as we can see in Fig.~\ref{fig:v2integrated}, it also fails to
reproduce the $\sigma$ dependence of the $k_T$-integrated $v_2$. The fact that we have a finite
$v_2$ for small cross sections for the fluid-dynamical calculation can be understood as follows. 
For these small cross sections, the constant-temperature freeze-out then happens at much
later times than the constant-Knudsen number freeze-out. At these large times, the produced elliptic flow is
large.

\begin{figure}
\includegraphics[width = 0.45\textwidth]{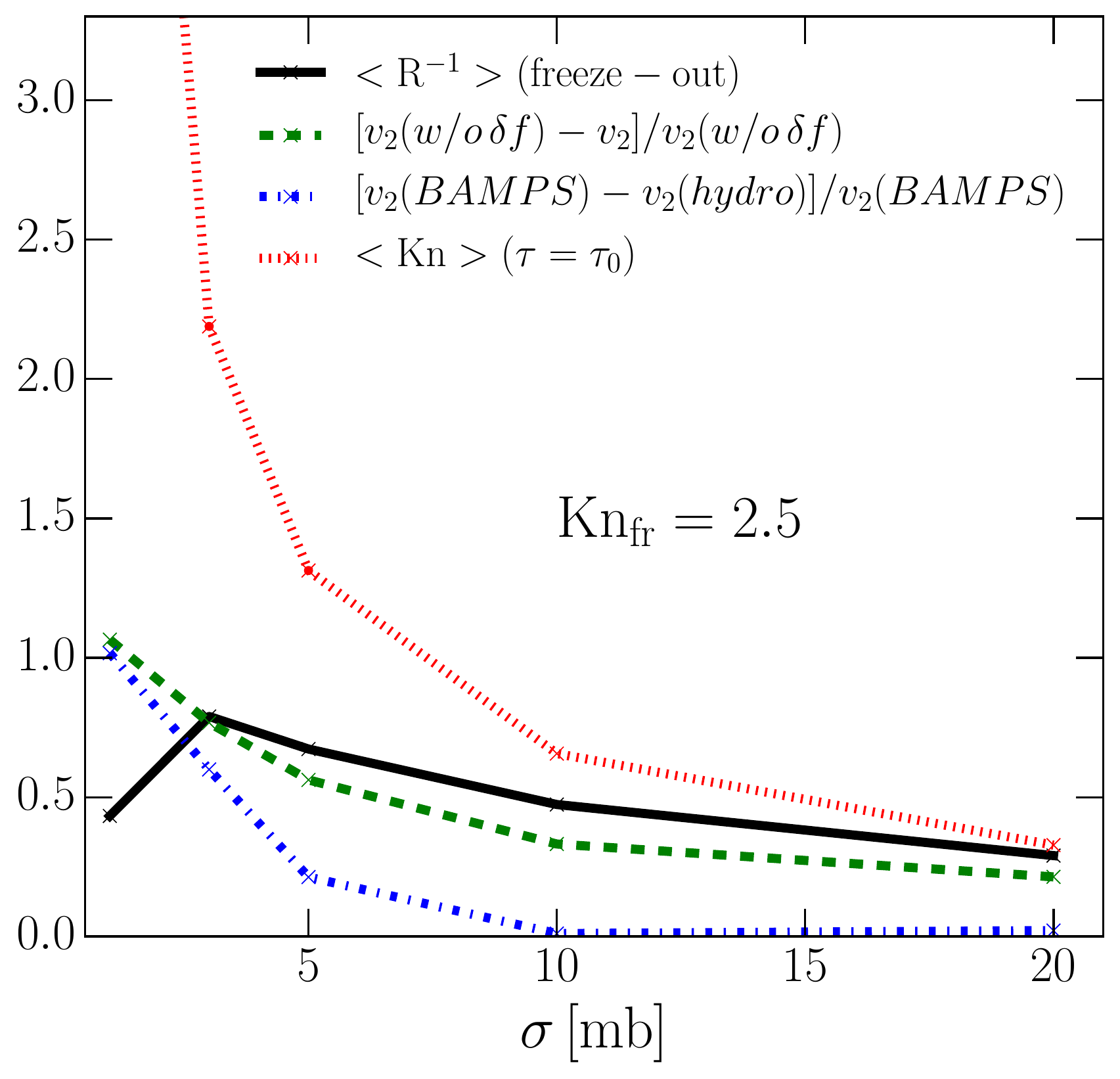}
\caption{(Color online) Average inverse Reynolds number on the freeze-out surface, average initial Knudsen number,
relative magnitude of $\delta f_{\mathbf{k}}$ in $v_2$, and relative difference of $v_2$ from fluid-dynamical and
BAMPS calculations.}
\label{fig:reynolds}
\end{figure}

Our main results are summarized in Fig.~\ref{fig:reynolds}, where we show the average inverse Reynolds number,
$\sqrt{|\pi^{\mu\nu}\pi_{\mu\nu}|}/p_0$, on the freeze-out boundary, and the average initial Knudsen number ${\rm Kn}$,
the magnitude of the $\delta f_{\mathbf{k}}$ contribution to $v_2$, and the relative difference of
$v_2$ between the fluid-dynamical
and BAMPS calculations as a function of cross section. The freeze-out condition is ${\rm Kn}_{\rm fr} = 2.5$.
The average inverse Reynolds number on the freeze-out surface is calculated by weighting it with the entropy flux
$s \Sigma_\mu u^\mu$ through the surface, in order to give those regions more weight from where most of the
particles are emerging. Similarly, the initial Knudsen number is calculated as an average over the transverse plane at
$\tau = \tau_0$ by weighting it with the entropy density $s(\tau, x, y)$.

Figure \ref{fig:reynolds} shows that the initial Knudsen number increases rapidly with decreasing
cross section, and exceeds one
between $\sigma = 10\mb $ and $5\mb$. This coincides with a similar sharp increase in the relative difference between the
fluid-dynamical and BAMPS calculations, indicating that the strict validity of fluid dynamics extends to ${\rm Kn} \sim 1$.
On the other hand, there is still a significant amount of $v_2$ generated between ${\rm Kn} = 1.5$ and $2.5$, as seen in
Fig.~\ref{fig:v2integrated}. This shows that ${\rm Kn} \sim 1$ is not a limit where the system decouples into free particles,
but in order to describe the dynamics of this type of events, the ${\rm Kn}=1 -  2$ phase needs still to be sufficiently
well described by fluid dynamics. With $\sigma = 5\mb$, most of the $v_2$ is generated during this phase, and fluid
dynamics can still give the correct $v_2$ within 20 \%.

Figure \ref{fig:reynolds} shows how the average inverse Reynolds number on the freeze-out boundary increases with
decreasing cross section. As a consequence also the relative contribution of $\delta f_{\mathbf{k}}$ corrections
to $v_2$ increases as well. The
region where the discrepancies between fluid dynamics and BAMPS start to appear, i.e., $\sigma = 5$ to $10\mb$,
is also a region where $\delta f_{\mathbf{k}}$ starts to contribute more than $50$ \% to the elliptic flow.
In our current set-up it is not
straightforward to study the applicability of fluid dynamics separately as a function of $\mathrm{R}^{-1}$ and
$\rm Kn$. However,
one can see from the figure that $\rm Kn$ and $\mathrm{R}^{-1}$ are independent dynamical quantities: the
Knudsen number stays
constant at the surface, but the Reynolds number varies with the cross section.

\section{Conclusions}
\label{sec:Conclusions}

In this work we addressed the question of the range of applicability of dissipative fluid dynamics to small systems
like those generated in p+p and p+A collisions. To this end, we solved the Boltzmann equation, using the
microscopic transport model BAMPS, for systems of massless particles
with boost invariance in longitudinal direction and different initial density distributions in the transverse plane.
The different scenarios included azimuthally symmetric Gaussian distributions with large (3\fm) and small (1\fm) widths,
and also an asymmetric distribution given by a thickness function provided by two overlapping Woods-Saxon distributions.
The evolution started in thermal and chemical equilibrium and was calculated for various values of the isotropic elastic
cross section from 20\mb down to 1\mb. For the same initializations, also Israel-Stewart type of dissipative
fluid-dynamical calculations, derived in the 14-moment approximation, were performed.

The ratio of microscopic over macroscopic length or time scales, the Knudsen number, is a critical measure for the
validity of a fluid-dynamical treatment. By varying the system size and the interaction cross sections in the way described
above, regions in space-time are probed where the Knudsen number ranges from values below 1 up to values
well above 1.

For a comparison between results from the microscopic and the fluid-dynamical treatment, one has to consider the
region of space-time where such a comparison is really meaningful. Technical reasons and 
computer-memory limitations within the BAMPS approach
restrict this region, because some numerical ``freeze-out'' is implicitly implemented by the choice of the
testparticle ansatz.

Comparing energy densities and fluid velocities, practically no differences can be seen between the two approaches,
even for small values of the cross section, even where Knudsen numbers are large. The components of the
shear-stress tensor turn out to be somewhat more sensitive to the Knudsen number. But also for small systems with
small cross sections, these profiles are still very similar.

In the case of azimuthally asymmetric initializations, additional quantities can be studied. The so-called momentum-space
eccentricity, $\varepsilon_p$, shows also a relatively small dependence on the Knudsen number, but already there more
significant deviations for a small cross section $\sigma \lesssim 5\mb$ can be seen. Overall, the situation is that if we
consider only the evolution of fluid-dynamical quantities, i.e., those appearing in $T^{\mu\nu}$ and $N^{\mu}$, fluid
dynamics is in an excellent agreement with the BAMPS calculations, even in the regions where the Knudsen number is
much larger than one. More sizable differences occur in regions where ${\rm Kn}\gtrsim 2 - 4$. 

A much more sensitive quantity is the transverse-momentum spectrum, and in particular the elliptic-flow coefficient $v_2$.
For these observables, the situation becomes more complicated, as the freeze-out has to be implemented separately
in fluid dynamics, taking also the numerical freeze-out in the BAMPS calculation into account. The
implementation requires additional modeling since the space-time evolution of the macroscopic fields ($T^{\mu\nu}$ and
$N^{\mu}$) needs to be converted back to the microscopic degrees of freedom. In this work, this conversion
was done within the 14-moment approximation to the local momentum distribution function.

The azimuthally averaged transverse-momentum spectra show very little sensitivity on the decoupling conditions,
and a good agreement between the two approaches is found. The main characteristics of the $k_T$-spectra is the
average $k_T$, which is generated quite early in the evolution. For the system considered here (massless particles),
further evolution practically does not affect the average $k_T$.

Elliptic flow, on the other hand, is strongly influenced by the decoupling condition. By using a constant-Knudsen number
decoupling criterion, ${\rm Kn}_{\rm fr} \sim 2.5$, we could well describe the BAMPS $v_2$ for large cross sections, and
observed a gradual break-down of the fluid-dynamical description when the cross section was decreased. A clear sign of
the failure of the fluid-dynamical approach is that, when the cross section was chosen too small, we could no longer find a
decoupling condition that would simultaneously describe the $k_T$-integrated $v_2$ and the low-$k_T$ part of the
$k_T$-differential $v_2(k_T)$. Furthermore, by varying the cross section the initial Knudsen number varies as well.
Thus we could infer that the strict validity of fluid dynamics extends up to ${\rm Kn} \sim 1$, but also observed that a
significant amount of $v_2$ can still be generated in the phase where ${\rm Kn} \sim 1 - 2$. Overall agreement with
 BAMPS still remained good, if this ${\rm Kn} \sim 1 - 2$ phase was sufficiently short compared to the whole evolution.

We also calculated the average inverse Reynolds numbers on the decoupling surface, and observed that they also
increase with decreasing cross section, and consequently the $\delta f_{\mathbf{k}}$
contribution to the elliptic flow grows larger, being
of order 50\proz{} when the fluid-dynamical description was starting to break down. In our set-up presented here, it was
not straightforward to study the influence of the Reynolds number and Knudsen numbers separately.
The influence of the Reynolds number on the applicability of fluid
dynamics could be tested by using non-equilibrium initial conditions,
but this is left for future work.

Obviously, the applicability of fluid dynamics depends on what kind of quantities we wish to calculate: for the
average $k_T$, and the space-time evolution of $T^{\mu\nu}$ the agreement was extending well beyond
${\rm Kn} \sim 1$, but for elliptic flow ${\rm Kn} \sim 1 - 2$ was the limit. In the future, we plan to extend this study
by considering more complicated and realistic initial conditions for A+A and p+A collisions that account for event-by-event
fluctuations of the initial densities. This would give access to higher harmonics beyond the elliptic-flow coefficient and offer 
the possibility to explore the question, whether the escape mechanism for smaller systems might indeed play a role 
for these harmonic-flow observables \cite{He:2015hfa}.
In order to improve the statistics, the reduction of the Boltzmann solver from (3+1) to (2+1) dimensions is necessary.
In addition, we want to emphasize that we will also investigate fluid-dynamical theories beyond the simple 
14-moment approximation, in order to systematically probe the applicability of different approximations.

\begin{acknowledgments}

The authors thank C.M.~Ko for reminding us of the escape mechanism.
This work was supported by the Bundesministerium f\"ur Bildung und
Forschung (BMBF), the Helmholtz International Center
for FAIR within the framework of the LOEWE program launched by the
State of Hesse, and by the Collaborative Research Center CRC-TR 211 ``Strong-interaction matter
under extreme conditions'' funded by DFG. Numerical computations have been performed at the
Center for Scientific Computing (CSC) at Frankfurt.
H.N.\ is supported by the European Union's Horizon 2020 research and innovation
programme under the Marie Sklodowska-Curie grant agreement no.\ 655285 and by the
Academy of Finland, project 297058. D.H.R.\ is partially supported by the High-end Foreign Experts project GDW20167100136 of the State
Administration of Foreign Experts Affairs of China. He greatly acknowledges the warm hospitality of the Department of
Physics of the University of Jyv\"askyl\"a, where part of this work was done.
\end{acknowledgments}

\end{document}